\documentclass{aa}  
\usepackage{txfonts}

\usepackage{graphicx}	% Including figure files
\usepackage{amsmath}	% Advanced maths commands
\usepackage{wrapfig}
\usepackage{multirow}
\usepackage{booktabs}
\usepackage{subcaption}
\usepackage{color}
\usepackage{longtable}
\usepackage{lscape}
\usepackage{multicol}

\usepackage{xspace}

\newcommand{\pks}{PKS\,0735+178\xspace}
\newcommand{\ic}{IceCube-211208A\xspace}

\begin{document}

   \title{The dynamics of the parsec-scale jet in the neutrino blazar \pks}

   \author{Yu-Sik Kim
          \inst{1}
          \and
          Jae-Young Kim\inst{1,2}\fnmsep\thanks{Corresponding author}
          }

   \institute{
   Department of Physics, Ulsan National Institute of Science and Technology (UNIST), Ulsan 44919, Republic of Korea \\ \email{jaeyoungkim@unist.ac.kr} \and
   Max-Planck-Institut f{\"u}r Radioastronomie, Auf dem H{\"u}gel 69, D-53121 Bonn, Germany
   }

   \date{Received; accepted}

% \abstract{}{}{}{}{} 
% 5 {} token are mandatory
 
  \abstract
   {Recent studies of individual track-like TeV-PeV IceCube neutrino events suggest that strongly jetted active galactic nuclei (AGNs), blazars, can be plausible sources of extragalactic high-energy neutrinos.
Although the broadband emission and neutrinos from such blazars can be modeled by hadronic jets with inverse Compton processes, various models show degeneracies. One of the reasons is the lack of high-resolution observations pinpointing the location and physical conditions of neutrino-emitting plasma.
Here, we present a very-long-baseline interferometry (VLBI) study of \pks that was recently associated with a high energy neutrino event \ic as well as alerts from other neutrino observatories.
We analyzed publicly available Very Long Baseline Array (VLBA) 15 and 43\,GHz data of \pks during 2020-2024, resolving the milliarcsecond-scale jet and tracing its time evolution in flux and structure, before and after the \ic event.
We find significant enhancements in the radio flux density ($\sim\times$ a few compared to its last quiescent state), apparent brightness temperature ($\sim\times10$), and synchrotron opacity at 15-43\,GHz of the VLBI nuclear region (spectral index $\alpha_{15}^{43}\sim-0.34$ to $\sim+0.07$) after \ic, strengthening the temporal correlation between the blazer and high-energy neutrino event. 
Furthermore, we find that the source ejected a new VLBI component, C2, from the VLBI core before the \ic event. C2 traveled further downstream at  $\sim4.2c$ apparent speed, which is close to the historical maximum speed for this object. 
C2 then passed a subluminally moving feature in the jet, C1, located at $\sim0.13\,$mas ($\sim0.77$\,pc) downstream the core at the time of \ic. 
The time of this apparent passage is statistically coincident with the time of \ic within 1$\sigma$ uncertainty, suggesting the location of this apparent passage to be a probable spatial origin of the \ic event.
Examination of the kinematic and emission properties of C1 and C2 reveal that it is difficult to find a single unique jet model that explains both the VLBI and broadband emission properties of \pks simultaneously.
Furthermore, we find that the large distance to C1 ($\sim$0.77\,pc) at the time of \ic makes it difficult to provide sufficient background radiation field for the photo-pion process in a single-zone jet model, making models that are less sensitive to the external radiation field more preferred.
}


   \maketitle
%
%________________________________________________________________

\section{Introduction}

   Neutrinos are fundamental, lightweight, and chargeless particles. 
   Therefore, astrophysical neutrinos can be observed with low opacity compared to electromagnetic observations at all other wavelengths and high-energy neutrinos can provide new insights into extreme physical processes in high-energy plasma that are otherwise inaccessible by other methods (see, e.g., \citealt{meszaros19,sharma24}). 
   So far, 275 individual TeV-PeV neutrinos of astrophysical origins have been reported by the IceCube Observatory \citep{icecube23}.
   In a limited number of cases, the counterparts of the neutrino events could be identified by temporal correlations based on monitoring and follow-up electromagnetic observations or spatial correlations using catalogs of astronomical sources at various wavelengths. 
   For example, the recent detection of the sub-PeV neutrino IceCube-170922A from the direction of a blazar TXS\,0506+056 and contemporaneous electromagnetic flares \citep{icecube18a,icecube18b} provided the first strong evidence that AGNs with especially powerful jets can be sources of extragalactic neutrinos.
   Also, studies employing statistical methods to probe the spatial and temporal correlations suggest that active galactic nuclei (AGNs) with strong jets, blazars, may be dominant sources of the high-energy neutrino events (e.g., \citealt{giommi20,plavin21,hovatta21}). 
   This suggests that the vicinity of supermassive black holes (SMBHs) or their powerful jets can provide the necessary particle acceleration, protons, and internal/external radiation fields for the production of multi-messenger (i.e., high-energy neutrinos and photons) signals (see, e.g., \citealt{dermer14,meszaros19} for a review).
   In December 2021, the IceCube observatory detected a 172\,TeV energy neutrino event \ic \citep{IceCubeColl2021} from the direction of a blazar \pks (SDSS J073807.39+174218.9, 4FGL J0738.1+1742) within the 90\% positional error region. 
   Similar high-energy neutrino events were detected close in time from other neutrino observatories (Baikal-GVD, \citealt{baikal21}; Baksan, \citealt{baksan21}; KM3Net, \citealt{km3net22}), which triggered follow-up observations of contemporaneous electromagnetic flares at optical, X-ray, and $\gamma-$ray energies (e.g., \citealt{sahakyan23}). All these results suggest that \pks is the most likely origin of the \ic event.
   Theoretical modeling of the observed broadband spectral energy density (SED) of \pks suggests that external inverse Compton scattering of incident photons by a lepto-hadronic jet best explains the neutrino and electromagnetic flares, compared to purely synchrotron self-Compton models \citep{acharyya23,prince24,bharathan24}.
   However, these studies, based on the source-integrated flux measurements, present common difficulties in finding a unique scenario of high-energy neutrino production. 
   More specifically, the studies show that purely leptonic jets can also well explain the electromagnetic spectrum. Also, hadronic jets often produce a too-large total jet power that exceeds the Eddington limit.
   The exact locations of the neutrino-emission regions and properties of seed photons are additional factors of uncertainty.
   Last but not least, the exact values of viewing angle, speed, particle energies, contents of the jet, and particle acceleration mechanisms during the \ic event are not yet well studied.
   For more details, we refer to discussions in \cite{sahakyan23,acharyya23}.

   In order to help resolve these ambiguities, here we conduct a study of analyzing time-series radio very-long-baseline (VLBI) data of \pks obtained during 2020-2024, including the time of \ic.
   Radio VLBI observations, thanks to their superb angular resolution, have played unique roles in identifying $\gamma$-ray emitting regions in jets of AGNs and tracing their flux and structural time evolution (see, e.g., \citealt{kim18a,kim20,paraschos23,kim23,jeong23,traianou24}).
   Also, high-resolution VLBI imaging can resolve unique fine-scale structures of the jets, such as large opening angle, edge-brightening, and jet curvatures (e.g., \citealt{pushkarev09,ros20,kim20}), which may be responsible for the generation of high-energy $\gamma-$ray emission as well as neutrinos (see also \citealt{ros20,britzen21}).
   In light of these advantages, we aim to pinpoint the exact location of the neutrino emission potentially related to \ic in \pks and constrain the structural, dynamical, and emission properties of the neutrino-emitting plasma blob. 
   Also, we investigate the evolution of multi-wavelength light curves of \pks in search of their possible temporal correlation with the VLBI-scale jet structural evolution and neutrinos on longer timescales than previously reported in the literature.

   This paper is organized as follows.
   In \S\ref{sec:data}, we describe the multi-messenger and radio VLBI data used in our work.
   In \S\ref{sec:analysis}, we present our analysis, especially focusing on the VLBI data. 
   The major results from our work are shown in \S\ref{sec:results}.
   We then discuss the physical implications of our findings in \S\ref{sec:discussions} and provide conclusions in \S\ref{sec:conclusions}.
   Throughout this paper, we adopt a redshift $z=0.42$ from \cite{nilsson12}, while other estimates of $z$ as large as $z\sim0.65$ are also reported \citep{falomo21}.
   We also assume a cosmology with $H_{0}=67.8\rm\,km\,s^{-1}\,Mpc^{-1}$, $\Omega_{m}=0.31,$ and $\Omega_{\Lambda}=0.69$ \citep{planck14}, which yields a 5.53\,pc/mas conversion scaling factor at the redshift of \pks.
%__________________________________________________________________

\begin{figure*}[t]
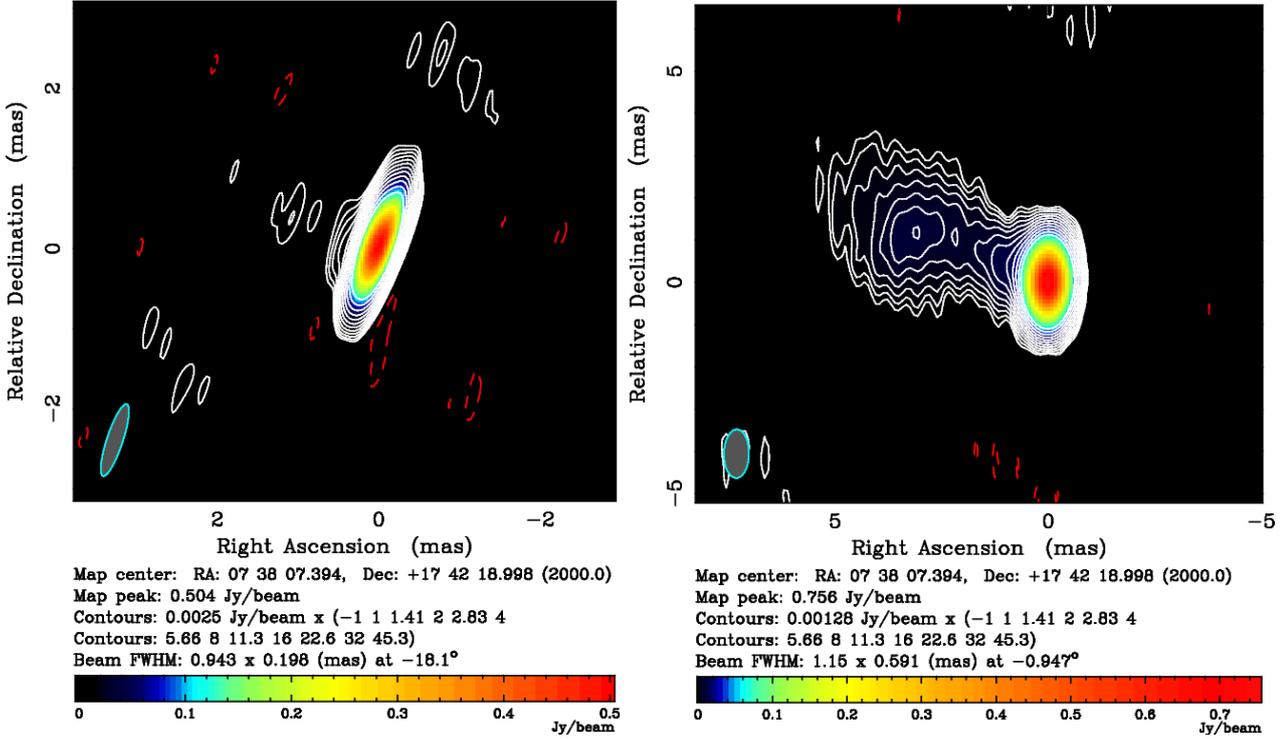

    \includegraphics[height=0.53\textwidth]{figs/BU.pdf}
    \includegraphics[height=0.53\textwidth]{figs/MOJAVE.pdf}
    \caption{VLBI images of \pks on 2021-12-10 and 2021-12-14 at 15 (left) and 43\,GHz (right), respectively.
    }
    \label{fig:representative_images}
\end{figure*}

\section{Multi-messenger Data of \pks}
\label{sec:data}

\subsection{Neutrinos from various neutrino observatories}
\label{subsec:neutrino}

   The IceCube observatory has issued an initial alert, \ic, of 172\,TeV energy on 2021 Dec 08 at UT 20:02:51.1 or MJD\,59556.84 \citep{IceCubeColl2021}, whose 90\% positional error region (statistical) marginally contained \pks.
   Afterward, multiple neutrino observatories reported additional detections of $\sim$GeV-TeV neutrinos after \ic, on delay timescales of a few hours to days \citep{baikal21,baksan21,km3net22}.
   In this study, we focus specifically on \ic and its arrival timing as a representative of all the neutrino events, because all these events occurred within a timescale of a week while our main interest is to discover weekly to monthly evolution of typically more slowly variable compact radio jet. 
   We refer to the literature (e.g., \citealt{sahakyan23,acharyya23,prince24}) for more details of the various lower-energy neutrino detections related to \ic and \pks.

\subsection{$\gamma$-rays from \textit{Fermi}/LAT}
\label{subsec:gamma}

   We made use of $\gamma-$ray data of \pks available from the Fermi-LAT Lightcurve Repository (LCR\footnote{\url{https://fermi.gsfc.nasa.gov/ssc/data/access/lat/LightCurveRepository/index.html}}; \citealt{abdollahi23}). 
   In the repository, we extracted weekly-averaged $0.1-100$\,GeV band $\gamma-$ray photon counts (in units of ph\,cm$^{-2}$\,s$^{-1}$) from the direction of the source 4FGL\,J0738.1+1742 \citep{4FGL} located at RA=$114.539^{\circ}$ and Dec=$17.707^{\circ}$, using minimum detection significance of TS=4 and fixed $\gamma-$ray photon index.
   The entire Fermi-LAT lightcurve of \pks spans a time range of $\sim2009-2024$. 
   The minimum, maximum, and median time cadences of observation are $\sim6$, $\sim28$, and $\sim7$ days, respectively.

\subsection{X-rays}
\label{subsec:x-ray}

   We have obtained $0.3-10.0$\,keV and $1.0-4.5$\,keV X-ray count rates, in the unit of counts\,$s^{-1}$, measured by the Neil Gehrels Swift Observatory \citep{gerhels04} around the timing of \ic.
   These data were partly taken from Table 1 of \cite{sahakyan23}; see references therein for the details of the data acquisition and processing.
   We have also obtained additional $0.3-10.0$\,keV X-ray count rates, also in counts\,$s^{-1}$, from the Swift X-Ray Telescope Monitoring of Fermi-LAT Gamma-Ray Sources of Interest program\footnote{\url{https://www.swift.psu.edu/monitoring/}} \citep{stroh13}.
   The minimum, maximum, and median cadences of the observations are $\sim0.075$\, $\sim3385$, and $\sim$7.23 days, respectively.

\subsection{Optical}
\label{subsec:optical}

   Flux densities of \pks at the optical $g$ and $V$ bands were obtained from a publicly available database from the All-sky Automated Survey for Supernovae (ASAS-SN) program\footnote{\url{https://www.astronomy.ohio-state.edu/asassn/}} \citep{shappee14,kochanek17}.
   The optical flux measurements of \pks span MJD\,55959-58452 (2012-02-01 to 2024-05-29), with minimum, maximum, and median cadences of $\sim0.8$, $\sim340$, and $\sim3$\,days.

\subsection{Radio VLBI}
\label{subsec:radio}

   We have obtained publicly available radio VLBI data of \pks from the 
   Monitoring Of Jets in Active Galactic Nuclei with VLBA Experiments (MOJAVE\footnote{\url{https://www.cv.nrao.edu/MOJAVE/}}; \citealt{mojave})
   and the 
   Blazars Entering the Astrophysical Multi-Messenger Era (BEAM-ME\footnote{\url{https://www.bu.edu/blazars/BEAM-ME.html}}; \citealt{jorstad16,jorstad17,weaver22}) programs. 
   They both provide fully calibrated VLBI data (both visibilities and images) from the Very Long Baseline Array (VLBA) observations at 15 and 43\,GHz, respectively.
   As for the images, we obtained the FITS map data of \pks at both 15 and 43\,GHz from the above databases and used the attached Clean Component (CC) table to extract the flux densities of the extended jet and the compact nuclear regions.
   We note that these CLEAN images were produced without strong prior conditions on the source size, and they account well for the total flux densities in the mas-scale jet constrained by the visibility amplitudes of the shortest baselines \citep{lister18,jorstad17}.
   As for the visibilities, we mainly used the calibrated UVFITS data at 43\,GHz and did not include the 15\,GHz observations, in order to analyze the time evolution of the source at the highest possible resolution by the visibility model-fitting technique.   
   Details of these analyses are described in \S\ref{sec:analysis}.
   To understand the long-term evolution of the jet until recent epochs in 2024, we have collected the 43\,GHz data from 2007 Jun 13 to 2024 Aug 31. Similarly, the 15\,GHz image data are obtained for a period of 2007 Mar 02 to 2024 May 17.

\begin{figure}[t]
\includegraphics[width=\columnwidth]{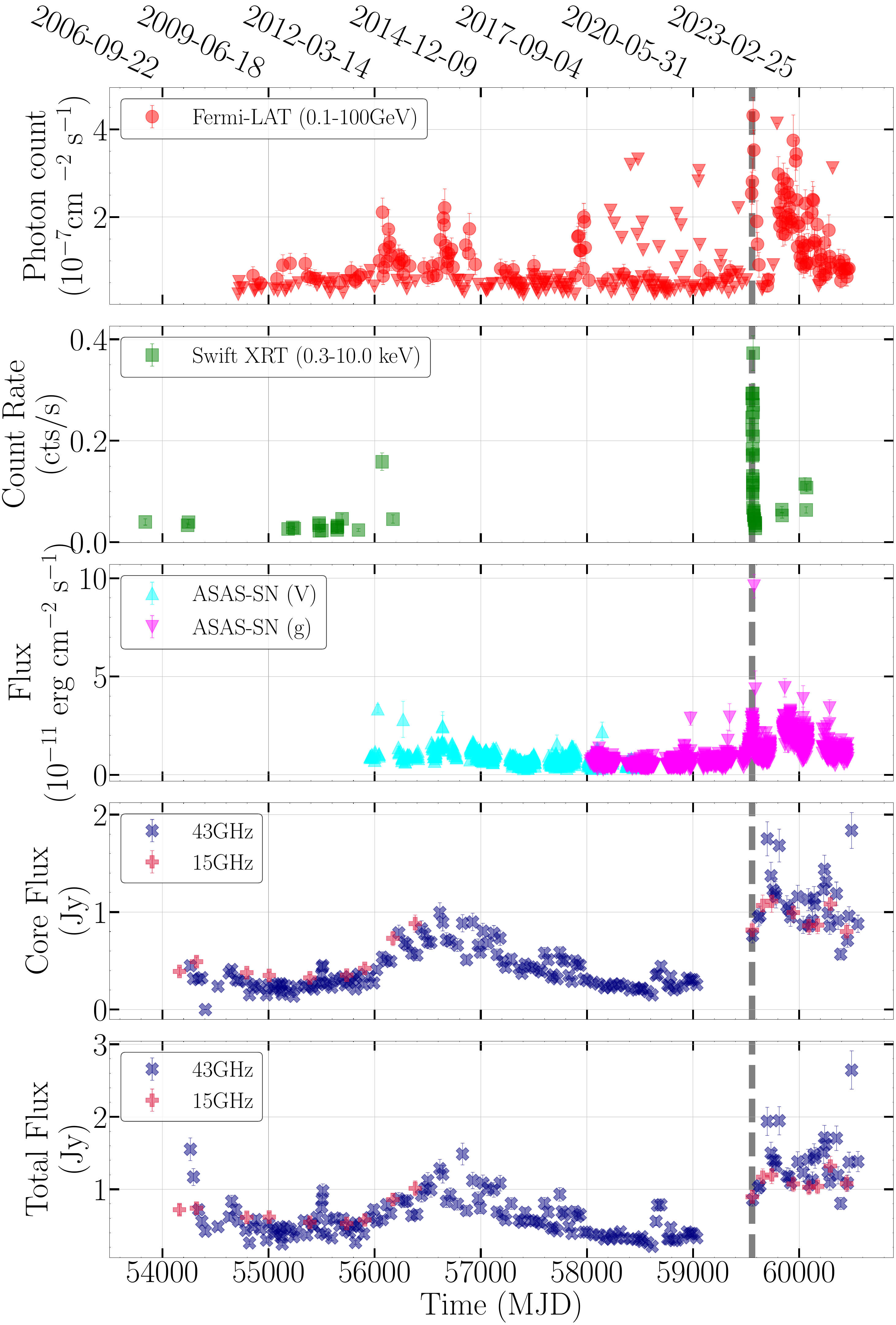}
\caption{
Multi-wavelength light curves of \pks.
From top to bottom: 
0.1-100\,GeV $\gamma-$ray photon flux from Fermi-LAT,  
0.3-10\,keV photon count from Swift/XRT,
optical $g$ and $V$ band flux densities from ASAS-SN, 
15 and 43\,GHz flux densities of the VLBI nuclear region from the VLBA (MOJAVE and BEAM-ME; see the text for the definition of the core), and the same but for the entire compact jet.
Vertical grey dashed lines mark the time of the \ic event.
}
\label{fig:lightcurve}
\end{figure}

\begin{figure}[t]
	\includegraphics[width=\columnwidth]{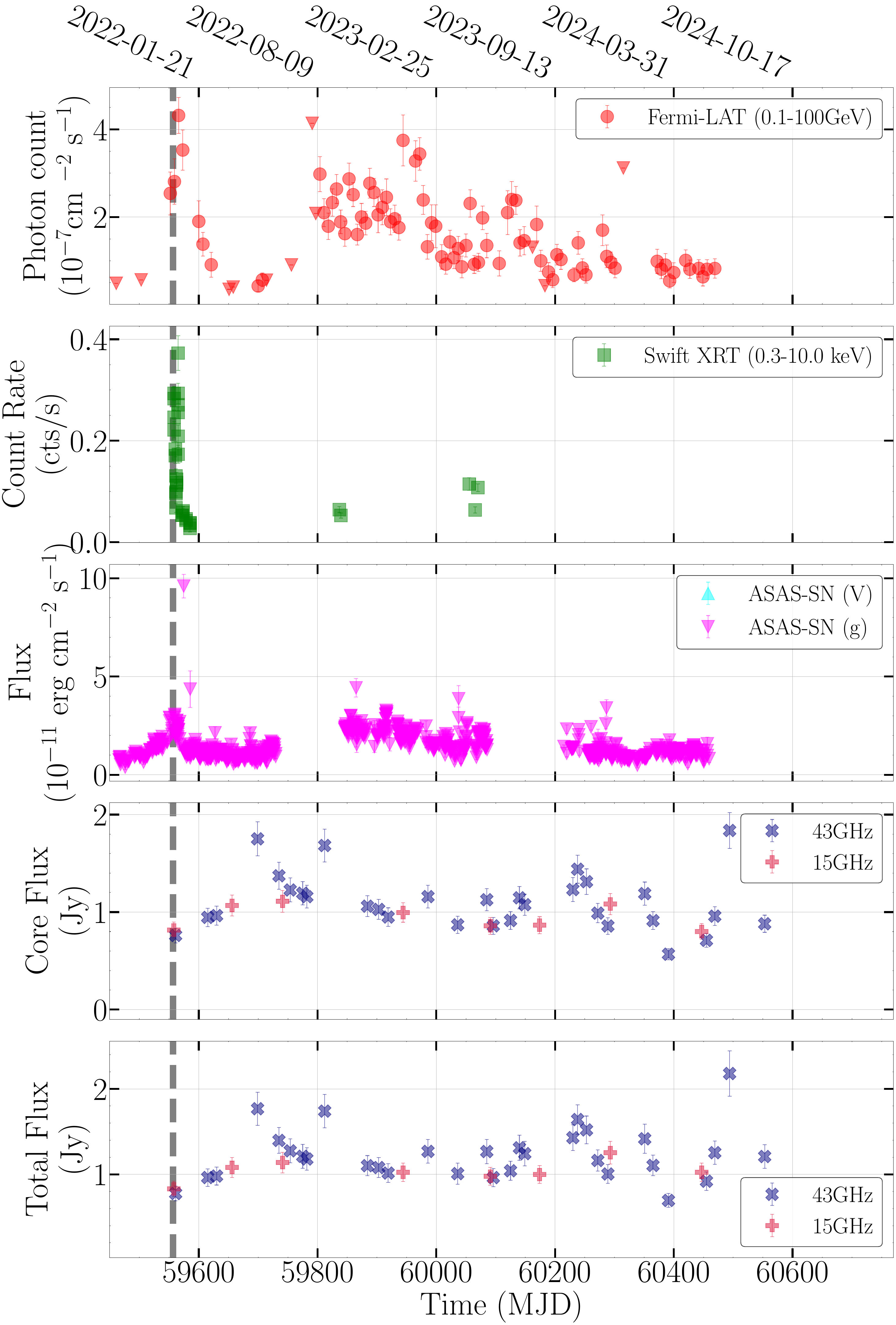}
    \caption{
   The same as Fig. \ref{fig:lightcurve} but highlighting more recent flux measurements after around MJD\,59400.
    }
    \label{fig:lightcurve_zoom}
\end{figure}

\section{Analysis of the VLBI data}
\label{sec:analysis}

\subsection{Image analysis}\label{subsec:image_analysis}
First of all, we defined the nuclear region and extended jet and measured their flux densities as follows.
At both frequencies, the nuclear region was defined as a circular region extending up to 0.4\,mas in radius (thus 0.8\,mas diameter) from the peak of the intensity in the image. 
The remaining extended jet is then defined as an extended region from 0.4 to 1.0\,mas distances from the peak of the intensity.
We chose the 0.4\,mas radius for the nuclear region because this is comparable to half of the VLBA beam size at 15\,GHz in the N-S direction.
Also, the 1.0\,mas distance cutoff for the extended jet was chosen based on visual inspections of the images of the jet emission over $>3\sigma$ image noises at 43\,GHz and to avoid including noisy CLEAN components close to the noise level in the accurate flux calculation.
For each region, we integrated the flux densities of the provided CLEAN components inside those areas and obtained representative flux densities at 15 and 43\,GHz.

To verify if the extended jet flux is largely affected by the somewhat arbitrary choice of the 1.0\,mas distance cutoff, we also computed the total flux densities using a larger jet size, 2.0\,mas. We find only an average increase in the total flux density of $\sim50\,$mJy for all the epochs, which is $\lesssim5$\% of change in the brightness of the source after \ic for the $>1$\,Jy source flux. The 5\% is also smaller than $\sim10$\% systematic flux uncertainties that we assumed in the later analysis. Thus, we consider that systematic uncertainties due to the cutoff in the jet size determination are negligible.

From these flux measurements, we also computed the 15-43\,GHz spectral index of the nuclear region, $\alpha_{15}^{43}$ ($S_{\nu}\propto \nu^{\alpha}$ where $S_{\nu}$ is the flux density and $\nu$ is the observing frequency).
Because of lack of VLBA 15\,GHz data around MJD\,56500--59500, we computed values of $\alpha$ for two different periods; MJD\,54000--56500 (well before \ic) and MJD\,59500--60000 (after \ic).
To align the timestamps of the 15 and 43\,GHz data, we time-averaged the flux measurements with time bin sizes of 295 and 123\,days for the earlier and later periods, respectively.
These bin sizes represent the mean cadences of the VLBA 15\,GHz observations in each period, which are more sparse in time than the 43\,GHz observations.
We then computed the spectral index as $\alpha_{15}^{43}=\log(S_{\rm 43\,GHz}/S_{\rm 15\,GHz})/\log(43/15)$, where $S_{\rm 15,43\,GHz}$ are the binned flux densities at 15 and 43\,GHz, respectively.
To estimate the uncertainties of $\alpha_{15}^{43}$, we assumed a conservative systematic 10\% flux uncertainties in $S_{\rm 15,43\,GHz}$ following \cite{mojave,jorstad17,weaver22}.
The final uncertainties of $\alpha$ were then obtained by the error propagation.

We also note an additional source of systematic uncertainty in $\alpha$; by setting the same peak-centered masks at both frequencies to compute the spectral indices at 14-43\ GHz, we implicitly assume the effect of possible relative shifts between 15 and 43\ GHz images to be small.   
In contrast, there could be a non-negligible frequency-dependent shift of the peak positions in the VLBI images of AGN jets, due to the opacity effect close to the base of the jet (i.e., ``coreshift''; \citealt{lobanov88}), which can be expected for a conically expanding relativistic jet \citep{blandford79}. 
To verify the presence and impact of the opacity effect in our calculations of the spectral indices, we estimated the relative offsets of the peak positions in the 15 and 43\,GHz images for several representative epochs before and after the \ic event, when there are contemporaneous dual-frequency VLBI datasets of the source within less than two weeks of time separation.
To estimate the amount of image alignment for each pair of 15 and 43\,GHz images, we modeled the visibilities of the 15 and 43\ GHz datasets using the Modelfit task in Difmap software (\citealt{shepherd97}; see also \S\ref{subsec:vis_fitting} for more details) and cross-identified extended jet features in both frequency datasets based on the jet morphologies. We then computed the spectral indices for each feature and used the centers of the most optically thin jet feature as the astrometric reference point. Confidence for the low opacity of such features were also obtained by their large fractional linear polarization of $\gtrsim10\%$.
By aligning the dual-frequency images, we found maximum relative offsets of the peak positions of $\lesssim0.15\,$mas out of six epochs before and after the \ic event.
This is approximately $\lesssim19$\,\% of the 0.8\,mas-diameter size, which is a small fraction. 
In order to quantify the impact of this small offset in the spectral indices of the nuclear region, we also added to the west direction $0.15$\,mas extra position shifts in the centers of the 0.8\,mas-diameter circular masks at 15\,GHz. We then repeated the same steps as above to compute revised spectral indices.
Comparing differences in the spectral indices with and without the shift of the 15\,GHz mask center positions, we found the mean value of the residuals to be $<\delta(\alpha_{15}^{43})> \sim0.02$.
This difference is significantly smaller than typical uncertainties $\alpha$ we estimated and was added in quadrature to the total error budget on $\alpha$.

\subsection{Gaussian source modeling}\label{subsec:vis_fitting}
In order to measure more detailed properties of the jet (e.g., brightness temperature, proper motions of the VLBI jet components), we primarily focused on the VLBA 43\,GHz data, because they were observed more regularly and also offered higher angular resolution to track motions close to the VLBI core. 
The whole model-fitting analysis was conducted as follows.
For each single epoch, we modeled the VLBA 43\,GHz visibilities using multiple circular Gaussian components in the visibility domain, where each component was described by four parameters, namely the flux density, $S$, distance from the core, $r$, the position angle, $PA$, and the full width at half maximum (FWHM), $d$. For the actual fitting, we used the Modelfit task implemented in the Difmap software \citep{shepherd97}, setting initial component model parameters by visual inspection of the dirty map and applying the Modelfit task iteratively until convergence is reached. 
We inspected the fit quality by computing the visibility-domain $\chi^{2}$ and visually inspecting the fit quality in the visibility amplitudes and phases, and also by searching for any significant residual flux over 3$\sigma$ level in the residual map. If significant extra emission was found from the inspection, additional Gaussian components were added and fitted again until no remaining flux at $>3\sigma$ level was found.
Also, following previous studies using the VLBA 43\,GHz data (e.g., \citealt{jorstad17}), we have fixed the minimum FWHM size of a component to be 1/5 of the geometric mean of the beam size, if the Modelfit task numerically enforced unrealistically small model sizes.

To trace the time evolution of the individual jet components, we needed to label the fitted VLBI components across various epochs with unique IDs.
The entire epochs were model-fitted following the steps outlined below,
 in order to reduce possible systematics in the labeling.
First, we manually model-fitted the first few epochs in order to obtain a reliable description of the global core and jet structure.
Then, we transferred the converged Gaussian components of a previous epoch to the next one as a new starting model for the iterative Modelfit task. 
By doing so, the total number of components was kept the same in adjacent epochs and the Modelfit task essentially fitted time-changes of the model parameters. 
This approach can reduce potential systematic uncertainties that could arise due to changes in the $(u,v)-$coverages and sensitivities from epoch to epoch
and also help more robust cross-identification of VLBI components across different epochs (see also other studies, e.g., \citealt{jorstad17}).
We note that if there would be a sudden and substantial structural change of the source occurs in certain epochs, for instance when a new VLBI component appears from the VLBI core, additional Gaussian components should be incorporated in the model-fitting. Such features were typically bright and clearly showed up in the above inspection procedures.  
We also note that sometimes, especially when the $(u,v)-$coverage is quite limited, we were forced to put additional Gaussian components, even if they did not have associated components in the previous epochs. In such cases, we simply kept their presence in the corresponding epochs, but did not provide them IDs for physical interpretations later on.
Also, if they did not appear to be of high significance in the following epochs, we removed them manually in the subsequent observations in order to maintain the total number of models as consistent as possible throughout the whole period.
After the last epoch was successfully fitted, we grouped and labeled the VLBI components based on the smoothness of time evolution of their positions, sizes, and fluxes in adjacent epochs.
While doing so, we defined the VLBI core as the reference point of the kinematic analysis.
We note that there are various possible definitions of the VLBI core, for instance based on the jet morphology, brightness temperature, optically thick spectrum, and time variability of the flux density (see, e.g., discussions in \citealt{hodgson17}).
Since we make use of only a single-frequency 43\,GHz data for the kinematic analysis, we chose the most upstream visible feature located at the west end of the jet with the highest brightness temperature as the core. We note that this feature shows the highest brightness temperature throughout the whole epochs analyzed in our study,
which is consistent with the properties of the VLBI core expected in the basic model of conically expanding relativistic jet \citep{blandford79}.
Also, we did not detect any significant emission to the west of the above-defined VLBI core across the analyzed epochs.

Overall, we found that three to five circular Gaussian components were sufficient to describe the jet morphology in all epochs and cross-identify individual VLBI components, especially after 2022 Apr 30 when the data were stably modeled with only three Gaussians. 
Also, the image root-mean-square (RMS) noise levels of the dirty maps after modeling were comparable to those of the CLEAN images.

However, we note that significant time-changing opacity in the base of the jet could also cause non-negligible drifts in the absolute position of the VLBI core (e.g., \citealt{plavin19}), especially during broadband electromagnetic flares in the source (e.g., \citealt{niinuma15}), leading to additional errors in the kinematics analysis of the downstream jet components. Since the VLBA 43\,GHz data used in our study lack phase referencing calibrations for astrometric image registrations \citep{jorstad17}, we followed an alternative method that inversely used traveling downstream jet components as probes of relative motion in the VLBI core over time. That is, we computed residuals of the motions of VLBI components from their linear fit models (see \citealt{lisakov17}). This calculation can alarm, if present, significant motions in the chosen reference point, by correlated group deviations of the residual motions--see also \citealt{kim20} for additional examples. 
To this end, we performed such an analysis for the two components identified at 43\,GHz after \ic (C1 and C2; see \S\ref{sec:results}), focusing on epochs where both components showed overall good fits for their linear models. The calculations of the motion residuals were then performed for both radial distance and RA/Dec positions, respectively. In all cases, we found overall $\lesssim0.03\,$mas of residual displacements of the jet component positions from their linear kinematic models, with the Pearson correlation coefficients of $<0.2$ for the residuals of C1 and C2. This indicates no statistically significant jittering of the core position over time at levels greater than $\sim0.03$\,mas. This is a much smaller value than the beam size of the VLBA observations at 43\,GHz.
For epochs after MJD\,60,000 where C2 only showed significant deceleration and therefore could not be analyzed in the above way, C1 still showed good linear fits, suggesting that these two components have intrinsic motions that are much larger than potential drifts of the VLBI core.  
Considering the above effects, we conclude that the proper motion of the VLBI core itself is not significant in our following analysis.

After the model-fitting and component ID assignment, the uncertainties of the model components were determined as follows.
Following \cite{casadio15,jorstad17}, we presumed that components with smaller sizes and higher flux densities would have smaller statistical errors in the model parameters, namely $S_{\nu}$, $r$, $PA$, and $d$. Therefore, we first computed the observed brightness temperature, $T_{\rm B}$ of each component as
\begin{equation}
    T_{\rm B} =
    1.22\times10^{12}\frac{ S_{\nu} }{ \nu^{2}d^{2} }
    (1+z)\,K
	\label{eq:brightness_temperature}
\end{equation}
\citep{kim18}
where $S_{\nu}$ is the flux density in Jy, $\nu$ is the observing frequency in GHz, $d$ is the FWHM size in mas, and $z=0.42$ is the redshift of \pks.
Statistical uncertainties of the other parameters were then computed by the following empirical relations \citep{casadio15,jorstad17}: 
\begin{equation}
\begin{aligned}
    \sigma_{X} &= 1.3 \times 10^{4} T_{\rm B}^{-0.6} \\
    \sigma_{Y} &= 2 \times \sigma_{X} \\
    \sigma_{S} &= 0.09 T_{\rm B}^{-0.1} \\
    \sigma_{d} &= 6.5 T_{\rm B}^{-0.25}
    \label{eq:uncertainties}
\end{aligned}
\end{equation}
\citep{casadio15,jorstad17,weaver22} where $\sigma_{X,Y}$ are the RA/Dec position errors in mas respectively, $\sigma_{S}$ is the flux density uncertainty in Jy, and $\sigma_{d}$ is the FWHM error in mas, respectively.
We note factor 2 in $\sigma_{Y}$ to account for $\sim\times2$ larger beam size in the N-S direction.
Uncertainty in $r$ and $PA$ -- $\sigma_{r}$ and $\sigma_{\rm PA}$, respectively -- were obtained by propagating these uncertainties.
Next, we added in quadrature systematic errors of 10\% for the flux density and 5\% for the sizes, respectively, to the above statistical errors in order to obtain the total error budget.
Because we adopted $T_{\rm B}$ as the primary observable in Eq. \ref{eq:uncertainties}, we only assumed systematic uncertainties in $S$ and $d$ to estimate representative uncertainties of $T_{\rm B}$.

To measure the proper motions and kinematics of each VLBI component, we modeled the radial distances of each unique component from the core as a function of time as 
\begin{equation}
r(t)=\mu_{\rm app}(t-t_{0})
\label{eq:r_t}
\end{equation}
where 
$r(t)=\sqrt{X(t)^{2}+Y(t)^{2}}$ is the radial distance of a component from the core, 
$t$ is the observing epoch in year, and
$\mu_{\rm app}$ is the apparent proper motions in mas/yr,
$t_{0}$ is the time, in year, of the ejection of a VLBI component from the VLBI core.
Using the $r(t)$ and $t$, we determined $\mu_{\rm app}$ and $t_{0}$ by linear regression using the curve\_fit package in Scipy \citep{scipy}.
The apparent jet speed in units of speed of light $c$, $\beta_{\rm app}$, was computed by
\begin{equation}
\beta_{\rm app}=\mu_{\rm app} D_{L}/(c(1+z))
\label{eq:beta_app}
\end{equation}
where $D_{L}=2371\,$Mpc is the luminosity distance to \pks and $c$ is the speed of light.

\section{Results}
\label{sec:results}

We show in Fig. \ref{fig:representative_images} the VLBI jet images of \pks at 15 and 43\,GHz close-in-time to the \ic event.
The source clearly shows a compact VLBI-core which dominates the overall flux and north-east oriented jet, which is similar to the general morphology of the source between 2007--2018 \citep{jorstad17,weaver22}.
However, the peak flux densities at 15 and 43\,GHz are significantly higher than the historical average values by factors of a few (see \citealt{jorstad17,lister19} and also \citealt{plavin23}).
A summary of the properties of the CLEAN images for all epochs is given in Table \ref{tab:properties_all_maps}.

The multi-wavelength light curves of the source are shown in Fig. \ref{fig:lightcurve} and a zoomed version after the \ic event in Fig. \ref{fig:lightcurve_zoom}. 
Two main features are noteworthy in the lightcurves: 
As other studies reported (e.g., \citealt{plavin23,prince24,bharathan24}), the source shows pronounced and nearly contemporaneous $\gamma$-ray, X-ray, and optical flares around the time of the \ic event.
However, the maxima of the VLBI radio flux densities at 15 and 43\,GHz occur with delay timescales of approximately $\sim 180$ and $\sim140$\,days respectively after \ic. 
We note that increased radio flux density is frequently seen in neutrino candidate AGNs close to other IceCube neutrino events (e.g., \citealt{plavin23}).
Another important point is that \pks still continues to be variable in $\gamma-$ray after $\sim$MJD\,59800. It might indicate a complex temporal correlation between neutrino and electromagnetic flares within the field of \ic due to a number of sources within the field and  poor angular resolutions of some of the involved observatories including the IceCube observatory itself (see, e.g., \citealt{garrappa19}, for the case of TXS\,0506+056), or source-intrinsic, post-neutrino evolution of the blazar \pks 
 itself (e.g., \citealt{cerruti19}).

\begin{figure}[t]
	\includegraphics[width=\columnwidth]{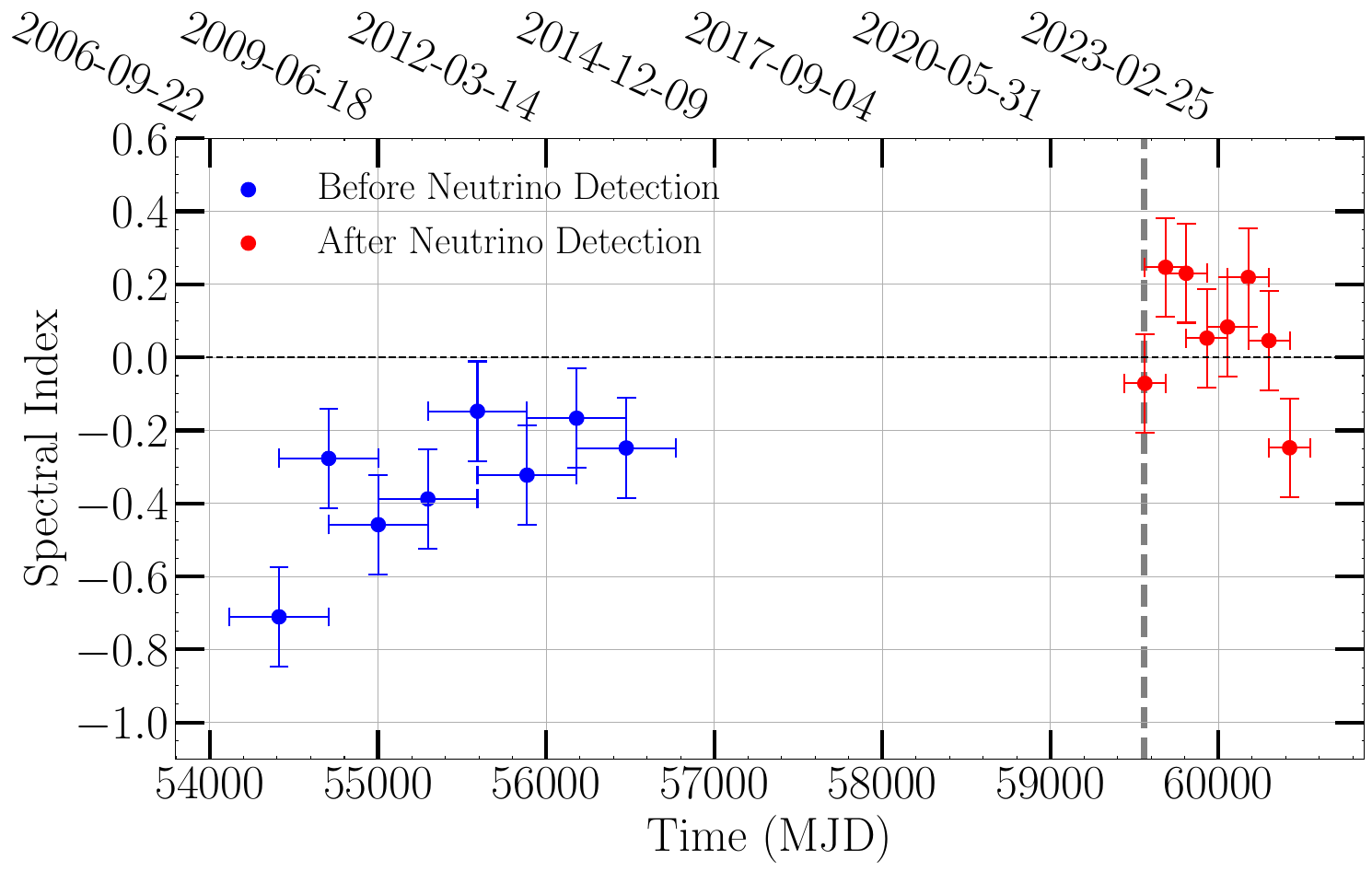}
    \caption{
    Spectral indices of the nuclear region at 15-43\,GHz. 
    Vertical dashed line indicates the timing of \ic and blue and red data points show the spectral indices before and after the neutrino events, respectively. 
    The horizontal error bars represent timescales for the binning at 15 and 43\,GHz.
    }
    \label{fig:spectral}
\end{figure}

Figure \ref{fig:spectral} shows the time evolution of the spectral index of the VLBI nuclear region at 15-43\,GHz window, $\alpha_{15}^{43}$, computed from the CLEAN images. While the VLBI core displayed relatively steep spectral indices before the \ic event (the mean and standard deviation spectral indices of $<\alpha_{15}^{43}>=-(0.34\pm0.17)$), 
the values of $\alpha_{15}^{43}$ changed substantially after the neutrino event, reaching $<\alpha_{15}^{43}>\sim0.07\pm0.16$. 
This most likely indicates a significant change in the synchrotron opacity of the nuclear region.

\begin{table}
\centering
\caption{Spectral indices of the nuclear region 15-43\,GHz before and after the \ic event.}
\begin{tabular}{cc}
\toprule
Time (MJD) & $\alpha_{15}^{43}$ \\
\midrule
\multicolumn{2}{c}{Before} \\
\hline
54411.50 & $-$0.71 $\pm$ 0.13 \\
54706.50 & $-$0.28 $\pm$ 0.13 \\
55001.50 & $-$0.46 $\pm$ 0.13 \\
55296.50 & $-$0.39 $\pm$ 0.13 \\
55591.50 & $-$0.15 $\pm$ 0.13 \\
55886.50 & $-$0.32 $\pm$ 0.13 \\
56181.50 & $-$0.17 $\pm$ 0.13 \\
56476.50 & $-$0.25 $\pm$ 0.13 \\
\midrule
\multicolumn{2}{c}{After} \\
\hline 
59561.50 & $-$0.07 $\pm$ 0.13 \\
59684.50 & 0.25 $\pm$ 0.13 \\
59807.50 & 0.23 $\pm$ 0.13 \\
59930.50 & 0.05 $\pm$ 0.13 \\
60053.50 & 0.08 $\pm$ 0.13 \\
60176.50 & 0.22 $\pm$ 0.13 \\
60299.50 & 0.05 $\pm$ 0.13 \\
60422.50 & $-$0.25 $\pm$ 0.13 \\
\bottomrule
\end{tabular}
\end{table}

\begin{figure*}
	\includegraphics[width=1.0\textwidth]{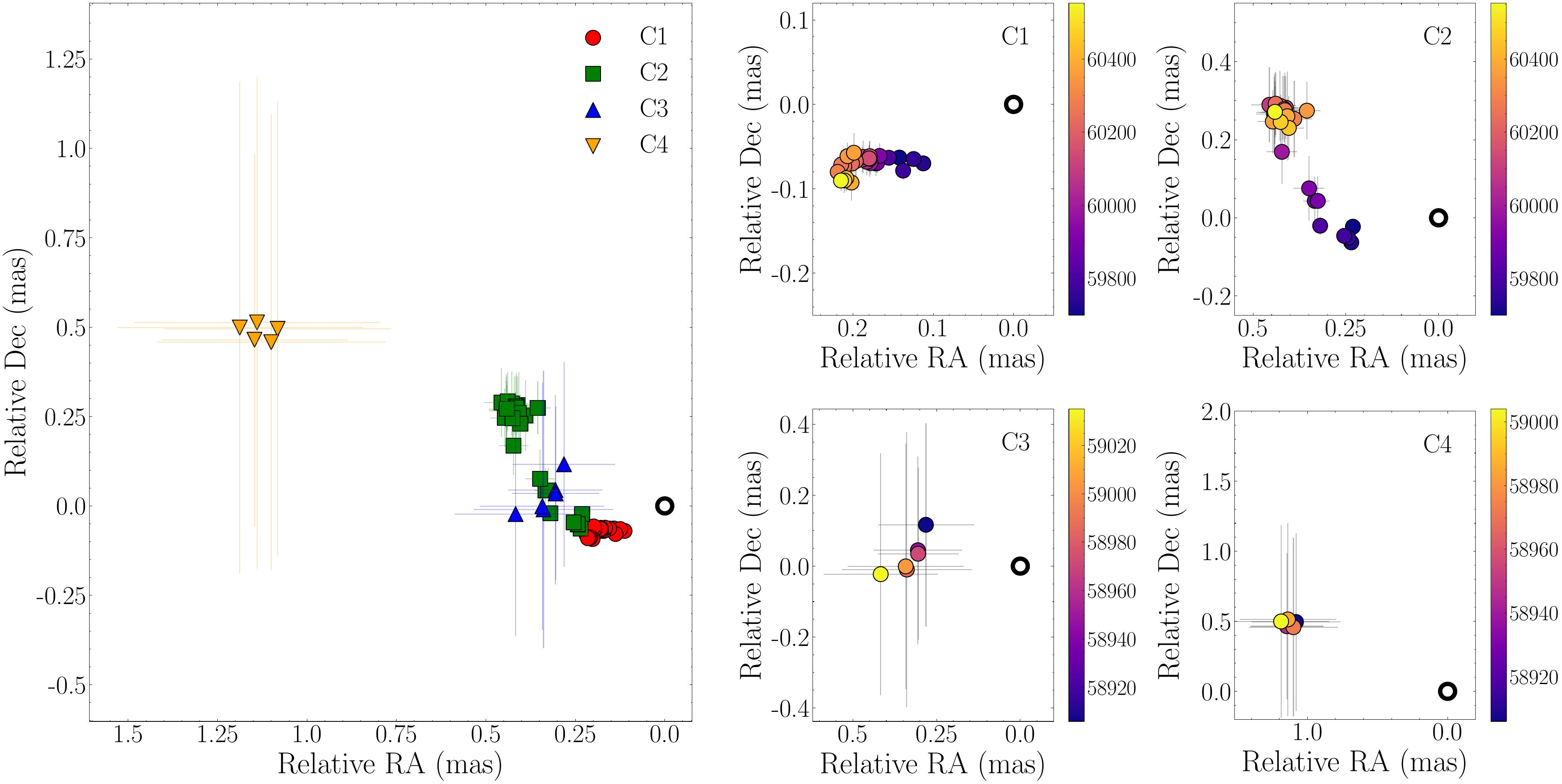}
    \caption{
    Positions of model-fitted components from the analysis of the VLBA 43\,GHz data of \pks during $\sim2020-2024$. 
    Large left panel: positions of all the components.
    Small right panels: detailed time evolution of the position for the individual components.
    In all panels, the empty black circle represents the location of the VLBI core (C0).
    }
    \label{fig:component_position}
\end{figure*}

\begin{figure}
	\includegraphics[width=\columnwidth]{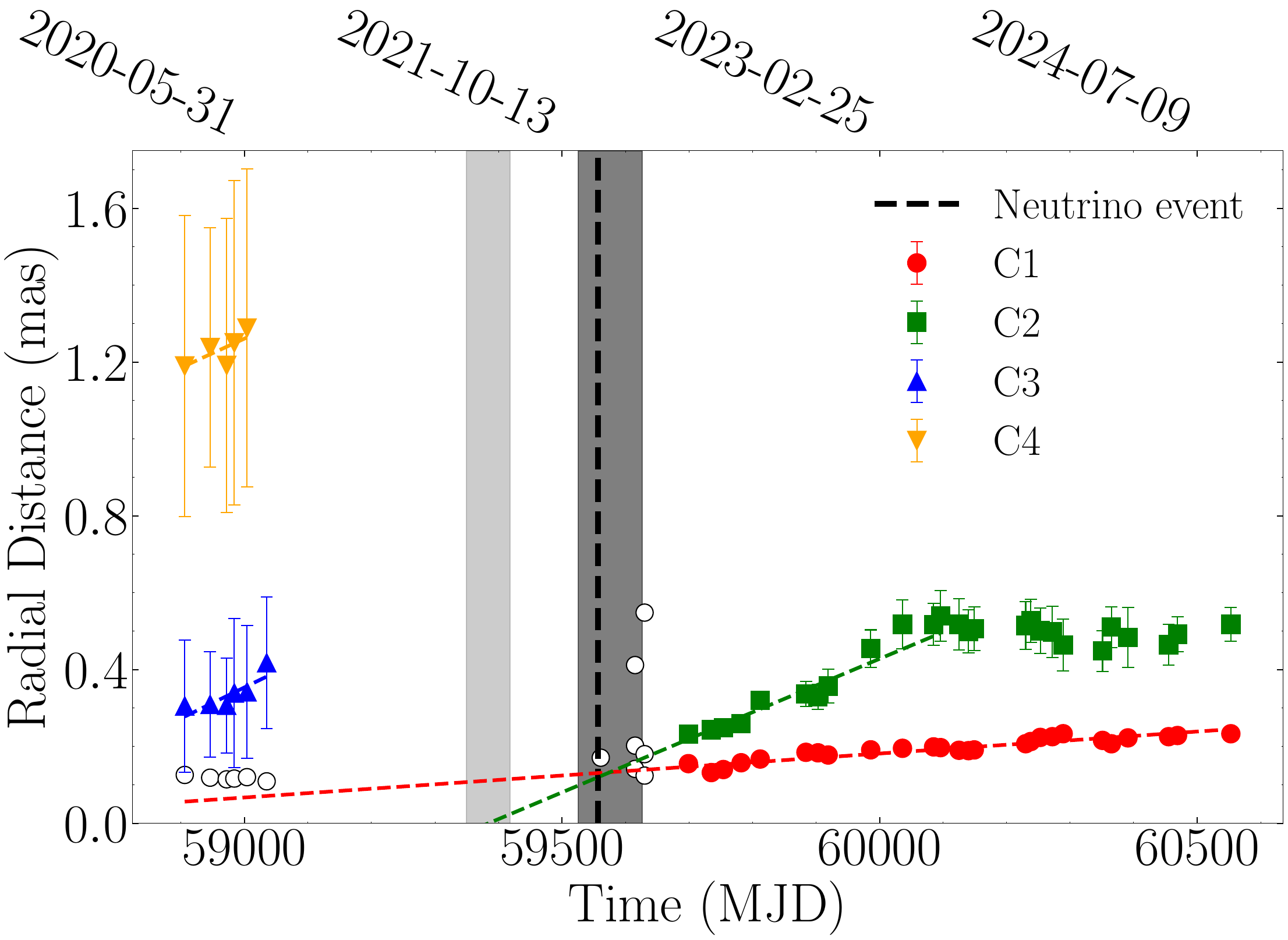}
\includegraphics[width=\columnwidth]{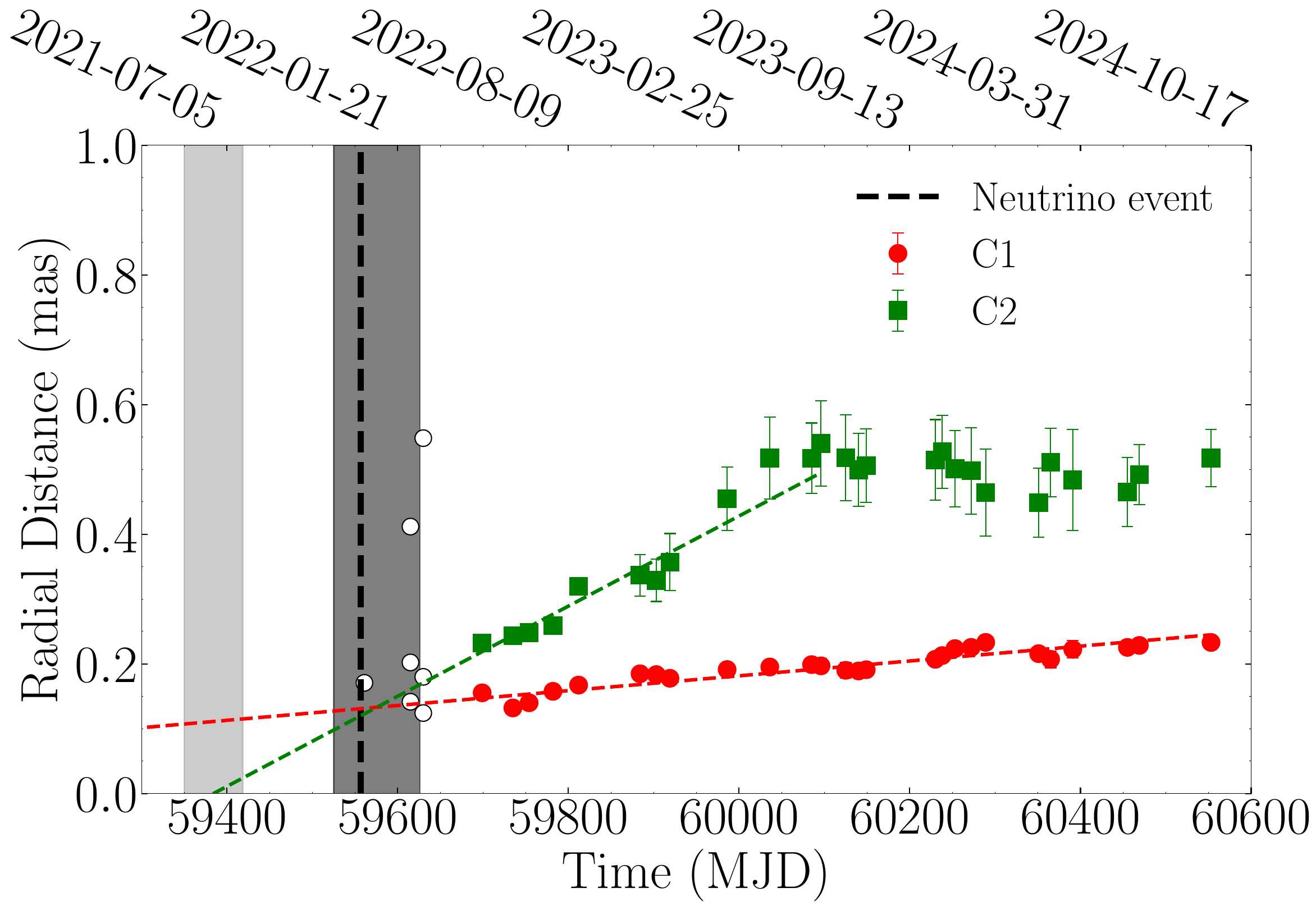}
    \caption{
    Radial distances $r(t)$ of the moving components C1 to C4.
    Top panel shows the distances for all the components.
    Bottom panel shows the same information but zoomed in smaller regions of time and radial distances to highlight C1 and C2.
    In both panels, empty circles represent model-fitted but not uniquely identified components (see \S\ref{subsec:vis_fitting}).
    Also, green and red dashed lines display linear models of $r(t)$ fitted to the data of C1 and C2, respectively.
    Light and dark grey regions indicate the timings of ejection of C2 from the VLBI core (C0), MJD\,$59384.1\pm34.5$, and passage of C1 through C2, MJD\,$59575.53\pm50.41$, respectively.
    Vertical dark dashed line corresponds to the timing of \ic.
    }
    \label{fig:distance}
\end{figure}

\begin{figure}
    \includegraphics[width=\columnwidth]{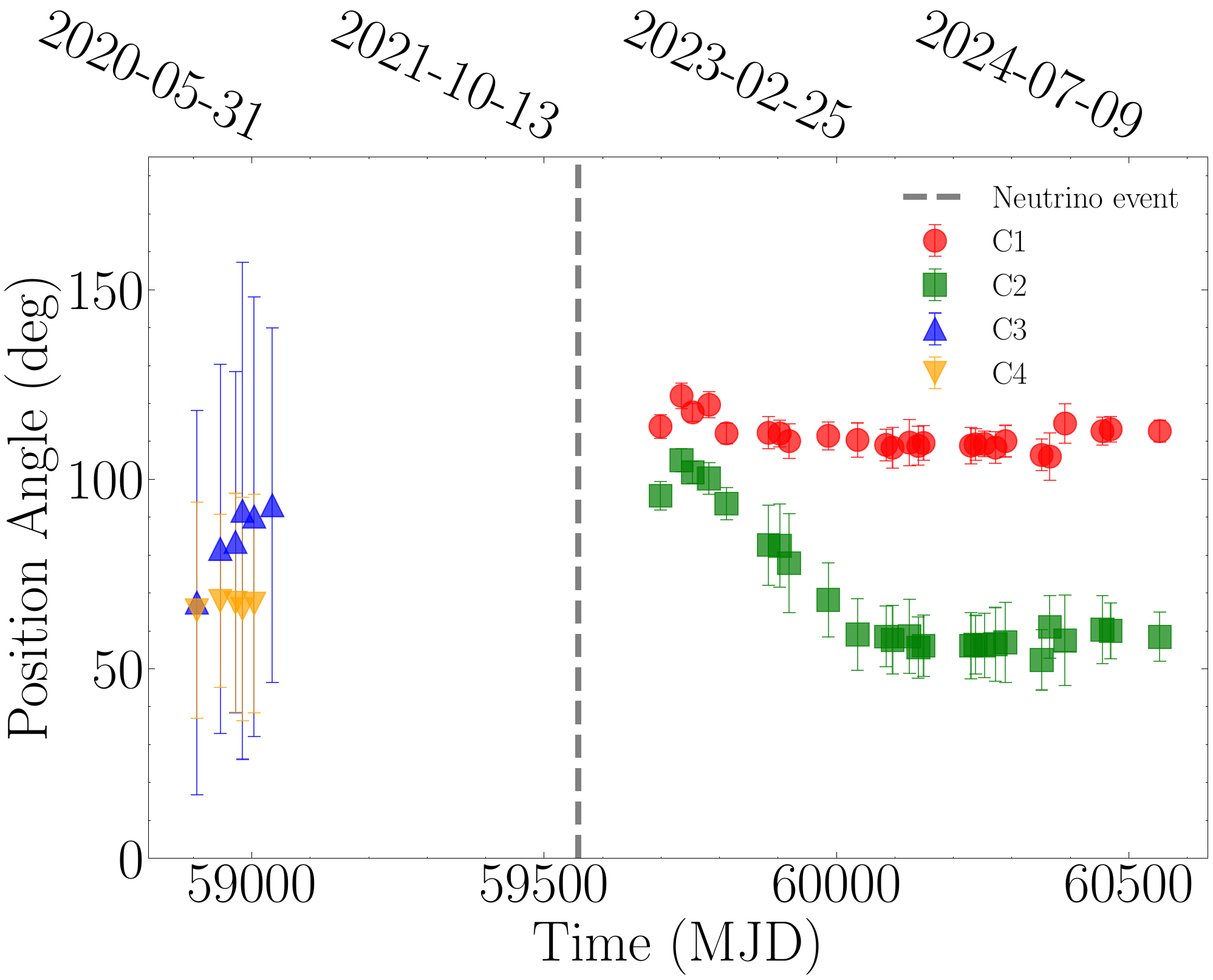}
    \caption{
    Position angles of C1 to C4 over time. Vertical dashed line corresponds to the timing of \ic.
    }
    \label{fig:PA}
\end{figure}

Below we present results based on the visibility model-fitting of the VLBA 43\,GHz data.
Details of the model components, including their IDs, fluxes, positions, and sizes for all the observed epochs, are shown in Fig. \ref{fig:modelfit} and Table \ref{tab:model_comp_all}.
Despite varying $(u,v)$-coverages and signal-to-noise values from epoch to epoch, we have been able to identify, following the procedures in \S\ref{subsec:vis_fitting}, five major components labeled as C0 to C4, with C0 corresponding to the VLBI core and taken as reference for the kinematic analysis.
The proper motions for C1 to C4 over time, in the RA/Dec, the radial core separation, and position angle are shown in Figs. \ref{fig:component_position}, \ref{fig:distance}, and \ref{fig:PA}, respectively.
The angular and apparent speeds of the components estimated by Eqs. \ref{eq:r_t} and \ref{eq:beta_app} are summarized in Table \ref{tab:speeds}.

As for C3 and C4, their motion speeds are consistent with zero within uncertainties. This is most probably due to the small number of epochs that we have analyzed to trace their motions.
On the other hand, the kinematics of C1 and C2 show several interesting features, which we report in detail below.
The component C1 moves at a significantly slower subluminal speed ($\sim0.042\pm0.002\,$mas/yr or $\sim0.77\pm0.039c$) compared to the fastest and median superluminal jet speeds reported for \pks ($0.257$ and $0.167$\,mas/yr at 15\,GHz; $\sim4.8c$ and $3.1c$, respectively \citep{lister19} and $0.242$ and $0.191$\,mas/yr at 43\,GHz; $\sim4.5c$ and $\sim3.6c$, respectively \citep{weaver22}).
Instead, the relatively slow speed of C1 is rather comparable to those of slow features reported by \cite{agudo06} for the inner $\lesssim1\,$mas jet of \pks during 1996 March to 2000 May ($0.023-0.024$\,mas/yr) when \pks was in a quiescent state.
Also, we note that \cite{weaver22} reported the presence of two stationary features at $\sim0.13$\,mas and $\sim0.31$\,mas downstream of the VLBI core at 43\,GHz during 2007 June to 2018 December (designated as A2 and A1 in their study, respectively). 
While some stationary jet features around times of significant electromagnetic flares show outward motions and then return to their original position (e.g., \citealt{jorstad17,weaver22}), 
the observed proper motion of the component C1 after 2021 does not match this behavior, that is, it only shows a monotonic outward motion. 
Therefore, here we presume that
those stationary components A1 and A2 in \cite{weaver22} were present before the \ic event but have disappeared in later epochs. 
We note that such stationary components have appeared and disappeared on timescales of a decade in jets of various AGNs \citep{weaver22}.
However, we cannot exclude other possibilities, such as C1 being a stationary feature that was displaced from its previous position and it might return to its original position after longer time. We discuss more details of these interpretations in \S\ref{sec:discussions}.

As for the component C2, it travels to the jet downstream at a significantly faster speed than C1 ($\sim0.25\pm0.02$\,mas/yr or $\sim4.7\pm0.4c$). 
In fact, this speed is close to the aforementioned maximum speeds reported for the jet in \pks.
Back-extrapolating the kinematics of C2, we find that the component has newly appeared from the core at the ejection time of $t_{0}={\rm MJD\,}59384.1\pm34.5$, which is earlier than the timing of \ic (MJD\,59556.84) by $172.7\pm34.5$\,days, nearly a $5\sigma$ statistical difference.
Furthermore, the position angle of C2 changes substantially over time, from $\sim105^{\circ}$ to $\sim50^{\circ}$ from the first epoch when the component was resolved and until $\sim$MJD\,$60,096$ after which C2 becomes nearly stationary with respect to C0 (see also Fig. \ref{fig:component_position}).
While \pks has previously displayed modulations of the jet position angle over time (e.g., \citealt{gomez01,agudo06,lister21,jorstad17,weaver22}), such a large PA change within less than two years has not been observed in \pks. Indeed, such a large drift of PA is also uncommon in the pc-scale jets of AGNs in general (see \citealt{lister21}).
Also, the apparent motion of C2 significantly decelerates after $\sim$MJD\,60,000 and then nearly stops moving outward, when the component reaches $\sim0.5$\,mas core separation ($\sim2.8\,$pc projected).
Although jet components close to the VLBI core or other nearby bright features could exhibit apparently non-radial, stationary, or inward-moving kinematics due to blending of the emitting features (e.g., \citealt{fromm13,lister21}), we note that C2 is relatively well separated from C1 and C0 at the later time. Thus we believe that the observed apparent deceleration is not an imaging or modeling artifact.
Indeed, such non-ballistic motions are observed not uncommonly in jets of AGNs with particularly small viewing angles (see, e.g., component B3 in CTA\,102 and B12 in 3C\,454.3 in \citealt{weaver22}, and component 19 in OJ\,287 in \citealt{lister21}). 
Such components could show apparently stationary kinematics on timescales of a few years, even if they previously displayed constant radial outward motions. 
Although physical reasons for these motions could be various, such as jet bending due to jet-interstellar medium (ISM) interaction or jet-internal helical motions (see, e.g., \citealt{traianou24} for a brief review), continued monitoring of those peculiar features also often reveal continued radial outward motions on longer timescales.
In this regard, we do not discuss those scenarios for C2 more deeply, as continued monitoring of the component may better reveal its origin.

Last and most strikingly, both the trajectories of C1 and C2 shown in Fig. \ref{fig:component_position} obtained from the linear motion fit and formal analysis of the core separation $r(t)$ of C2 strongly suggests an apparent passage of C2 through C1 at MJD\,$59575.53\pm50.41$ at a core separation distance of $0.13\pm0.015\,$mas.
This is $18.69\pm50.41$ days difference compared to the time of \ic (MJD\,59556.84), suggesting strong temporal correlation between this apparent passage and the \ic event and that the corresponding region can be a possible spatial origin of \ic within the compact jet of the blazer \pks.

\begin{figure}
 \includegraphics[width=\columnwidth]{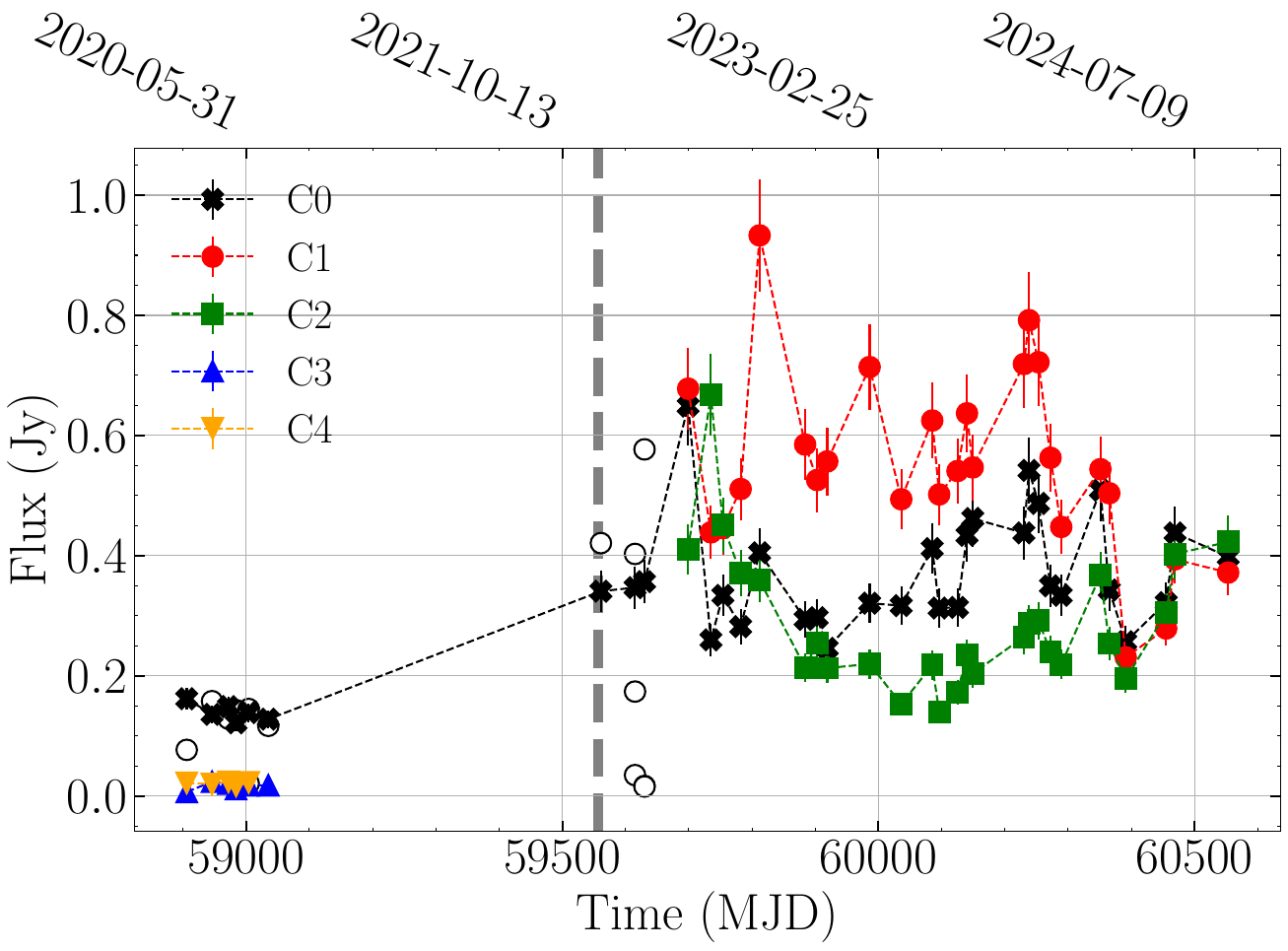}
 \includegraphics[width=\columnwidth]{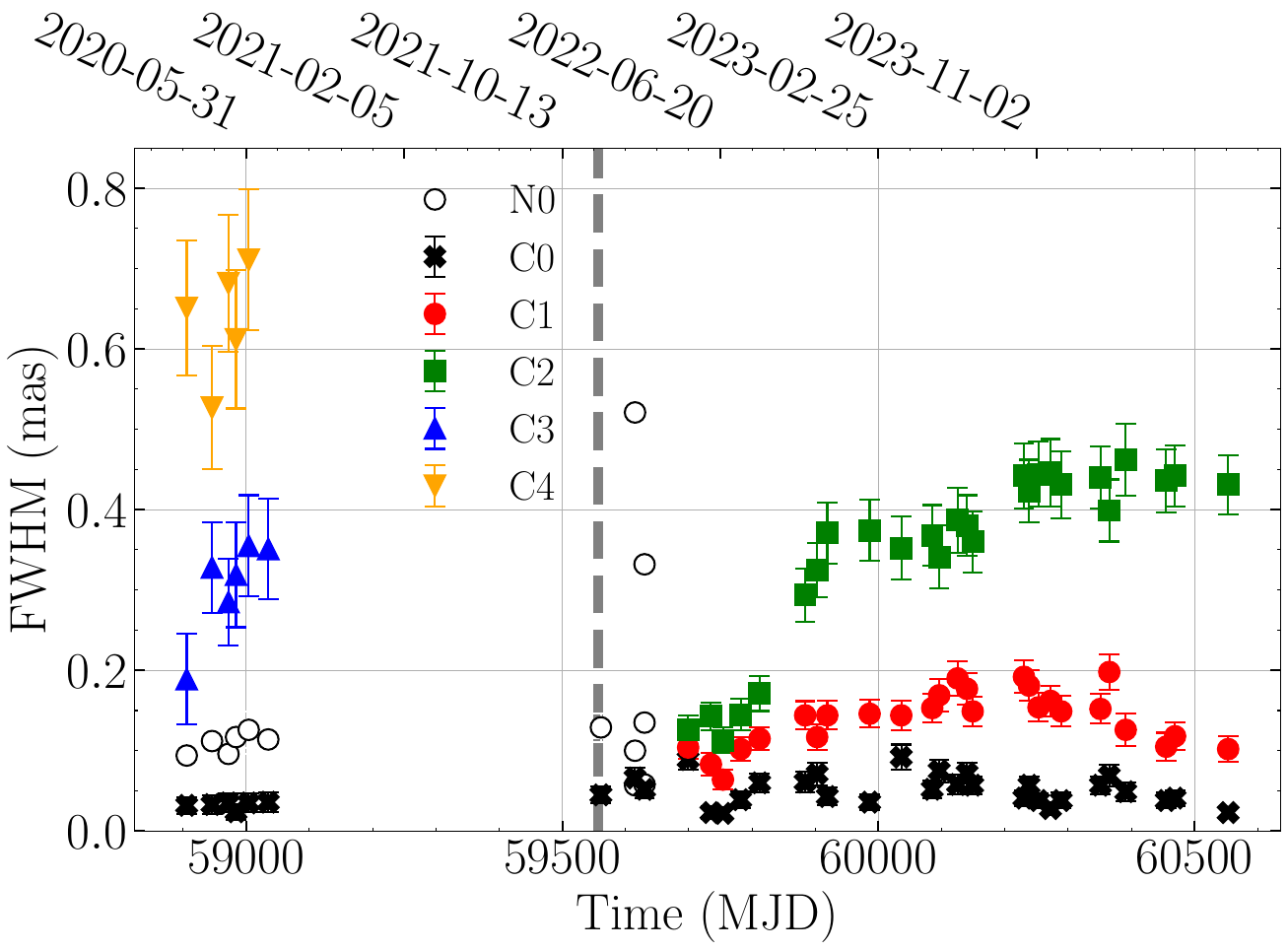}
 \includegraphics[width=\columnwidth]{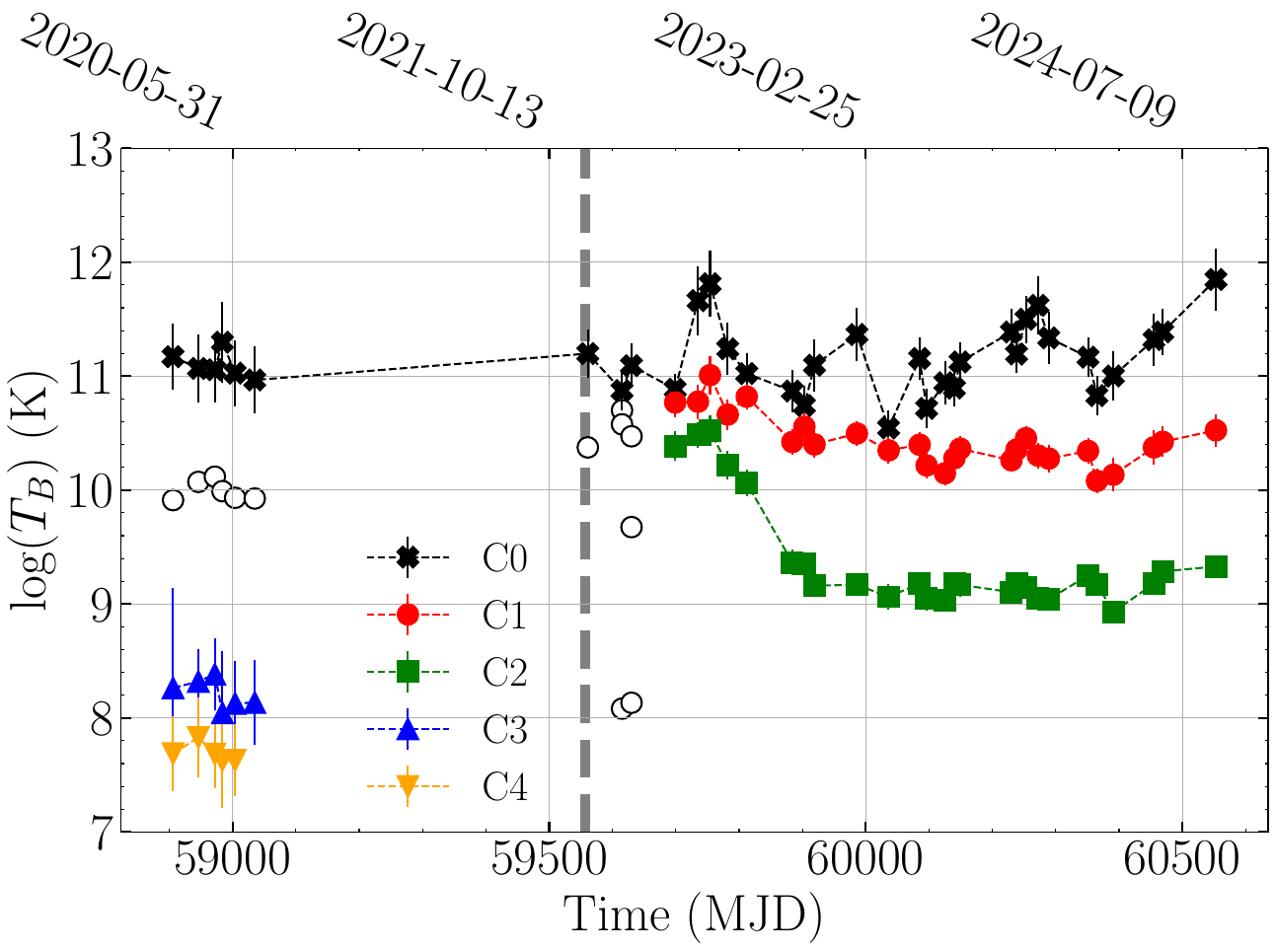}
    \caption{Flux densities (top), FWHM sizes (middle), and brightness temperatures (bottom) of the model-fitted components C0 to C4.
    In all panels, vertical dashed lines indicate the timing of \ic.
    }
    \label{fig:flux_size_tb}
\end{figure}

In Fig. \ref{fig:flux_size_tb} we show the time evolution of the flux densities, FWHM sizes, and brightness temperatures $T_{\rm B}$ of C0 to C4. 
At the time of the \ic event (MJD\,59556.84), C0  showed factors $\sim2-3$ higher flux densities compared to the latest quiescent period at $\sim$MJD\,$59,000$, and even more flux increase later on.
However, the value of $T_{\rm B}$ did not change significantly at this time.
Instead, the $T_{\rm B}$ of C0 significantly increase later by factors $\sim\times10$, with the maxima appearing around $\sim$MJD\,59750, which is $\sim200$\,days after the \ic event. 
After this time, C1 becomes brighter than C0, dominating the VLBI-scale flux density of the jet in \pks.
Interestingly, the flux density of C2 begins to rise significantly after $\sim$MJD\,60,096, coincident with the deceleration and position angle change of the component. This might indicate suggests certain physical evolution of the component, such as interaction with the ISM or jet bending towards the line of sight and increased Doppler beaming effect.

\begin{figure}
    \includegraphics[width=\columnwidth]{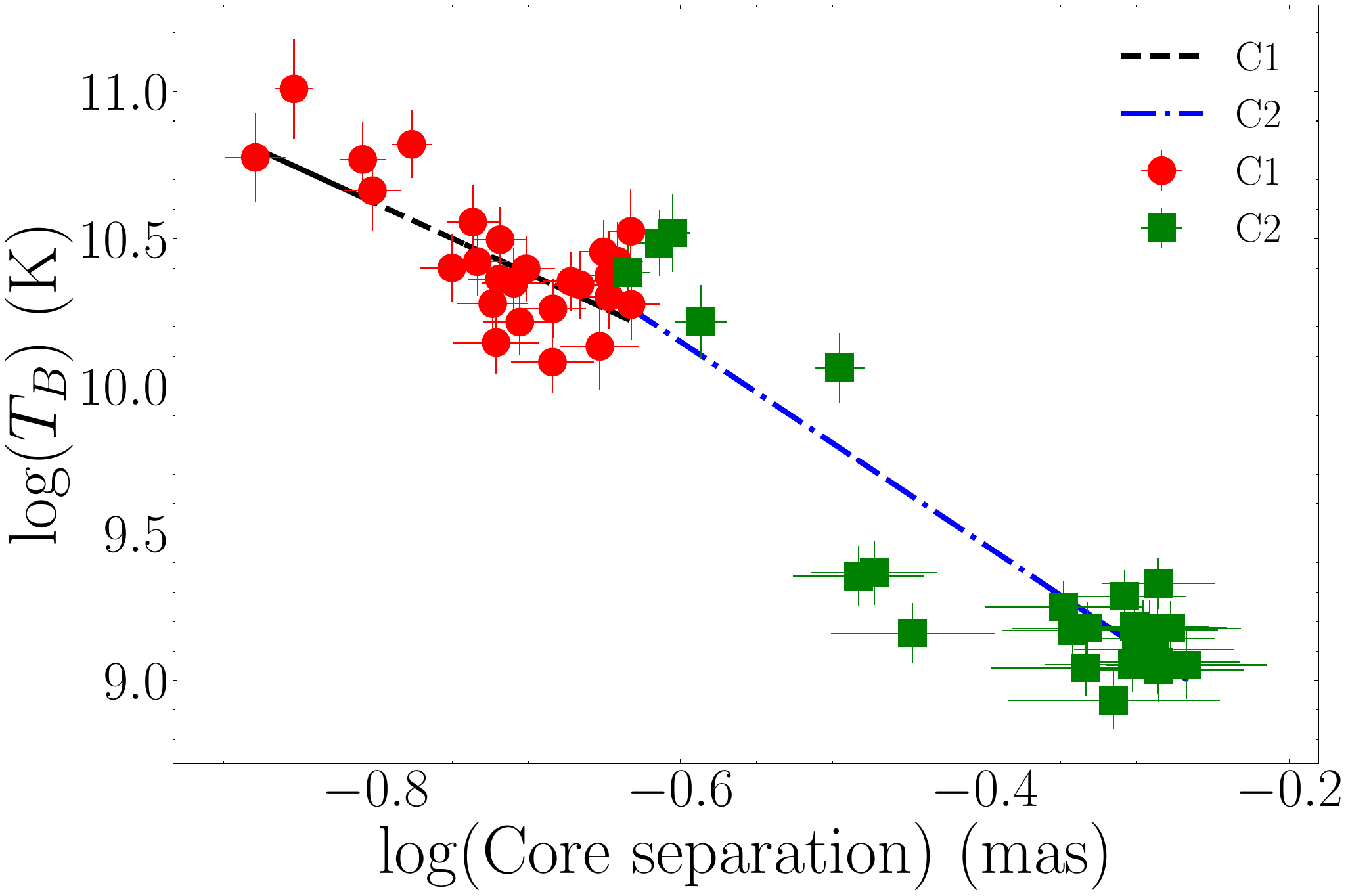}
    \caption{
    Log brightness temperatures versus radial distances of the jet features C1 and C2 after the \ic event.
    The slopes of the fitted lines are $-(2.37 \pm0.53)$ and $-(3.45\pm0.38)$ for C1 and C2, respectively.
    }
    \label{fig:tb_vs_radial_distance}
\end{figure}

We have also measured the variation of $T_{\rm B}$ versus the radial core distances for C1 and C2, which can be a useful diagnostics of various evolutionary stages of traveling jet components, namely Compton, synchrotron, and adiabatic energy loss (see \citealt{marscher92,fromm11,schinzel12}).
Specifically, we fit a model of 
\begin{equation}
T_{\rm B}(r)\propto r^{-\epsilon} 
\label{eq:tb_vs_r}
\end{equation}
\citep{schinzel12} where
$r$ is the core separation in milliarcsecond and $\epsilon$ is a free parameter to be fitted.
The results are shown in Fig. \ref{fig:tb_vs_radial_distance}.
The values of $\epsilon$ are $2.37\pm0.53$ and $3.45\pm0.38$ for C1 and C2, respectively.
For both components, the derived values of $\epsilon$ lie closely to $\epsilon_{C}=1.25$ and $\epsilon_{a}=3.17$, which are expected values from respectively the Compton-dominated or adiabatic energy losses in a jet dominated by the transverse magnetic field, a typical synchrotron spectral index of $\alpha=-0.5$, and a constant Doppler factor, in contrast to $\epsilon_{S}=4.17$ which is expected for a pure synchrotron energy loss case for the same jet \citep{schinzel12}.

\begin{table}
    \centering
    \caption{
    Proper motions and apparent speeds of each Gaussian component.
    }
    \label{tab:speeds}
    \begin{tabular}{ccc}
        \hline
        ID & Proper motion (mas/yr) & $\beta_{\rm app}$\\
        \hline
        C1 & 0.042 $\pm$ 0.002 & 0.777  $\pm$ 0.039 \\
        C2 & 0.254 $\pm$ 0.023 & 4.720  $\pm$ 0.433 \\
        C3 & 0.294 $\pm$ 0.611 & 5.475 $\pm$ 11.36 \\
        C4 & 0.276 $\pm$ 1.910 & 5.136 $\pm$ 35.51 \\
        \hline
    \end{tabular}
\end{table}

Lastly, we compute the jet Lorentz factor $\Gamma$, Doppler factor $\delta$, viewing angle $\theta$, and their limiting values using the apparent jet speeds and time evolution of the flux densities for individual components.
First of all, the minimum bulk Lorentz factor, $\Gamma_{\rm min}$, can be computed straightforwardly as follows:
\begin{equation}
    \Gamma_{\rm min} =\sqrt{\beta_{\rm app}^{2}+1}
	\label{eq:Lorentz_factor}
\end{equation}
\citep{boettcher12,paraschos23}. 
We find $\Gamma_{\rm min}=1.27\pm0.02$ and $4.82\pm0.42$ for C1 and C2, respectively. 
Then, the corresponding minimum critical viewing angles are:
\begin{equation}
    \theta_{c} = \arcsin{(\frac{1}{\Gamma_{\rm min}})}.
	\label{eq:critical_angle}
\end{equation}
\citep{boettcher12}.
For C2 which shows higher $\Gamma_{\rm min}$, we find $\theta_{c}\sim(11.9\pm1.06)^{\circ}$.
This value is comparable to the results of \cite{agudo06} who derived $\theta\leq9^{\circ}$ for the period 1996-2000.
We note that \cite{weaver22} obtained $\theta\sim1.09^{\circ}-3.36^{\circ}$ for the two most reliable jet features in their analysis of the VLBA 43\,GHz data of \pks during 2007-2018.

\begin{figure}
    \centering
        \includegraphics[width=\columnwidth]{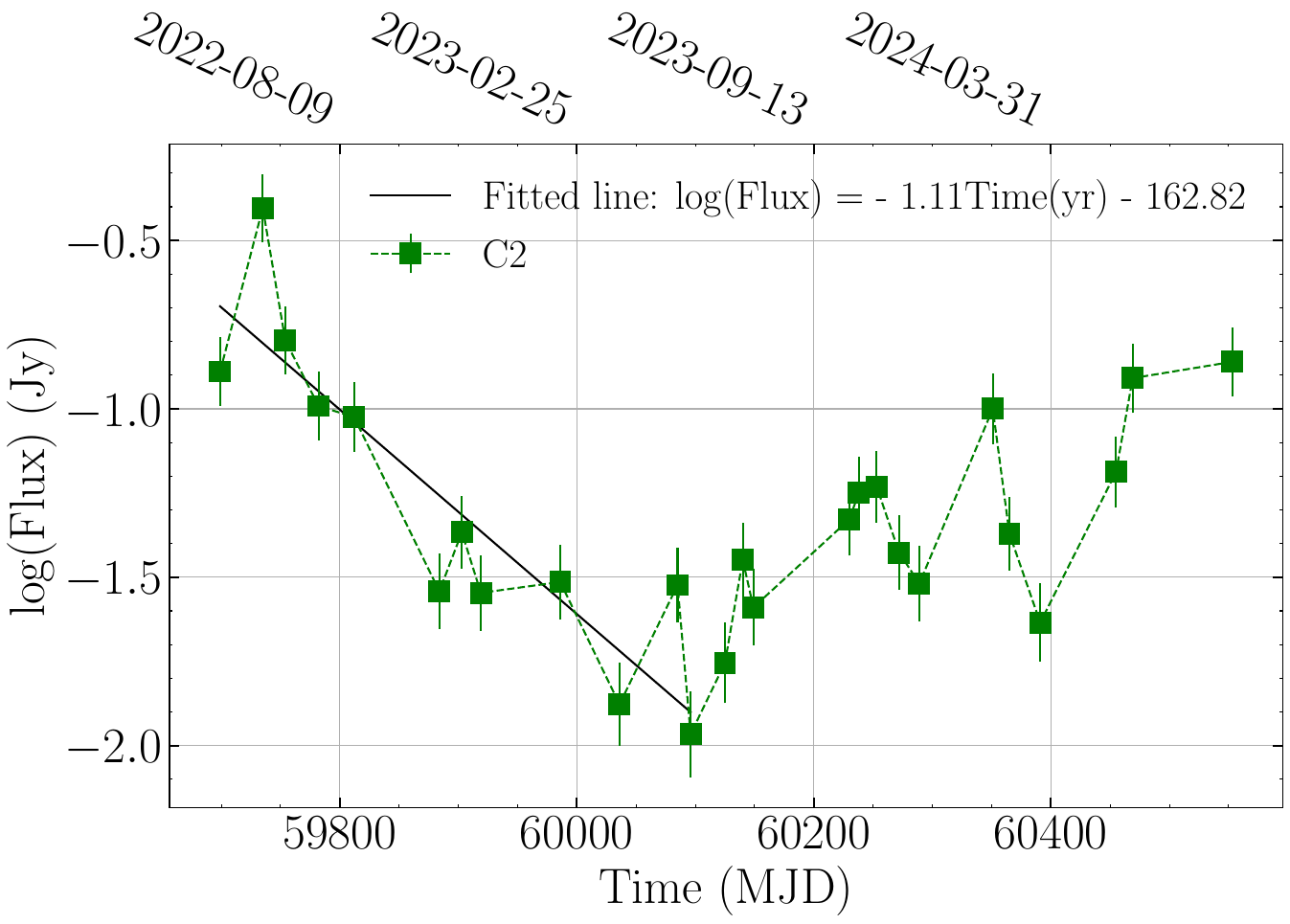}
    \caption{Flux density evolution of the component C2 versus time. 
    A linear model fitted to the data is shown in solid black line.}
    \label{fig:log_flux_time}
\end{figure}

In order to estimate the Doppler factors without information about the counterjet emission, we make use of the time variability of C2. More specifically, we compute the variability Doppler factor, variability Lorentz factor, and corresponding viewing angle. 
Following \cite{jorstad17,weaver22}, the variability Doppler factor can be expressed as
\begin{equation}
\delta_{\rm var}=\frac{15.8\,d\,D_{L}}{\tau_{\rm var}(1+z)}    
\label{eq:var_doppler}
\end{equation}
where 
$d$ is the angular size of the component in mas (i.e., FWHM size of Gaussian component), 
$D_{L}$ is the luminosity distance in Gpc, 
$\tau_{\rm var}=dt/\ln(S_{\rm max}/S_{\rm min})$ is the Doppler boosting-corrected variability timescale of the superluminal jet feature in year, and
$dt$ is the timescale, in year, passed between maximum and minimum flux densities $S_{\rm max}$ and $S_{\rm min}$.
Further following \cite{jorstad17}, we estimate $\tau_{\rm var}$ by fitting a model of the flux variation $\ln(S(t)/S_{0})=k(t-t_{\rm max})$ to the data, where $S(t)$ is the flux density as a function of time $t$, $t_{\rm max}>t$ is the epoch when $S_{\rm max}$ occurs, and $S_{0}$ and $k$ are free parameters to be determined by least-square fitting of the model to the data.
Once $k$ is determined, it follows that $\tau_{\rm var}=|1/k|$.
By following the above procedures to the data of C2, which shows a clear decay of flux (see the top panel of Fig. \ref{fig:flux_size_tb}), we find 
$k=-(1.11\pm0.17)$ (see Fig. \ref{fig:log_flux_time}) and accordingly $\tau_{\rm var}\sim0.90\pm0.14$\,yr. 
We note that Eq. \ref{eq:var_doppler} is meaningful only if $\tau_{\rm var}$ is shorter than that of the light crossing time of the emitting component.
For a mean FWHM size of $0.26\,$mas of C2 during which C2 showed gradual flux decrease, the light-crossing time at the distance of \pks is $\sim4.7\,$years. 
This is a significantly longer time than $\tau_{\rm var}$.

Plugging into Eq. \ref{eq:var_doppler} a mean and standard deviation of the angular size of C2 $<d>=0.26\pm0.10\,$mas during the period when C2 displayed constant radial speed (Fig. \ref{fig:distance}), $D_{L}=2.36\,$Gpc, and $z=0.42$, we obtain
$\delta_{\rm var}=7.59\pm3.15$.
This value is in broad agreement with a lower limit obtained by the MOJAVE VLBA 15\,GHz observations ($\delta>4.8$; \citealt{homan21}) and also $\delta_{\rm var}=12.2\pm3.3$ derived by \cite{weaver22} at 43\,GHz using the VLBA.
However, it should be noted that $\delta\sim7-8$ is still significantly smaller than $\delta\sim30$ adopted in the literature for the modeling of the emission of photons and neutrino in \pks for \ic (see, e.g., \citealt{sahakyan23,prince24}).

We then compute the variability Lorentz factor, $\Gamma_{\rm var}$, and variability viewing angle, $\theta_{\rm var}$. Following \cite{jorstad17}, 
\begin{equation}
\Gamma_{\rm var}=\frac{\beta_{\rm app}^{2}+\delta_{\rm var}^{2}+1}{2\delta_{\rm var}},
\label{eq:var_lorenz}
\end{equation}
\begin{equation}
\theta_{\rm var}=\tan^{-1}\frac{2\beta_{\rm app}}{\beta_{\rm app}^{2}+\delta_{\rm var}^{2}-1}.
\label{eq:var_angle}
\end{equation}
Using the above $\delta_{\rm var}$ and $\beta_{\rm app}=4.72\pm0.43$ of C2, we find
$\Gamma_{\rm var}=5.33\pm0.97$ and $\theta_{\rm var}=6.83^{\circ}\pm4.11^{\circ}$.
These values are in agreement with $\Gamma_{\rm min}\sim4.8$ and $\theta_{c}\sim12^{\circ}$ obtained by Eqs. \ref{eq:Lorentz_factor} and \ref{eq:critical_angle}.
Also, the value of $\theta_{\rm var}$ is consistent within uncertainties with $\theta=3.4^{\circ}\pm1.6^{\circ}$ that has been estimated by \cite{weaver22} for the B3 component in their study, although another smaller angle of $\theta=1.1^{\circ}\pm0.3^{\circ}$ has also been estimated for another traveling component B4.

\section{Discussions}
\label{sec:discussions}

\subsection{Dynamical nature of C1 and C2}\label{subsec:discussion1}

Although we find a statistical connection between the apparent passage of C2 through C1 with the \ic event, the dynamical nature of C1 and C2 needs to be characterized, in order to understand the particle acceleration processes responsible for the \ic event itself and the contemporaneous multi-wavelength flares. 
The C2 component exhibits a highly superluminal speed ($\beta_{\rm app}\sim4.7$, $\Gamma_{\rm var}\sim5.3$), near its historical maximum. 
Since patterns such as instabilities propagate slower than plasma motions \citep{hardee00,perucho04}, this $\Gamma$ in principle can be a lower limit to the true jet speed. Alternatively, the large superluminal speed can also represent a traveling shock that is created by a strong pressure perturbation at the jet base and propagates downstream, at a faster speed than that of the bulk jet plasma. Nevertheless, statistical studies show that fastest apparent motions of AGN jets correlate with apparent radio luminosities, and thus well represent the true plasma speeds (\citealt{lister09}). Based on these considerations,  we expect that the historically large $\beta_{\rm app}$ of C2 should reflect, to some extent, the true speed of the jet. 
Conversely, the slow $\beta_{\rm app}\sim0.77c$ of C1 is well below the historic median jet speeds ($3.1-3.6c$ at 15–43\,GHz; \S\ref{sec:results}), indicating that the feature likely corresponds to patterns such as 
slowly traveling shocks (\citealt{marscher85,daly88,gomez97}), 
Kelvin–Helmholtz pinch‐mode instabilities (e.g., trailing shocks; \citealt{mertens16,fuentes23,gomez01_2,agudo01,kadler08,beutchert18}), 
or standing recollimation shocks in temporarily oscillating motions (\citealt{fromm13,cohen14,jorstad17,plavin19,weaver22}).

\begin{figure}
\includegraphics[width=\columnwidth]{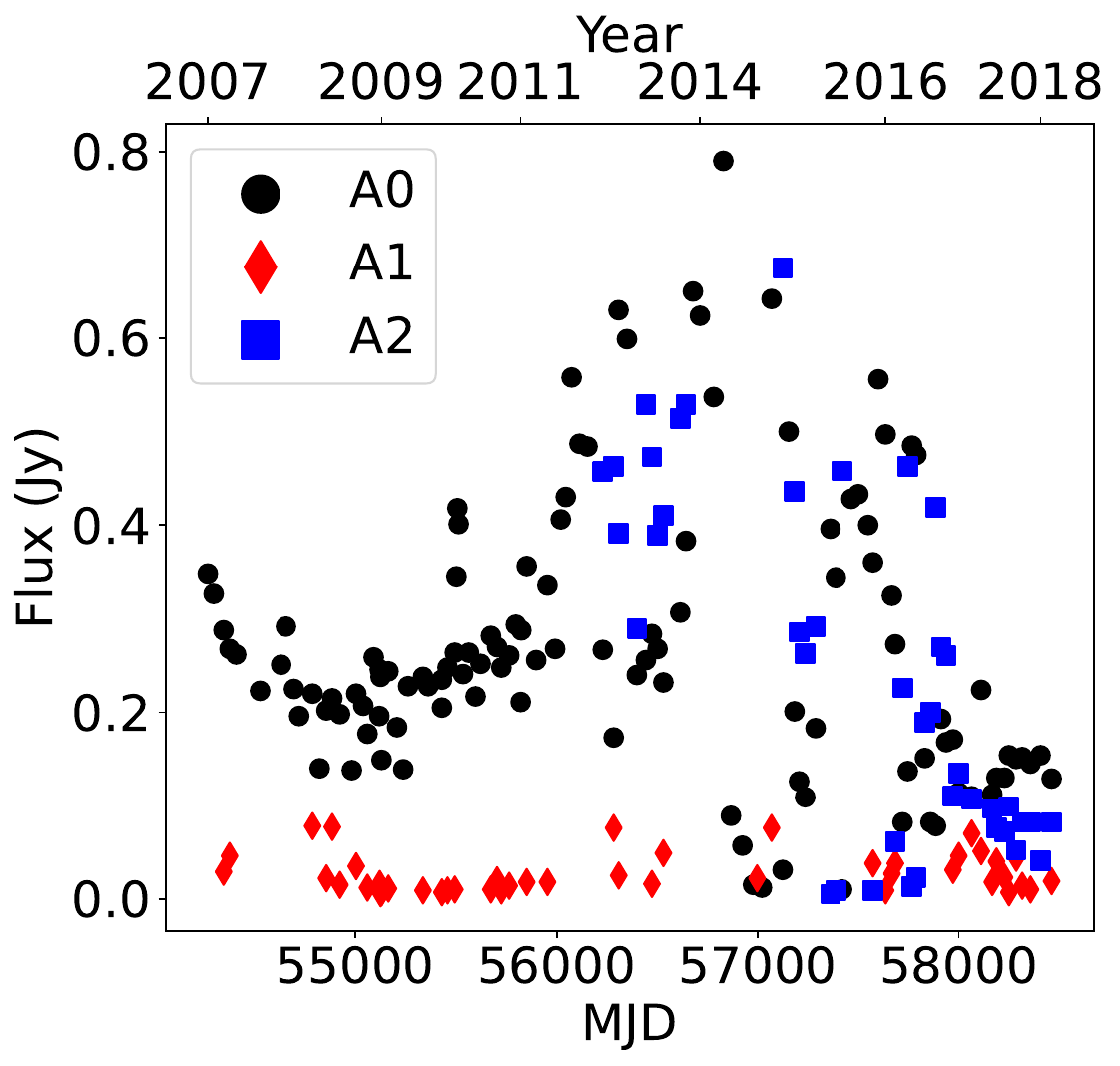}
\caption{
Flux densities of the VLBI core (A0) and stationary components (A1 and A2) downstream of the jet in \pks, that were identified before 2018 as reported in \cite{weaver22}.
}
\label{fig:a0a1a2_flux}
\end{figure}

We examine each of the above scenarios in detail, and show that a single scenario is unlikely to fully explain the emission and kinematic properties of C1. 
(1) {\it a moving shock originating from the core:} by back-extrapolating the linear fit to the radial separations of C1 versus time to close to the ejection epoch from C0 (Fig.\,\ref{fig:distance}), we find no corresponding emitting feature at the predicted core distance, apart from an unlabeled component at $\sim0.3$\,mas. This suggests C1 did not originate from the core and could not be a traveling shock from the jet base. 
(2) {\it trailing shock formed by C2:} simulations and observations of a propagating strong pressure perturbation (e.g., a strong superluminal component) shows the formation of a series of slowly moving, trailing shocks behind (e.g., \citealt{gomez01_2,kadler08,jorstad16,beutchert18}).
This in principle can explain the kinematics of C1. However, the simulations also predict the trailing shocks to remain fainter than the core or major traveling component (\citealt{agudo01,gomez01_2,kadler08,beutchert18}), leaving the high flux of C1 unexplained.
(3) {\it temporarily displaced standing shock structure:} the flux density of C1 remains larger than C0 for most of the epochs studied in this work (Fig.\ref{fig:flux_size_tb}). Such large ratio is uncommon for moving components at 43\,GHz (\citealt{weaver22}), but resembles those of quasi‐stationary features during major radio flares in other AGN jets (\citealt{hodgson17,lisakov17}) as well as in \pks itself in the earlier epochs (the A2 component; see \citealt{jorstad17,weaver22} and Fig. \ref{fig:a0a1a2_flux}). 
This is also in agreement with the generation of stronger radio flares involving recollimation shocks (e.g., \citealt{fromm16}).
The trajectory of C1 also shows a hint of inward motion close to the last epochs analyzed in our study (see C1 in Fig~\ref{fig:component_position}), so C1 might be considered a standing shock but displaced from its previous position (e.g., A2). 
However, we find no statistically significant deviation of the motions of C1 from the linear outward fit (Fig.~\ref{fig:distance}), and thus longer monitoring is required to confirm or falsify this hypothesis.

We note that this complexity is not too surprising, given that the jet in \pks has significantly changed in its emissivity and dynamics compared to the past quiescent periods (cf. \citealt{gabuzda94,gomez01,agudo01,jorstad16,weaver22}).
In fact, \pks has undergone a major outburst close to the time of \ic, reaching its brightest state at all wavelengths since $\sim2006$ (Fig. \ref{fig:lightcurve})
and modeling the broadband SED close to the time of \ic requires exceptionally large jet power in excess of the Eddington limit (e.g., \citealt{prince24}).
Furthermore, we do not detect the previously known two quasi-stationary features A1 and A2, which can be direct evidence for a significant change in the jet structure.
Observationally, their non-detection could also be explained by the limited dynamic ranges of recent VLBA 43\,GHz observations since 2020, because A1 and A2 became considerably faint until $\sim2018$, reaching $\lesssim100\,$mJy flux densities (Fig. \ref{fig:a0a1a2_flux}), while C1 is clearly much brighter (Fig. \ref{fig:flux_size_tb}). However, this also implies that the stationary features A1 and A2 are still significantly fainter than C1 and C2 in our study, still leading to the conclusion of significant change in the jet emissivity related to the variations in either the jet mass, magnetic field, or particle energy.

Therefore, continued monitoring of the source at comparable or higher resolution, and also follow-up analysis of the linear polarimetric properties of C1 and C2, such as its degree of linear polarization and the electric vector position angle, will allow us to better understand their nature for quantitative particle accelerations modeling (e.g., \citealt{cawthorne06,marscher14,beutchert18,kim19}).

\subsection{Implications for the external inverse Compton models}\label{subsec:ext_photon}

\begin{figure}
\includegraphics[width=\columnwidth]{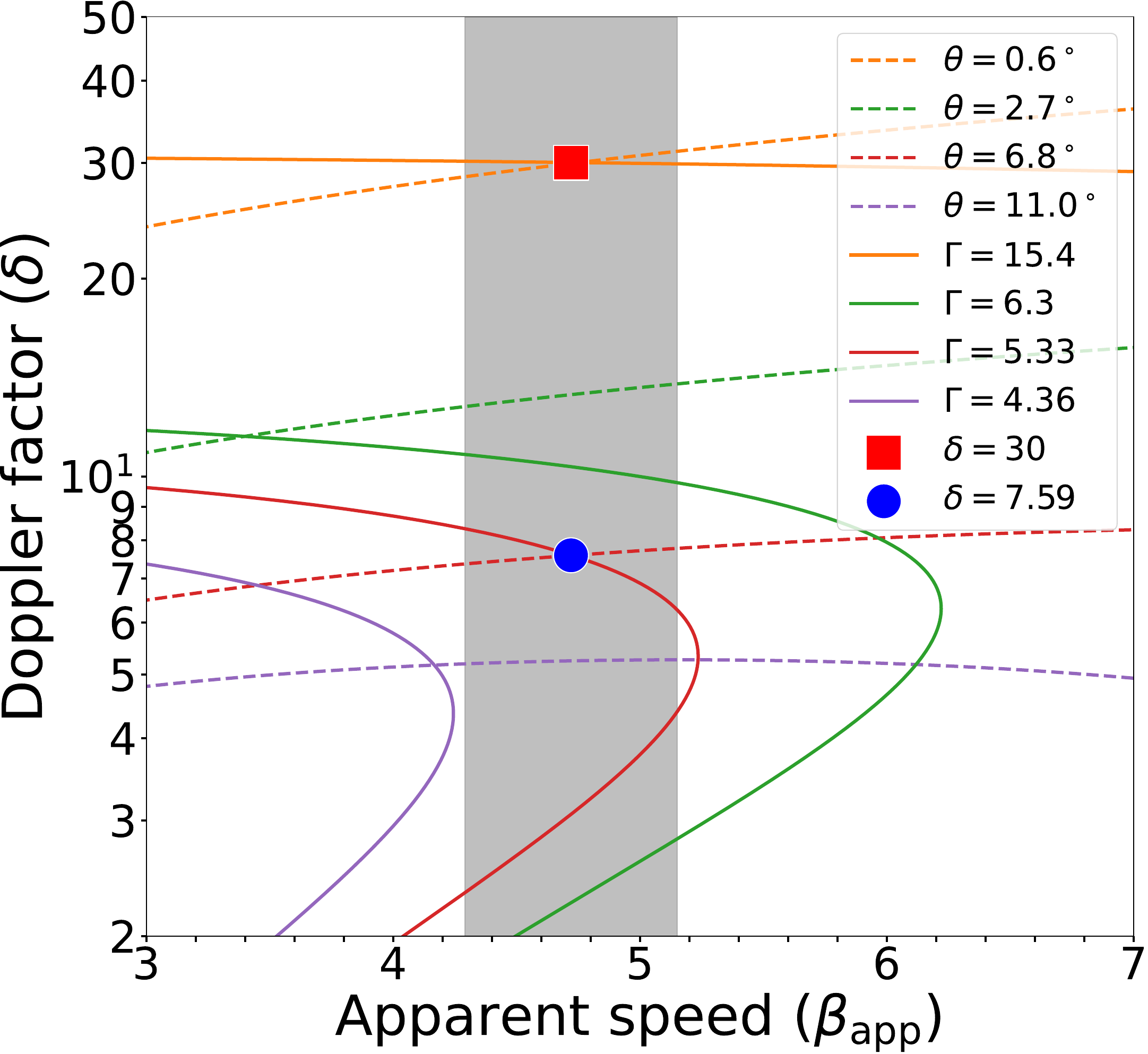}
\caption{
    Relations between observed apparent speeds and Doppler factors for various ranges of the Lorentz factors and viewing angles.
    The gray shaded area denotes the range of observed $\beta_{\rm app}$ for C2.
    $\delta=30$ indicates the value of the Doppler factor used in the literature for theoretical modeling \citep{sahakyan23,acharyya23,prince24} and $\delta=7.59$ corresponds to a value estimated for C2.
}
\label{fig:doppler_vs_bapp}
\end{figure}

The source of external background photon fields around the apparent C1-C2 overlap region is also important not only for the production of high-energy neutrinos via the proton-photon interactions (see, e.g., \citealt{padovani19}) but also for the contemporaneous gamma-to-optical flares in the source
(e.g., \citealt{sahakyan23,acharyya23,prince24}).
Here we primarily consider the external inverse Compton model for a lepto-hadronic jet, given various issues demonstrated by the synchrotron self-Compton models employed in previous works to explain the \ic event in association with \pks \citep{sahakyan23,acharyya23,prince24}.

In the external inverse-Compton models, the distance between the central engine and the site of high-energy photons and neutrinos is a crucial parameter. 
We estimate the distance between the apparent C1-C2 overlap region and possible sources of external radiation fields \,as follows. 
On 2021 Dec 08 (i.e., the time of \ic), the component C1 is estimated to be $\sim0.13\pm0.01$\,mas away from the VLBI core (C0), according to the linear motion fit (Fig. \ref{fig:distance}). Using the 5.53\,pc/mas scaling factor at the distance of \pks and $\theta=6.83^{\circ}\pm4.11^{\circ}$, the de-projected distance between the core and C1 is $\sim6.05\pm3.65\,$pc. 
The central engine can be located further upstream of the 43\,GHz VLBI core due to the coreshift effect (see also \S\ref{subsec:radio}), 
given significant 15-43\,GHz nuclear opacity in the inner 0.4\,mas region of the jet in \pks after the detection of \ic.
Thus the above $\sim6.05\pm3.65\,$pc distance can be considered as a lower limit.

We note that this distance is significantly longer than that of the broad line region (BLR) from the central engine of \pks, $R_{\rm BLR}=0.79\times10^{17}\,$cm\,$\sim0.026\,$pc (e.g., \citealt{prince24}) by a factor of at least $\sim230$, making it challenging for the BLR to be the primary source of external UV photon fields.
Another source of seed photons could be the dusty molecular torus around the central SMBH, which could typically extend up to several hundreds of parsecs (see, e.g., \citealt{jaffe04,carilli19}). 
However, AGN torii will mainly provide infrared photons with their maximum dust sublimation temperatures of $\sim1000-2000\,$K occurring near the central BH (e.g., \citealt{kishimoto07}), while radiation field at UV energies is considered to be necessary to explain high-energy flares in \pks especially at $\gamma$-ray energies \citep{sahakyan23,acharyya23,prince24}.

Therefore, additional sources of external background radiation field should be considered, or models that are less sensitive to the radiative environment would be more appropriate for \ic.
One of such models is the spine-sheath scenario, in which slow sheath layer surrounding the central relativistic spine of a two-flow jet (e.g., \citealt{ghisellini05}). 
In particular, the relative velocity difference between the two flows can greatly amplify the photon field around the central spine, leading to productions of $>$100\,TeV neutrinos, provided the central spine carries abundant high-energy cosmic rays \citep{tavecchio14,tavecchio15}. 
Although detailed calculations of the broadband SEDs is beyond the scope of our work, we briefly comment on two other neutrino AGNs for which the spine-sheath model also showed some relevance, and suggest that additional studies should be worthwhile.
For instance, \citet{ros20} found a newly developed edge-brightened structure in the inner jet of neutrino blazar TXS\,0506+056 after the IceCube-170922A event.
Also, \citet{britzen21} reported a peculiar arc-like sheath in PKS\,1502+106 close to the time of an IceCube neutrino event that could provide the necessary seed photons for a passing jet component, for the inverse Compton processes to occur. 
To investigate the relevance of the spine-sheath scenario for \pks, higher-resolution imaging of \pks in search of limb-brightened jet structure would be important.

\subsection{The Doppler factor crisis in \pks}\label{subsec:doppler_crisis}

Lastly, we discuss the large discrepancy in the Doppler factors in \pks constrained by our radio VLBI data analysis (i.e., $\delta_{\rm var}\sim7.6\pm3.2$ in C2 which likely represents the underlying jet flow; see \S\ref{subsec:discussion1}) and a larger value of $\delta\sim30$ that has been widely adopted for the emission modeling in the literature.
In fact, such a discrepancy is well-known as the `Doppler crisis' (see, e.g., references in \citealt{ros20}), and was also noted in earlier studies of \pks concerning \ic (see, e.g., \citealt{sahakyan23}).

Besides the aforementioned spine-sheath model, another scenario which could reconcile the discrepancy is the bending of the jet and its deceleration after the apparent C1-C2 overlap region.
This is motivated by the fact that C2 displays a large change in the apparent position angle over time by $\sim90^{\circ}$ (Fig. \ref{fig:PA}).
One possible way to change the apparent jet position angles is to vary the intrinsic jet viewing angle along the jet propagation.
This could, in turn, introduce large changes in the jet Doppler factor during a short period of time, resembling a flare from the frame of observer (e.g., \citealt{raiteri17}), in contrast to our calculations based on a single time-averaged $\delta$ for C2 during a long period of the years $\sim2022-2024$.
Below we adopt a toy model in which the jet Doppler factor was as large as $\delta\sim30$ at the time of \ic, and later it decreases to $\delta\sim7.6$ with increasing jet viewing angle (i.e., jet bending away from the observer).
For this, we show in Fig. \ref{fig:doppler_vs_bapp} the relation between the Doppler factor $\delta=1/(\Gamma(1-\beta\cos\theta))$ and $\beta_{\rm app}$ for various $\Gamma$ and $\theta$. 
For $\delta\sim30$ and a fixed $\beta_{\rm app}=4.72\pm0.43$, we find $\theta\sim0.6^{\circ}$ and $\Gamma\sim15$ are required.
We note that somewhat less relativistic values of $\Gamma\sim7-10$ have been reported in the jet of \pks by \cite{weaver22} during 2007-2018.
To achieve a lower $\delta_{\rm var}=7.59\pm3.15$ with the same $\beta_{\rm app}$, smaller $\Gamma_{\rm var}=5.53\pm0.97$ and larger $\theta_{\rm var}=6.83^{\circ}\pm4.11^{\circ}$ are necessary, implying not only the jet bending but also significant jet deceleration.
We note that  significant decelerations are also observed in jets of blazars in general, on scales of $\sim10-100$\,pc (projected; \citealt{homan15}).
As for the jet viewing angle, although the physical reason for the jet bending could be various (see, e.g., discussions in \citealt{kim20}), from the observational perspectives \cite{conway93} and \cite{singal16} note that a change in the apparent position angle by $\sim90^{\circ}$ could result from only a few degrees of change in the intrinsic $\theta$ for large bulk Lorentz factor of $\Gamma\sim10$, thus compatible with the observed kinematics of C2. In this perspective, the change of $\theta\sim0.6{^\circ}\rightarrow6.8^{\circ}$ is physically not impossible. 
Additional joint analysis, especially using the temporal evolution of the radio synchrotron spectra, would be able to examine if the above-mentioned temporal change of the Doppler factor could reproduce the synchrotron peak frequency and peak flux density (see, e.g., \citealt{fromm11}).

\section{Conclusions}
\label{sec:conclusions}

In this paper, we studied in detail the VLBI-scale radio flux, spectral, and structural changes of the jet in \pks around the time of the \ic high energy neutrino event, in order to localize the site of the neutrino emission and constrain physical properties of the jet, which can shed light on the production of the extragalactic TeV-PeV neutrinos from AGNs.
Our main conclusions can be summarized as follows.
\begin{enumerate}
    \item As pointed out by various studies (e.g., \citealt{sahakyan23,acharyya23,prince24}), \pks was in a highly elevated state in all electromagnetic wavelengths around the time of \ic. At 15 and 43\,GHz, the compact mas-scale VLBI nuclear region accounts for the  overall evolution of the radio light curves.
    \item The VLBI nuclear region shows a significantly flatter synchrotron spectrum after the detection of \ic (mean values of $<\alpha_{15}^{43}>\sim-0.34 \rightarrow 0.07$ from before to after the \ic event), suggesting increased synchrotron opacity and therefore higher energies of emitting particles in the compact nuclear region.
    \item We find a subluminal component (C1) at $\sim0.1-0.2$\,mas downstream of the VLBI core (C0) after the \ic event, and the appearance of a new VLBI component (C2) from the core before the time of \ic. 
    Remarkably, the C2 component apparently passed through C1 at the time of \ic within 1$\sigma$ statistical uncertainty, suggesting that the location where this apparent passage occurs can be a probable spatial origin of the high-energy neutrino.
    \item In particular, the C2 component travels at an apparent speed of $\beta_{\rm app}\sim4.8$, which is comparable to the previously measured maximum speeds in the jet of \pks  \citep{lister19,weaver22} and thus suggest its origin to be related to the underlying bulk relativistic jet flow.
    The Doppler and Lorentz factors of C2 are estimated from the flux variability, finding $\delta_{\rm var}\sim7.6$ and $\Gamma_{\rm var}\sim5.3$, resulting in  
    a slightly large viewing angle of $\theta_{\rm var}\sim6.8^{\circ}$. Our derived $\delta$ and $\Gamma$ are somewhat smaller than those reported about \pks in the literature \citep{weaver22}.
    On the other hand, the component C2 shows significant apparent bending in its sky-projected trajectory by $\sim50^{\circ}$ followed by the component deceleration and significant re-brightening after $\sim$MJD\,60,000.
    This behavior might be explained by a helical jet motion and the local beaming effect, but continued monitoring of C2 is needed for confirmation.
    \item On the other hand, 
    the subluminal speed of C1 below the median jet speed suggests its pattern-like origin, such as traveling shock, instabilities, or standing shock temporarily displaced from its stable position.
    However, we find that no single scenario can fully explain both the bright emission as well as the jet kinematics.
    This complexity is not too surprising, considering that the source was undergoing a strong outburst close to the time of \ic, which could have also accompanied significant changes in the jet structure, compared to its previous quiescent periods.
    \item 
    We find that the location of the apparent C1-C2 overlap region is too far from the central engine, $\gtrsim6.5\,$pc (de-projected, using the above jet viewing angle). The large distance challenges external inverse-Compton scattering models in the literature (e.g., \citealt{prince24}), in which the BLR provides sufficient external photon field. 
    Therefore, we suggest that other sources of external photons are required, or models that are less sensitive to the external photon field should be considered (e.g., spine-sheath model; \citealt{tavecchio15}). 
    \item Lastly, to reconcile the discrepancy in the observed small and required large Doppler factors (i.e., $\delta_{\rm var}\sim7.6$ from our observational study versus $\delta\sim30$ used by models in the literature; e.g., \citealt{sahakyan23}), 
    we suggest a toy model, in which C2 had initially more relativistic speed ($\Gamma\sim15$) and smaller viewing-angle ($\theta\sim0.6^{\circ}$)--sufficient to reproduce $\delta\sim30$--but the jet speed decreased and bent away from observer to larger $\theta$, decreasing $\delta$ to observed values.
    Apparently large position angle change of $\sim50^{\circ}$ in C2 could support this toy model.
\end{enumerate}

Overall, our study has demonstrated the importance of high-resolution, time-monitoring VLBI observations of neutrino blazar candidates to resolve, localize, and track neutrino-emitting plasma and its physics. 
Follow-up studies with quantitative modeling of the SED of \pks using boundary conditions obtained by our study will provide more realistic scenarios of the \ic event and contemporaneous multi-wavelength flares.
Also, continued higher-resolution and polarimetric VLBI observations of neutrino blazar candidates will be helpful in probing their particle acceleration mechanisms, magnetic field properties, and the transverse jet structure (e.g., the spine-sheath configuration) which can be related to the origins of the seed photons and jet dynamics.

\begin{acknowledgements}
% Entry for the table of contents, for this guide only
\addcontentsline{toc}{section}{Acknowledgements}
% Personal
We thank the anonymous referee for helpful comments which improved the manuscript.
J.-Y.K. is grateful to Yuri Y. Kovalev for insightful suggestions and discussions, and Svetlana G. Jorstad for comments about the VLBA 43\,GHz data of \pks.
% Funding
Y.-S.K. and J.-Y.K. are supported for this research by the National Research Foundation of Korea (NRF) grant funded by the Korean government (Ministry of Science and ICT; grant no. 2022R1C1C1005255, RS-2022-NR071771, RS-2022-00197685).
% BU
This study makes use of VLBA data from the VLBA-BU Blazar Monitoring Program (BEAM-ME and VLBA-BU-BLAZAR;
http://www.bu.edu/blazars/BEAM-ME.html), funded by NASA through the Fermi Guest Investigator Program. The VLBA is an instrument of the National Radio Astronomy Observatory. The National Radio Astronomy Observatory is a facility of the National Science Foundation operated by Associated Universities, Inc.
% MOJAVE
This research has made use of data from the MOJAVE database that is maintained by the MOJAVE team \citep{lister18}.
% Others
We acknowledge the use of the Astrophysics Data System (ADS), the Fermi-LAT Light Curve Repository (LCR), and the ASAS-SN sky server.

\end{acknowledgements}

%-------------------------------------------------------------------

\bibliographystyle{aa}
\bibliography{0735}

\begin{thebibliography}{92}
\expandafter\ifx\csname natexlab\endcsname\relax\def\natexlab#1{#1}\fi

\bibitem[{{Abbasi} {et~al.}(2023){Abbasi}, {Ackermann}, {Adams}, {Agarwalla}, {Aguilar}, {Ahlers}, {Alameddine}, {Amin}, {Andeen}, {Anton}, {Arg{\"u}elles}, {Ashida}, {Athanasiadou}, {Axani}, {Bai}, {Balagopal V}, {Baricevic}, {Barwick}, {Basu}, {Bay}, {Beatty}, {Becker}, {Becker Tjus}, {Beise}, {Bellenghi}, {BenZvi}, {Berley}, {Bernardini}, {Besson}, {Binder}, {Bindig}, {Blaufuss}, {Blot}, {Bontempo}, {Book}, {Boscolo Meneguolo}, {B{\"o}ser}, {Botner}, {B{\"o}ttcher}, {Bourbeau}, {Braun}, {Brinson}, {Brostean-Kaiser}, {Burley}, {Busse}, {Butterfield}, {Campana}, {Carloni}, {Carnie-Bronca}, {Chattopadhyay}, {Chau}, {Chen}, {Chen}, {Chirkin}, {Choi}, {Clark}, {Classen}, {Coleman}, {Collin}, {Connolly}, {Conrad}, {Coppin}, {Correa}, {Countryman}, {Cowen}, {Dave}, {De Clercq}, {DeLaunay}, {Delgado}, {Dembinski}, {Deng}, {Deoskar}, {Desai}, {Desiati}, {de Vries}, {de Wasseige}, {DeYoung}, {Diaz}, {D{\'\i}az-V{\'e}lez}, {Dittmer}, {Domi}, {Dujmovic}, {DuVernois}, {Ehrhardt}, {Eller}, {Engel}, {Erpenbeck}, {Evans},
  {Evenson}, {Fan}, {Fang}, {Farrag}, {Fazely}, {Fedynitch}, {Feigl}, {Fiedlschuster}, {Finley}, {Fischer}, {Fox}, {Franckowiak}, {Friedman}, {Fritz}, {F{\"u}rst}, {Gaisser}, {Gallagher}, {Ganster}, {Garcia}, {Gerhardt}, {Ghadimi}, {Glaser}, {Glauch}, {Gl{\"u}senkamp}, {Goehlke}, {Gonzalez}, {Goswami}, {Grant}, {Gray}, {Griffin}, {Griswold}, {G{\"u}nther}, {Gutjahr}, {Haack}, {Hallgren}, {Halliday}, {Halve}, {Halzen}, {Hamdaoui}, {Ha Minh}, {Hanson}, {Hardin}, {Harnisch}, {Hatch}, {Haungs}, {Helbing}, {Hellrung}, {Henningsen}, {Heuermann}, {Heyer}, {Hickford}, {Hidvegi}, {Hill}, {Hill}, {Hoffman}, {Hoshina}, {Hou}, {Huber}, {Hultqvist}, {H{\"u}nnefeld}, {Hussain}, {Hymon}, {In}, {Ishihara}, {Jacquart}, {Janik}, {Jansson}, {Japaridze}, {Jayakumar}, {Jeong}, {Jin}, {Jones}, {Kang}, {Kang}, {Kang}, {Kappes}, {Kappesser}, {Kardum}, {Karg}, {Karl}, {Karle}, {Katz}, {Kauer}, {Kelley}, {Zathul}, {Kheirandish}, {Kiryluk}, {Klein}, {Kochocki}, {Koirala}, {Kolanoski}, {Kontrimas}, {K{\"o}pke}, {Kopper}, {Koskinen},
  {Koundal}, {Kovacevich}, {Kowalski}, {Kozynets}, {Kruiswijk}, {Krupczak}, {Kumar}, {Kun}, {Kurahashi}, {Lad}, {Lagunas Gualda}, {Lamoureux}, {Larson}, {Lauber}, {Lazar}, {Lee}, {Leonard DeHolton}, {Leszczy{\'n}ska}, {Lincetto}, {Liu}, {Liubarska}, {Lohfink}, {Love}, {Lozano Mariscal}, {Lu}, {Lucarelli}, {Ludwig}, {Luszczak}, {Lyu}, {Madsen}, {Mahn}, {Makino}, {Manao}, {Mancina}, {Marie Sainte}, {Mari{\c{s}}}, {Marka}, {Marka}, {Marsee}, {Martinez-Soler}, {Maruyama}, {Mayhew}, {McElroy}, {McNally}, {Mead}, {Meagher}, {Mechbal}, {Medina}, {Meier}, {Merckx}, {Merten}, {Micallef}, {Montaruli}, {Moore}, {Morii}, {Morse}, {Moulai}, {Mukherjee}, {Naab}, {Nagai}, {Nakos}, {Naumann}, {Necker}, {Neumann}, {Niederhausen}, {Nisa}, {Noell}, {Nowicki}, {Obertacke Pollmann}, {O'Dell}, {Oehler}, {Oeyen}, {Olivas}, {Orsoe}, {Osborn}, {O'Sullivan}, {Pandya}, {Park}, {Parker}, {Paudel}, {Paul}, {P{\'e}rez de los Heros}, {Peterson}, {Philippen}, {Pieper}, {Pizzuto}, {Plum}, {Pont{\'e}n}, {Popovych}, {Prado Rodriguez}, {Pries},
  {Procter-Murphy}, {Przybylski}, {Rack-Helleis}, {Rawlins}, {Rechav}, {Rehman}, {Reichherzer}, {Renzi}, {Resconi}, {Reusch}, {Rhode}, {Richman}, {Riedel}, {Roberts}, {Robertson}, {Rodan}, {Roellinghoff}, {Rongen}, {Rott}, {Ruhe}, {Ruohan}, {Ryckbosch}, {Safa}, {Saffer}, {Salazar-Gallegos}, {Sampathkumar}, {Sanchez Herrera}, {Sandrock}, {Santander}, {Sarkar}, {Sarkar}, {Savelberg}, {Savina}, {Schaufel}, {Schieler}, {Schindler}, {Schl{\"u}ter}, {Schl{\"u}ter}, {Schmidt}, {Schneider}, {Schr{\"o}der}, {Schumacher}, {Schwefer}, {Sclafani}, {Seckel}, {Seunarine}, {Shah}, {Sharma}, {Shefali}, {Shimizu}, {Silva}, {Skrzypek}, {Smithers}, {Snihur}, {Soedingrekso}, {S{\o}gaard}, {Soldin}, {Sommani}, {Spannfellner}, {Spiczak}, {Spiering}, {Stamatikos}, {Stanev}, {Stezelberger}, {St{\"u}rwald}, {Stuttard}, {Sullivan}, {Taboada}, {Ter-Antonyan}, {Thiesmeyer}, {Thompson}, {Thwaites}, {Tilav}, {Tollefson}, {T{\"o}nnis}, {Toscano}, {Tosi}, {Trettin}, {Tung}, {Turcotte}, {Twagirayezu}, {Ty}, {Unland Elorrieta}, {Upadhyay},
  {Upshaw}, {Valtonen-Mattila}, {Vandenbroucke}, {van Eijndhoven}, {Vannerom}, {van Santen}, {Vara}, {Veitch-Michaelis}, {Venugopal}, {Verpoest}, {Veske}, {Walck}, {Watson}, {Weaver}, {Weigel}, {Weindl}, {Weldert}, {Wendt}, {Werthebach}, {Weyrauch}, {Whitehorn}, {Wiebusch}, {Willey}, {Williams}, {Wolf}, {Wolf}, {Wrede}, {Xu}, {Yanez}, {Yildizci}, {Yoshida}, {Yu}, {Yu}, {Yuan}, {Zhang}, \& {Zhelnin}}]{icecube23}
{Abbasi}, R., {Ackermann}, M., {Adams}, J., {et~al.} 2023, \apjs, 269, 25

\bibitem[{{Abdollahi} {et~al.}(2020){Abdollahi}, {Acero}, {Ackermann}, {Ajello}, {Atwood}, {Axelsson}, {Baldini}, {Ballet}, {Barbiellini}, {Bastieri}, {Becerra Gonzalez}, {Bellazzini}, {Berretta}, {Bissaldi}, {Blandford}, {Bloom}, {Bonino}, {Bottacini}, {Brandt}, {Bregeon}, {Bruel}, {Buehler}, {Burnett}, {Buson}, {Cameron}, {Caputo}, {Caraveo}, {Casandjian}, {Castro}, {Cavazzuti}, {Charles}, {Chaty}, {Chen}, {Cheung}, {Chiaro}, {Ciprini}, {Cohen-Tanugi}, {Cominsky}, {Coronado-Bl{\'a}zquez}, {Costantin}, {Cuoco}, {Cutini}, {D'Ammando}, {DeKlotz}, {de la Torre Luque}, {de Palma}, {Desai}, {Digel}, {Di Lalla}, {Di Mauro}, {Di Venere}, {Dom{\'\i}nguez}, {Dumora}, {Fana Dirirsa}, {Fegan}, {Ferrara}, {Franckowiak}, {Fukazawa}, {Funk}, {Fusco}, {Gargano}, {Gasparrini}, {Giglietto}, {Giommi}, {Giordano}, {Giroletti}, {Glanzman}, {Green}, {Grenier}, {Griffin}, {Grondin}, {Grove}, {Guiriec}, {Harding}, {Hayashi}, {Hays}, {Hewitt}, {Horan}, {J{\'o}hannesson}, {Johnson}, {Kamae}, {Kerr}, {Kocevski}, {Kovac'evic'},
  {Kuss}, {Landriu}, {Larsson}, {Latronico}, {Lemoine-Goumard}, {Li}, {Liodakis}, {Longo}, {Loparco}, {Lott}, {Lovellette}, {Lubrano}, {Madejski}, {Maldera}, {Malyshev}, {Manfreda}, {Marchesini}, {Marcotulli}, {Mart{\'\i}-Devesa}, {Martin}, {Massaro}, {Mazziotta}, {McEnery}, {Mereu}, {Meyer}, {Michelson}, {Mirabal}, {Mizuno}, {Monzani}, {Morselli}, {Moskalenko}, {Negro}, {Nuss}, {Ojha}, {Omodei}, {Orienti}, {Orlando}, {Ormes}, {Palatiello}, {Paliya}, {Paneque}, {Pei}, {Pe{\~n}a-Herazo}, {Perkins}, {Persic}, {Pesce-Rollins}, {Petrosian}, {Petrov}, {Piron}, {Poon}, {Porter}, {Principe}, {Rain{\`o}}, {Rando}, {Razzano}, {Razzaque}, {Reimer}, {Reimer}, {Remy}, {Reposeur}, {Romani}, {Saz Parkinson}, {Schinzel}, {Serini}, {Sgr{\`o}}, {Siskind}, {Smith}, {Spandre}, {Spinelli}, {Strong}, {Suson}, {Tajima}, {Takahashi}, {Tak}, {Thayer}, {Thompson}, {Tibaldo}, {Torres}, {Torresi}, {Valverde}, {Van Klaveren}, {van Zyl}, {Wood}, {Yassine}, \& {Zaharijas}}]{4FGL}
{Abdollahi}, S., {Acero}, F., {Ackermann}, M., {et~al.} 2020, \apjs, 247, 33

\bibitem[{{Abdollahi} {et~al.}(2023){Abdollahi}, {Ajello}, {Baldini}, {Ballet}, {Bastieri}, {Becerra Gonzalez}, {Bellazzini}, {Berretta}, {Bissaldi}, {Bonino}, {Brill}, {Bruel}, {Burns}, {Buson}, {Cameron}, {Caputo}, {Caraveo}, {Cibrario}, {Ciprini}, {Cristarella Orestano}, {Crnogorcevic}, {Cutini}, {D'Ammando}, {De Gaetano}, {Digel}, {Di Lalla}, {Di Venere}, {Dom{\'\i}nguez}, {Ramazani}, {Fegan}, {Ferrara}, {Fiori}, {Fleischhack}, {Franckowiak}, {Fukazawa}, {Fusco}, {Gammaldi}, {Gargano}, {Garrappa}, {Gasbarra}, {Gasparrini}, {Giglietto}, {Giordano}, {Giroletti}, {Green}, {Grenier}, {Guiriec}, {Gustafsson}, {Hays}, {Horan}, {Hou}, {J{\'o}hannesson}, {Kerr}, {Kocevski}, {Kuss}, {Latronico}, {Li}, {Liodakis}, {Longo}, {Loparco}, {Lorusso}, {Lott}, {Lovellette}, {Lubrano}, {Maldera}, {Manfreda}, {Mart{\'\i}-Devesa}, {Mazziotta}, {Mereu}, {Meyer}, {Michelson}, {Mizuno}, {Monzani}, {Morselli}, {Moskalenko}, {Negro}, {Omodei}, {Orlando}, {Ormes}, {Paneque}, {Panzarini}, {Perkins}, {Persic}, {Pesce-Rollins},
  {Pillera}, {Porter}, {Principe}, {Racusin}, {Rain{\`o}}, {Rando}, {Rani}, {Razzano}, {Razzaque}, {Reimer}, {Reimer}, {S{\'a}nchez-Conde}, {Parkinson}, {Scargle}, {Scotton}, {Serini}, {Sgr{\`o}}, {Siskind}, {Spandre}, {Spinelli}, {Suson}, {Tajima}, {Thompson}, {Torres}, {Valverde}, {Venters}, {Wadiasingh}, {Wagner}, \& {Wood}}]{abdollahi23}
{Abdollahi}, S., {Ajello}, M., {Baldini}, L., {et~al.} 2023, \apjs, 265, 31

\bibitem[{{Acharyya} {et~al.}(2023){Acharyya}, {Adams}, {Archer}, {Bangale}, {Bartkoske}, {Batista}, {Benbow}, {Brill}, {Buckley}, {Christiansen}, {Chromey}, {Errando}, {Falcone}, {Feng}, {Foote}, {Fortson}, {Furniss}, {Gallagher}, {Hanlon}, {Hanna}, {Hervet}, {Hinrichs}, {Hoang}, {Holder}, {Humensky}, {Jin}, {Kaaret}, {Kertzman}, {Kherlakian}, {Kieda}, {Kleiner}, {Korzoun}, {Kumar}, {Lang}, {Lundy}, {Maier}, {McGrath}, {Millard}, {Millis}, {Mooney}, {Moriarty}, {Mukherjee}, {O'Brien}, {Ong}, {Pohl}, {Pueschel}, {Quinn}, {Ragan}, {Reynolds}, {Ribeiro}, {Roache}, {Sadeh}, {Sadun}, {Saha}, {Santander}, {Sembroski}, {Shang}, {Splettstoesser}, {Talluri}, {Tucci}, {Vassiliev}, {Weinstein}, {Williams}, {Wong}, {Woo}, {Aharonian}, {Aschersleben}, {Backes}, {Martins}, {Batzofin}, {Becherini}, {Berge}, {Bernl{\"o}hr}, {Bi}, {B{\"o}ttcher}, {Boisson}, {Bolmont}, {de Bony de Lavergne}, {Borowska}, {Bouyahiaoui}, {Bradascio}, {Breuhaus}, {Brose}, {Brun}, {Bruno}, {Bulik}, {Burger-Scheidlin}, {Caroff}, {Casanova},
  {Cecil}, {Celic}, {Cerruti}, {Chand}, {Chandra}, {Chen}, {Chibueze}, {Chibueze}, {Cotter}, {Dai}, {Mbarubucyeye}, {Djannati-Ata{\"\i}}, {Dmytriiev}, {Doroshenko}, {Einecke}, {Ernenwein}, {de Clairfontaine}, {Filipovic}, {Fontaine}, {F{\"u}{\ss}ling}, {Funk}, {Gabici}, {Ghafourizadeh}, {Giavitto}, {Glawion}, {Glicenstein}, {Goswami}, {Grolleron}, {Haerer}, {Hinton}, {Holch}, {Holler}, {Horns}, {Jamrozy}, {Jankowsky}, {Joshi}, {Jung-Richardt}, {Kasai}, {Katarzy{\'n}ski}, {Khatoon}, {Kh{\'e}lifi}, {Klepser}, {Klu{\'z}niak}, {Kosack}, {Kostunin}, {Lang}, {Le Stum}, {Lemi{\`e}re}, {Lenain}, {Leuschner}, {Lohse}, {Luashvili}, {Lypova}, {Mackey}, {Malyshev}, {Marandon}, {Marchegiani}, {Marcowith}, {Mart{\'\i}-Devesa}, {Marx}, {Mitchell}, {Moderski}, {Mohrmann}, {Montanari}, {Moulin}, {Murach}, {Nakashima}, {Niemiec}, {Noel}, {O'Brien}, {Olivera-Nieto}, {de Ona Wilhelmi}, {Ostrowski}, {Panny}, {Panter}, {Peron}, {Prokhorov}, {P{\"u}hlhofer}, {Punch}, {Quirrenbach}, {Reichherzer}, {Reimer}, {Reimer}, {Ren},
  {Renaud}, {Rieger}, {Rudak}, {Ruiz-Velasco}, {Sahakian}, {Santangelo}, {Sasaki}, {Sch{\"a}fer}, {Sch{\"u}ssler}, {Schutte}, {Schwanke}, {Shapopi}, {Specovius}, {Spencer}, {Stawarz}, {Steenkamp}, {Steinmassl}, {Sushch}, {Suzuki}, {Takahashi}, {Tanaka}, {Terrier}, {van Eldik}, {Vecchi}, {Veh}, {Venter}, {Vink}, {White}, {Wierzcholska}, {Wong}, {Zacharias}, {Zargaryan}, {Zdziarski}, {Zech}, {Zouari}, {{\.Z}ywucka}, {Mori}, \& {H.~E.~S.~S. Collaboration}}]{acharyya23}
{Acharyya}, A., {Adams}, C.~B., {Archer}, A., {et~al.} 2023, \apj, 954, 70

\bibitem[{{Agudo} {et~al.}(2006){Agudo}, {G{\'o}mez}, {Gabuzda}, {Marscher}, {Jorstad}, \& {Alberdi}}]{agudo06}
{Agudo}, I., {G{\'o}mez}, J.~L., {Gabuzda}, D.~C., {et~al.} 2006, \aap, 453, 477

\bibitem[{{Agudo} {et~al.}(2001){Agudo}, {G{\'o}mez}, {Mart{\'\i}}, {Ib{\'a}{\~n}ez}, {Marscher}, {Alberdi}, {Aloy}, \& {Hardee}}]{agudo01}
{Agudo}, I., {G{\'o}mez}, J.-L., {Mart{\'\i}}, J.-M., {et~al.} 2001, \apjl, 549, L183

\bibitem[{{Beuchert} {et~al.}(2018){Beuchert}, {Kadler}, {Perucho}, {Gro{\ss}berger}, {Schulz}, {Agudo}, {Casadio}, {G{\'o}mez}, {Gurwell}, {Homan}, {Kovalev}, {Lister}, {Markoff}, {Molina}, {Pushkarev}, {Ros}, {Savolainen}, {Steinbring}, {Thum}, \& {Wilms}}]{beutchert18}
{Beuchert}, T., {Kadler}, M., {Perucho}, M., {et~al.} 2018, \aap, 610, A32

\bibitem[{{Bharathan} {et~al.}(2024){Bharathan}, {Stalin}, {Sahayanathan}, {Bhattacharyya}, \& {Mathew}}]{bharathan24}
{Bharathan}, A.~M., {Stalin}, C.~S., {Sahayanathan}, S., {Bhattacharyya}, S., \& {Mathew}, B. 2024, \mnras, 529, 3503

\bibitem[{{Blandford} \& {K{\"o}nigl}(1979)}]{blandford79}
{Blandford}, R.~D. \& {K{\"o}nigl}, A. 1979, \apj, 232, 34

\bibitem[{{Boettcher} {et~al.}(2012){Boettcher}, {Harris}, \& {Krawczynski}}]{boettcher12}
{Boettcher}, M., {Harris}, D.~E., \& {Krawczynski}, H. 2012, {Relativistic Jets from Active Galactic Nuclei}

\bibitem[{{Britzen} {et~al.}(2021){Britzen}, {Zaja{\v{c}}ek}, {Popovi{\'c}}, {Fendt}, {Tramacere}, {Pashchenko}, {Jaron}, {P{\'a}nis}, {Petrov}, {Aller}, \& {Aller}}]{britzen21}
{Britzen}, S., {Zaja{\v{c}}ek}, M., {Popovi{\'c}}, L.~{\v{C}}., {et~al.} 2021, \mnras, 503, 3145

\bibitem[{{Carilli} {et~al.}(2019){Carilli}, {Perley}, {Dhawan}, \& {Perley}}]{carilli19}
{Carilli}, C.~L., {Perley}, R.~A., {Dhawan}, V., \& {Perley}, D.~A. 2019, \apjl, 874, L32

\bibitem[{{Casadio} {et~al.}(2015){Casadio}, {G{\'o}mez}, {Jorstad}, {Marscher}, {Larionov}, {Smith}, {Gurwell}, {L{\"a}hteenm{\"a}ki}, {Agudo}, {Molina}, {Bala}, {Joshi}, {Taylor}, {Williamson}, {Arkharov}, {Blinov}, {Borman}, {Di Paola}, {Grishina}, {Hagen-Thorn}, {Itoh}, {Kopatskaya}, {Larionova}, {Larionova}, {Morozova}, {Rastorgueva-Foi}, {Sergeev}, {Tornikoski}, {Troitsky}, {Thum}, \& {Wiesemeyer}}]{casadio15}
{Casadio}, C., {G{\'o}mez}, J.~L., {Jorstad}, S.~G., {et~al.} 2015, \apj, 813, 51

\bibitem[{{Cawthorne}(2006)}]{cawthorne06}
{Cawthorne}, T.~V. 2006, \mnras, 367, 851

\bibitem[{{Cerruti} {et~al.}(2019){Cerruti}, {Zech}, {Boisson}, {Emery}, {Inoue}, \& {Lenain}}]{cerruti19}
{Cerruti}, M., {Zech}, A., {Boisson}, C., {et~al.} 2019, \mnras, 483, L12

\bibitem[{{Cohen} {et~al.}(2014){Cohen}, {Meier}, {Arshakian}, {Homan}, {Hovatta}, {Kovalev}, {Lister}, {Pushkarev}, {Richards}, \& {Savolainen}}]{cohen14}
{Cohen}, M.~H., {Meier}, D.~L., {Arshakian}, T.~G., {et~al.} 2014, \apj, 787, 151

\bibitem[{{Conway} \& {Murphy}(1993)}]{conway93}
{Conway}, J.~E. \& {Murphy}, D.~W. 1993, \apj, 411, 89

\bibitem[{{Daly} \& {Marscher}(1988)}]{daly88}
{Daly}, R.~A. \& {Marscher}, A.~P. 1988, \apj, 334, 539

\bibitem[{{Dermer} {et~al.}(2014){Dermer}, {Murase}, \& {Inoue}}]{dermer14}
{Dermer}, C.~D., {Murase}, K., \& {Inoue}, Y. 2014, Journal of High Energy Astrophysics, 3, 29

\bibitem[{{Dzhilkibaev} {et~al.}(2021){Dzhilkibaev}, {Suvorova}, \& {Baikal-GVD Collaboration}}]{baikal21}
{Dzhilkibaev}, Z.~A., {Suvorova}, O., \& {Baikal-GVD Collaboration}. 2021, The Astronomer's Telegram, 15112, 1

\bibitem[{{Falomo} {et~al.}(2021){Falomo}, {Treves}, \& {Paiano}}]{falomo21}
{Falomo}, R., {Treves}, A., \& {Paiano}, S. 2021, The Astronomer's Telegram, 15132, 1

\bibitem[{{Filippini} {et~al.}(2022){Filippini}, {Illuminati}, {Heijboer}, {Gatius}, {Muller}, {Dornic}, {Huang}, {Le Stum}, {Palacios Gonz{\'a}lez}, {Celli}, {Zegarelli}, {Coniglione}, {Samtleben}, {Kovalev}, \& {Plavin}}]{km3net22}
{Filippini}, F., {Illuminati}, G., {Heijboer}, A., {et~al.} 2022, The Astronomer's Telegram, 15290, 1

\bibitem[{{Fromm} {et~al.}(2016){Fromm}, {Perucho}, {Mimica}, \& {Ros}}]{fromm16}
{Fromm}, C.~M., {Perucho}, M., {Mimica}, P., \& {Ros}, E. 2016, \aap, 588, A101

\bibitem[{{Fromm} {et~al.}(2011){Fromm}, {Perucho}, {Ros}, {Savolainen}, {Lobanov}, {Zensus}, {Aller}, {Aller}, {Gurwell}, \& {L{\"a}hteenm{\"a}ki}}]{fromm11}
{Fromm}, C.~M., {Perucho}, M., {Ros}, E., {et~al.} 2011, \aap, 531, A95

\bibitem[{{Fromm} {et~al.}(2013){Fromm}, {Ros}, {Perucho}, {Savolainen}, {Mimica}, {Kadler}, {Lobanov}, {Lister}, {Kovalev}, \& {Zensus}}]{fromm13}
{Fromm}, C.~M., {Ros}, E., {Perucho}, M., {et~al.} 2013, \aap, 551, A32

\bibitem[{{Fuentes} {et~al.}(2023){Fuentes}, {G{\'o}mez}, {Mart{\'\i}}, {Perucho}, {Zhao}, {Lico}, {Lobanov}, {Bruni}, {Kovalev}, {Chael}, {Akiyama}, {Bouman}, {Sun}, {Cho}, {Traianou}, {Toscano}, {Dahale}, {Foschi}, {Gurvits}, {Jorstad}, {Kim}, {Marscher}, {Mizuno}, {Ros}, \& {Savolainen}}]{fuentes23}
{Fuentes}, A., {G{\'o}mez}, J.~L., {Mart{\'\i}}, J.~M., {et~al.} 2023, Nature Astronomy, 7, 1359

\bibitem[{{Gabuzda} {et~al.}(1994){Gabuzda}, {Wardle}, {Roberts}, {Aller}, \& {Aller}}]{gabuzda94}
{Gabuzda}, D.~C., {Wardle}, J.~F.~C., {Roberts}, D.~H., {Aller}, M.~F., \& {Aller}, H.~D. 1994, \apj, 435, 128

\bibitem[{{Garrappa} {et~al.}(2019){Garrappa}, {Buson}, {Franckowiak}, {Fermi-LAT Collaboration}, {Shappee}, {Beacom}, {Dong}, {Holoien}, {Kochanek}, {Prieto}, {Stanek}, {Thompson}, {ASAS-SN Collaboration}, {Aartsen}, {Ackermann}, {Adams}, {Aguilar}, {Ahlers}, {Ahrens}, {Alispach}, {Andeen}, {Anderson}, {Ansseau}, {Anton}, {Arg{\"u}elles}, {Auffenberg}, {Axani}, {Backes}, {Bagherpour}, {Bai}, {Barbano}, {Barwick}, {Baum}, {Bay}, {Beatty}, {Becker}, {Becker Tjus}, {BenZvi}, {Berley}, {Bernardini}, {Besson}, {Binder}, {Bindig}, {Blaufuss}, {Blot}, {Bohm}, {B{\"o}rner}, {B{\"o}ser}, {Botner}, {Bourbeau}, {Bourbeau}, {Bradascio}, {Braun}, {Bretz}, {Bron}, {Brostean-Kaiser}, {Burgman}, {Busse}, {Carver}, {Chen}, {Cheung}, {Chirkin}, {Clark}, {Classen}, {Collin}, {Conrad}, {Coppin}, {Correa}, {Cowen}, {Cross}, {Dave}, {de Andr{\'e}}, {De Clercq}, {DeLaunay}, {Dembinski}, {Deoskar}, {De Ridder}, {Desiati}, {de Vries}, {de Wasseige}, {de With}, {DeYoung}, {Diaz}, {D{\'\i}az-V{\'e}lez}, {Dujmovic}, {Dunkman},
  {Dvorak}, {Eberhardt}, {Ehrhardt}, {Eller}, {Evenson}, {Fahey}, {Fazely}, {Felde}, {Filimonov}, {Finley}, {Franckowiak}, {Friedman}, {Fritz}, {Gaisser}, {Gallagher}, {Ganster}, {Garrappa}, {Gerhardt}, {Ghorbani}, {Glauch}, {Gl{\"u}senkamp}, {Goldschmidt}, {Gonzalez}, {Grant}, {Griffith}, {G{\"u}nder}, {G{\"u}nd{\"u}z}, {Haack}, {Hallgren}, {Halve}, {Halzen}, {Hanson}, {Hebecker}, {Heereman}, {Helbing}, {Hellauer}, {Henningsen}, {Hickford}, {Hignight}, {Hill}, {Hoffman}, {Hoffmann}, {Hoinka}, {Hokanson-Fasig}, {Hoshina}, {Huang}, {Huber}, {Hultqvist}, {H{\"u}nnefeld}, {Hussain}, {In}, {Iovine}, {Ishihara}, {Jacobi}, {Japaridze}, {Jeong}, {Jero}, {Jones}, {Kang}, {Kappes}, {Kappesser}, {Karg}, {Karl}, {Karle}, {Katz}, {Kauer}, {Keivani}, {Kelley}, {Kheirandish}, {Kim}, {Kintscher}, {Kiryluk}, {Kittler}, {Klein}, {Koirala}, {Kolanoski}, {K{\"o}pke}, {Kopper}, {Kopper}, {Koskinen}, {Kowalski}, {Krings}, {Kr{\"u}ckl}, {Kulacz}, {Kunwar}, {Kurahashi}, {Kyriacou}, {Labare}, {Lanfranchi}, {Larson}, {Lauber},
  {Lazar}, {Leonard}, {Leuermann}, {Liu}, {Lohfink}, {Lozano Mariscal}, {Lu}, {Lucarelli}, {L{\"u}nemann}, {Luszczak}, {Madsen}, {Maggi}, {Mahn}, {Makino}, {Mallot}, {Mancina}, {Mari{\c{s}}}, {Maruyama}, {Mase}, {Maunu}, {Meagher}, {Medici}, {Medina}, {Meier}, {Meighen-Berger}, {Menne}, {Merino}, {Meures}, {Miarecki}, {Micallef}, {Moment{\'e}}, {Montaruli}, {Moore}, {Moulai}, {Nagai}, {Nahnhauer}, {Nakarmi}, {Naumann}, {Neer}, {Niederhausen}, {Nowicki}, {Nygren}, {Obertacke Pollmann}, {Olivas}, {O'Murchadha}, {O'Sullivan}, {Palczewski}, {Pandya}, {Pankova}, {Park}, {Peiffer}, {P{\'e}rez de los Heros}, {Pieloth}, {Pinat}, {Pizzuto}, {Plum}, {Price}, {Przybylski}, {Raab}, {Raissi}, {Rameez}, {Rauch}, {Rawlins}, {Rea}, {Reimann}, {Relethford}, {Renzi}, {Resconi}, {Rhode}, {Richman}, {Robertson}, {Rongen}, {Rott}, {Ruhe}, {Ryckbosch}, {Rysewyk}, {Safa}, {Sanchez Herrera}, {Sandrock}, {Sandroos}, {Santander}, {Sarkar}, {Sarkar}, {Satalecka}, {Schaufel}, {Schlunder}, {Schmidt}, {Schneider}, {Schneider},
  {Schumacher}, {Sclafani}, {Seckel}, {Seunarine}, {Silva}, {Snihur}, {Soedingrekso}, {Soldin}, {Song}, {Spiczak}, {Spiering}, {Stachurska}, {Stamatikos}, {Stanev}, {Stasik}, {Stein}, {Stettner}, {Steuer}, {Stezelberger}, {Stokstad}, {St{\"o}{\ss}l}, {Strotjohann}, {Stuttard}, {Sullivan}, {Sutherland}, {Taboada}, {Tenholt}, {Ter-Antonyan}, {Terliuk}, {Tilav}, {Tomankova}, {T{\"o}nnis}, {Toscano}, {Tosi}, {Tselengidou}, {Tung}, {Turcati}, {Turcotte}, {Turley}, {Ty}, {Unger}, {Unland Elorrieta}, {Usner}, {Vandenbroucke}, {Van Driessche}, {van Eijk}, {van Eijndhoven}, {Vanheule}, {van Santen}, {Vraeghe}, {Walck}, {Wallace}, {Wallraff}, {Wandkowsky}, {Watson}, {Weaver}, {Weiss}, {Weldert}, {Wendt}, {Werthebach}, {Westerhoff}, {Whelan}, {Whitehorn}, {Wiebe}, {Wiebusch}, {Wille}, {Williams}, {Wills}, {Wolf}, {Wood}, {Wood}, {Woschnagg}, {Wrede}, {Xu}, {Xu}, {Xu}, {Yanez}, {Yodh}, {Yoshida}, {Yuan}, \& {IceCube Collaboration}}]{garrappa19}
{Garrappa}, S., {Buson}, S., {Franckowiak}, A., {et~al.} 2019, \apj, 880, 103

\bibitem[{{Gehrels} {et~al.}(2004){Gehrels}, {Chincarini}, {Giommi}, {Mason}, {Nousek}, {Wells}, {White}, {Barthelmy}, {Burrows}, {Cominsky}, {Hurley}, {Marshall}, {M{\'e}sz{\'a}ros}, {Roming}, {Angelini}, {Barbier}, {Belloni}, {Campana}, {Caraveo}, {Chester}, {Citterio}, {Cline}, {Cropper}, {Cummings}, {Dean}, {Feigelson}, {Fenimore}, {Frail}, {Fruchter}, {Garmire}, {Gendreau}, {Ghisellini}, {Greiner}, {Hill}, {Hunsberger}, {Krimm}, {Kulkarni}, {Kumar}, {Lebrun}, {Lloyd-Ronning}, {Markwardt}, {Mattson}, {Mushotzky}, {Norris}, {Osborne}, {Paczynski}, {Palmer}, {Park}, {Parsons}, {Paul}, {Rees}, {Reynolds}, {Rhoads}, {Sasseen}, {Schaefer}, {Short}, {Smale}, {Smith}, {Stella}, {Tagliaferri}, {Takahashi}, {Tashiro}, {Townsley}, {Tueller}, {Turner}, {Vietri}, {Voges}, {Ward}, {Willingale}, {Zerbi}, \& {Zhang}}]{gerhels04}
{Gehrels}, N., {Chincarini}, G., {Giommi}, P., {et~al.} 2004, \apj, 611, 1005

\bibitem[{{Ghisellini} {et~al.}(2005){Ghisellini}, {Tavecchio}, \& {Chiaberge}}]{ghisellini05}
{Ghisellini}, G., {Tavecchio}, F., \& {Chiaberge}, M. 2005, \aap, 432, 401

\bibitem[{{Giommi} {et~al.}(2020){Giommi}, {Glauch}, {Padovani}, {Resconi}, {Turcati}, \& {Chang}}]{giommi20}
{Giommi}, P., {Glauch}, T., {Padovani}, P., {et~al.} 2020, \mnras, 497, 865

\bibitem[{{G{\'o}mez} {et~al.}(2001{\natexlab{a}}){G{\'o}mez}, {Guirado}, {Agudo}, {Marscher}, {Alberdi}, {Marcaide}, \& {Gabuzda}}]{gomez01}
{G{\'o}mez}, J.~L., {Guirado}, J.~C., {Agudo}, I., {et~al.} 2001{\natexlab{a}}, \mnras, 328, 873

\bibitem[{{G{\'o}mez} {et~al.}(2001{\natexlab{b}}){G{\'o}mez}, {Marscher}, {Alberdi}, {Jorstad}, \& {Agudo}}]{gomez01_2}
{G{\'o}mez}, J.-L., {Marscher}, A.~P., {Alberdi}, A., {Jorstad}, S.~G., \& {Agudo}, I. 2001{\natexlab{b}}, \apjl, 561, L161

\bibitem[{{G{\'o}mez} {et~al.}(1997){G{\'o}mez}, {Mart{\'\i}}, {Marscher}, {Ib{\'a}{\~n}ez}, \& {Alberdi}}]{gomez97}
{G{\'o}mez}, J.~L., {Mart{\'\i}}, J.~M., {Marscher}, A.~P., {Ib{\'a}{\~n}ez}, J.~M., \& {Alberdi}, A. 1997, \apjl, 482, L33

\bibitem[{{Hardee}(2000)}]{hardee00}
{Hardee}, P.~E. 2000, \apj, 533, 176

\bibitem[{{Hodgson} {et~al.}(2017){Hodgson}, {Krichbaum}, {Marscher}, {Jorstad}, {Rani}, {Marti-Vidal}, {Bach}, {Sanchez}, {Bremer}, {Lindqvist}, {Uunila}, {Kallunki}, {Vicente}, {Fuhrmann}, {Angelakis}, {Karamanavis}, {Myserlis}, {Nestoras}, {Chidiac}, {Sievers}, {Gurwell}, \& {Zensus}}]{hodgson17}
{Hodgson}, J.~A., {Krichbaum}, T.~P., {Marscher}, A.~P., {et~al.} 2017, \aap, 597, A80

\bibitem[{{Homan} {et~al.}(2021){Homan}, {Cohen}, {Hovatta}, {Kellermann}, {Kovalev}, {Lister}, {Popkov}, {Pushkarev}, {Ros}, \& {Savolainen}}]{homan21}
{Homan}, D.~C., {Cohen}, M.~H., {Hovatta}, T., {et~al.} 2021, \apj, 923, 67

\bibitem[{{Homan} {et~al.}(2015){Homan}, {Lister}, {Kovalev}, {Pushkarev}, {Savolainen}, {Kellermann}, {Richards}, \& {Ros}}]{homan15}
{Homan}, D.~C., {Lister}, M.~L., {Kovalev}, Y.~Y., {et~al.} 2015, \apj, 798, 134

\bibitem[{{Hovatta} {et~al.}(2021){Hovatta}, {Lindfors}, {Kiehlmann}, {Max-Moerbeck}, {Hodges}, {Liodakis}, {L{\"a}hteem{\"a}ki}, {Pearson}, {Readhead}, {Reeves}, {Suutarinen}, {Tammi}, \& {Tornikoski}}]{hovatta21}
{Hovatta}, T., {Lindfors}, E., {Kiehlmann}, S., {et~al.} 2021, \aap, 650, A83

\bibitem[{{IceCube Collaboration}(2021)}]{IceCubeColl2021}
{IceCube Collaboration}. 2021, GRB Coordinates Network, 31191, 1

\bibitem[{{IceCube Collaboration} {et~al.}(2018{\natexlab{a}}){IceCube Collaboration}, {Aartsen}, {Ackermann}, {Adams}, {Aguilar}, {Ahlers}, {Ahrens}, {Al Samarai}, {Altmann}, {Andeen}, {Anderson}, {Ansseau}, {Anton}, {Arg{\"u}elles}, {Auffenberg}, {Axani}, {Bagherpour}, {Bai}, {Barron}, {Barwick}, {Baum}, {Bay}, {Beatty}, {Becker Tjus}, {Becker}, {BenZvi}, {Berley}, {Bernardini}, {Besson}, {Binder}, {Bindig}, {Blaufuss}, {Blot}, {Bohm}, {B{\"o}rner}, {Bos}, {B{\"o}ser}, {Botner}, {Bourbeau}, {Bourbeau}, {Bradascio}, {Braun}, {Brenzke}, {Bretz}, {Bron}, {Brostean-Kaiser}, {Burgman}, {Busse}, {Carver}, {Cheung}, {Chirkin}, {Christov}, {Clark}, {Classen}, {Coenders}, {Collin}, {Conrad}, {Coppin}, {Correa}, {Cowen}, {Cross}, {Dave}, {Day}, {de Andr{\'e}}, {De Clercq}, {DeLaunay}, {Dembinski}, {De Ridder}, {Desiati}, {de Vries}, {de Wasseige}, {de With}, {DeYoung}, {D{\'\i}az-V{\'e}lez}, {di Lorenzo}, {Dujmovic}, {Dumm}, {Dunkman}, {Dvorak}, {Eberhardt}, {Ehrhardt}, {Eichmann}, {Eller}, {Evenson}, {Fahey},
  {Fazely}, {Felde}, {Filimonov}, {Finley}, {Flis}, {Franckowiak}, {Friedman}, {Fritz}, {Gaisser}, {Gallagher}, {Gerhardt}, {Ghorbani}, {Glauch}, {Gl{\"u}senkamp}, {Goldschmidt}, {Gonzalez}, {Grant}, {Griffith}, {Haack}, {Hallgren}, {Halzen}, {Hanson}, {Hebecker}, {Heereman}, {Helbing}, {Hellauer}, {Hickford}, {Hignight}, {Hill}, {Hoffman}, {Hoffmann}, {Hoinka}, {Hokanson-Fasig}, {Hoshina}, {Huang}, {Huber}, {Hultqvist}, {H{\"u}nnefeld}, {Hussain}, {In}, {Iovine}, {Ishihara}, {Jacobi}, {Japaridze}, {Jeong}, {Jero}, {Jones}, {Kalaczynski}, {Kang}, {Kappes}, {Kappesser}, {Karg}, {Karle}, {Katz}, {Kauer}, {Keivani}, {Kelley}, {Kheirandish}, {Kim}, {Kim}, {Kintscher}, {Kiryluk}, {Kittler}, {Klein}, {Koirala}, {Kolanoski}, {K{\"o}pke}, {Kopper}, {Kopper}, {Koschinsky}, {Koskinen}, {Kowalski}, {Krings}, {Kroll}, {Kr{\"u}ckl}, {Kunwar}, {Kurahashi}, {Kuwabara}, {Kyriacou}, {Labare}, {Lanfranchi}, {Larson}, {Lauber}, {Leonard}, {Lesiak-Bzdak}, {Leuermann}, {Liu}, {Lozano Mariscal}, {Lu}, {L{\"u}nemann}, {Luszczak},
  {Madsen}, {Maggi}, {Mahn}, {Mancina}, {Maruyama}, {Mase}, {Maunu}, {Meagher}, {Medici}, {Meier}, {Menne}, {Merino}, {Meures}, {Miarecki}, {Micallef}, {Moment{\'e}}, {Montaruli}, {Moore}, {Morse}, {Moulai}, {Nahnhauer}, {Nakarmi}, {Naumann}, {Neer}, {Niederhausen}, {Nowicki}, {Nygren}, {Obertacke Pollmann}, {Olivas}, {O'Murchadha}, {O'Sullivan}, {Palczewski}, {Pandya}, {Pankova}, {Peiffer}, {Pepper}, {P{\'e}rez de los Heros}, {Pieloth}, {Pinat}, {Plum}, {Price}, {Przybylski}, {Raab}, {R{\"a}del}, {Rameez}, {Rauch}, {Rawlins}, {Rea}, {Reimann}, {Relethford}, {Relich}, {Resconi}, {Rhode}, {Richman}, {Robertson}, {Rongen}, {Rott}, {Ruhe}, {Ryckbosch}, {Rysewyk}, {Safa}, {S{\"a}lzer}, {Sanchez Herrera}, {Sandrock}, {Sandroos}, {Santander}, {Sarkar}, {Sarkar}, {Satalecka}, {Schlunder}, {Schmidt}, {Schneider}, {Schoenen}, {Sch{\"o}neberg}, {Schumacher}, {Sclafani}, {Seckel}, {Seunarine}, {Soedingrekso}, {Soldin}, {Song}, {Spiczak}, {Spiering}, {Stachurska}, {Stamatikos}, {Stanev}, {Stasik}, {Stein}, {Stettner},
  {Steuer}, {Stezelberger}, {Stokstad}, {St{\"o}{\ss}l}, {Strotjohann}, {Stuttard}, {Sullivan}, {Sutherland}, {Taboada}, {Tatar}, {Tenholt}, {Ter-Antonyan}, {Terliuk}, {Tilav}, {Toale}, {Tobin}, {Toennis}, {Toscano}, {Tosi}, {Tselengidou}, {Tung}, {Turcati}, {Turley}, {Ty}, {Unger}, {Usner}, {Vandenbroucke}, {Van Driessche}, {van Eijk}, {van Eijndhoven}, {Vanheule}, {van Santen}, {Vogel}, {Vraeghe}, {Walck}, {Wallace}, {Wallraff}, {Wandler}, {Wandkowsky}, {Waza}, {Weaver}, {Weiss}, {Wendt}, {Werthebach}, {Westerhoff}, {Whelan}, {Whitehorn}, {Wiebe}, {Wiebusch}, {Wille}, {Williams}, {Wills}, {Wolf}, {Wood}, {Wood}, {Woschnagg}, {Xu}, {Xu}, {Xu}, {Yanez}, {Yodh}, {Yoshida}, {Yuan}, {Fermi-LAT Collaboration}, {Abdollahi}, {Ajello}, {Angioni}, {Baldini}, {Ballet}, {Barbiellini}, {Bastieri}, {Bechtol}, {Bellazzini}, {Berenji}, {Bissaldi}, {Blandford}, {Bonino}, {Bottacini}, {Bregeon}, {Bruel}, {Buehler}, {Burnett}, {Burns}, {Buson}, {Cameron}, {Caputo}, {Caraveo}, {Cavazzuti}, {Charles}, {Chen}, {Cheung},
  {Chiang}, {Chiaro}, {Ciprini}, {Cohen-Tanugi}, {Conrad}, {Costantin}, {Cutini}, {D'Ammando}, {de Palma}, {Digel}, {Di Lalla}, {Di Mauro}, {Di Venere}, {Dom{\'\i}nguez}, {Favuzzi}, {Franckowiak}, {Fukazawa}, {Funk}, {Fusco}, {Gargano}, {Gasparrini}, {Giglietto}, {Giomi}, {Giommi}, {Giordano}, {Giroletti}, {Glanzman}, {Green}, {Grenier}, {Grondin}, {Guiriec}, {Harding}, {Hayashida}, {Hays}, {Hewitt}, {Horan}, {J{\'o}hannesson}, {Kadler}, {Kensei}, {Kocevski}, {Krauss}, {Kreter}, {Kuss}, {La Mura}, {Larsson}, {Latronico}, {Lemoine-Goumard}, {Li}, {Longo}, {Loparco}, {Lovellette}, {Lubrano}, {Magill}, {Maldera}, {Malyshev}, {Manfreda}, {Mazziotta}, {McEnery}, {Meyer}, {Michelson}, {Mizuno}, {Monzani}, {Morselli}, {Moskalenko}, {Negro}, {Nuss}, {Ojha}, {Omodei}, {Orienti}, {Orlando}, {Palatiello}, {Paliya}, {Perkins}, {Persic}, {Pesce-Rollins}, {Piron}, {Porter}, {Principe}, {Rain{\`o}}, {Rando}, {Rani}, {Razzano}, {Razzaque}, {Reimer}, {Reimer}, {Renault-Tinacci}, {Ritz}, {Rochester}, {Saz Parkinson},
  {Sgr{\`o}}, {Siskind}, {Spandre}, {Spinelli}, {Suson}, {Tajima}, {Takahashi}, {Tanaka}, {Thayer}, {Thompson}, {Tibaldo}, {Torres}, {Torresi}, {Tosti}, {Troja}, {Valverde}, {Vianello}, {Vogel}, {Wood}, {Wood}, {Zaharijas}, {MAGIC Collaboration}, {Ahnen}, {Ansoldi}, {Antonelli}, {Arcaro}, {Baack}, {Babi{\'c}}, {Banerjee}, {Bangale}, {Barres de Almeida}, {Barrio}, {Becerra Gonz{\'a}lez}, {Bednarek}, {Bernardini}, {Berti}, {Bhattacharyya}, {Biland}, {Blanch}, {Bonnoli}, {Carosi}, {Carosi}, {Ceribella}, {Chatterjee}, {Colak}, {Colin}, {Colombo}, {Contreras}, {Cortina}, {Covino}, {Cumani}, {Da Vela}, {Dazzi}, {De Angelis}, {De Lotto}, {Delfino}, {Delgado}, {Di Pierro}, {Dom{\'\i}nguez}, {Dominis Prester}, {Dorner}, {Doro}, {Einecke}, {Elsaesser}, {Fallah Ramazani}, {Fern{\'a}ndez-Barral}, {Fidalgo}, {Foffano}, {Pfrang}, {Fonseca}, {Font}, {Franceschini}, {Fruck}, {Galindo}, {Gallozzi}, {Garc{\'\i}a L{\'o}pez}, {Garczarczyk}, {Gaug}, {Giammaria}, {Godinovi{\'c}}, {Gora}, {Guberman}, {Hadasch}, {Hahn}, {Hassan},
  {Hayashida}, {Herrera}, {Hose}, {Hrupec}, {Inoue}, {Ishio}, {Konno}, {Kubo}, {Kushida}, {Lelas}, {Lindfors}, {Lombardi}, {Longo}, {L{\'o}pez}, {Maggio}, {Majumdar}, {Makariev}, {Maneva}, {Manganaro}, {Mannheim}, {Maraschi}, {Mariotti}, {Mart{\'\i}nez}, {Masuda}, {Mazin}, {Minev}, {M}, {Mirzoyan}, {Moralejo}, {Moreno}, {Moretti}, {Nagayoshi}, {Neustroev}, {Niedzwiecki}, {Nievas Rosillo}, {Nigro}, {Nilsson}, {Ninci}, {Nishijima}, {Noda}, {Nogu{\'e}s}, {Paiano}, {Palacio}, {Paneque}, {Paoletti}, {Paredes}, {Pedaletti}, {Peresano}, {Persic}, {Prada Moroni}, {Prandini}, {Puljak}, {Rodriguez Garcia}, {Reichardt}, {Rhode}, {Rib{\'o}}, {Rico}, {Righi}, {Rugliancich}, {Saito}, {Satalecka}, {Schweizer}, {Sitarek}, {{\v{S}}nidaric {\textasciiacute}}, {Sobczynska}, {Stamerra}, {Strzys}, {Suri{\'c}}, {Takahashi}, {Tavecchio}, {Temnikov}, {Terzi{\'c}}, {Teshima}, {Torres-Alb{\`a}}, {Treves}, {Tsujimoto}, {Vanzo}, {Vazquez Acosta}, {Vovk}, {Ward}, {Will}, {S}, {Zaric {\textasciiacute}}, {AGILE Team}, {Lucarelli},
  {Tavani}, {Piano}, {Donnarumma}, {Pittori}, {Verrecchia}, {Barbiellini}, {Bulgarelli}, {Caraveo}, {Cattaneo}, {Colafrancesco}, {Costa}, {Di Cocco}, {Ferrari}, {Gianotti}, {Giuliani}, {Lipari}, {Mereghetti}, {Morselli}, {Pacciani}, {Paoletti}, {Parmiggiani}, {Pellizzoni}, {Picozza}, {Pilia}, {Rappoldi}, {Trois}, {Vercellone}, {Vittorini}, {ASAS-SN Team}, {Stanek}, {Franckowiak}, {Kochanek}, {Beacom}, {Thompson}, {Holoien}, {Dong}, {Prieto}, {Shappee}, {Holmbo}, {HAWC Collaboration}, {Abeysekara}, {Albert}, {Alfaro}, {Alvarez}, {Arceo}, {Arteaga-Vel{\'a}zquez}, {Avila Rojas}, {Ayala Solares}, {Becerril}, {Belmont-Moreno}, {Bernal}, {Caballero-Mora}, {Capistr{\'a}n}, {Carrami{\~n}ana}, {Casanova}, {Castillo}, {Cotti}, {Cotzomi}, {Couti{\~n}o de Le{\'o}n}, {De Le{\'o}n}, {De la Fuente}, {Diaz Hernandez}, {Dichiara}, {Dingus}, {DuVernois}, {D{\'\i}az-V{\'e}lez}, {Ellsworth}, {Engel}, {Fiorino}, {Fleischhack}, {Fraija}, {Garc{\'\i}a-Gonz{\'a}lez}, {Garfias}, {Gonz{\'a}lez Mu{\~n}oz}, {Gonz{\'a}lez}, {Goodman},
  {Hampel-Arias}, {Harding}, {Hernandez}, {Hona}, {Hueyotl-Zahuantitla}, {Hui}, {H{\"u}ntemeyer}, {Iriarte}, {Jardin-Blicq}, {Joshi}, {Kaufmann}, {Kunde}, {Lara}, {Lauer}, {Lee}, {Lennarz}, {Le{\'o}n Vargas}, {Linnemann}, {Longinotti}, {Luis-Raya}, {Luna-Garc{\'\i}a}, {Malone}, {Marinelli}, {Martinez}, {Martinez-Castellanos}, {Mart{\'\i}nez-Castro}, {Mart{\'\i}nez-Huerta}, {Matthews}, {Miranda-Romagnoli}, {Moreno}, {Mostaf{\'a}}, {Nayerhoda}, {Nellen}, {Newbold}, {Nisa}, {Noriega-Papaqui}, {Pelayo}, {Pretz}, {P{\'e}rez-P{\'e}rez}, {Ren}, {Rho}, {Rivi{\`e}re}, {Rosa-Gonz{\'a}lez}, {Rosenberg}, {Ruiz-Velasco}, {Ruiz-Velasco}, {Salesa Greus}, {Sandoval}, {Schneider}, {Schoorlemmer}, {Sinnis}, {Smith}, {Springer}, {Surajbali}, {Tibolla}, {Tollefson}, {Torres}, {Villase{\~n}or}, {Weisgarber}, {Werner}, {Yapici}, {Gaurang}, {Zepeda}, {Zhou}, {{\'A}lvarez}, {H.~E.~S.~S. Collaboration}, {Abdalla}, {Ang{\"u}ner}, {Armand}, {Backes}, {Becherini}, {Berge}, {B{\"o}ttcher}, {Boisson}, {Bolmont}, {Bonnefoy}, {Bordas},
  {Brun}, {B{\"u}chele}, {Bulik}, {Caroff}, {Carosi}, {Casanova}, {Cerruti}, {Chakraborty}, {Chandra}, {Chen}, {Colafrancesco}, {Davids}, {Deil}, {Devin}, {Djannati-Ata{\"\i}}, {Egberts}, {Emery}, {Eschbach}, {Fiasson}, {Fontaine}, {Funk}, {F{\"u}{\ss}ling}, {Gallant}, {Gat{\'e}}, {Giavitto}, {Glawion}, {Glicenstein}, {Gottschall}, {Grondin}, {Haupt}, {Henri}, {Hinton}, {Hoischen}, {Holch}, {Huber}, {Jamrozy}, {Jankowsky}, {Jankowsky}, {Jouvin}, {Jung-Richardt}, {Kerszberg}, {Kh{\'e}lifi}, {King}, {Klepser}, {Kluz {\textasciiacute}niak}, {Komin}, {Kraus}, {Lefaucheur}, {Lemi{\`e}re}, {Lemoine-Goumard}, {Lenain}, {Leser}, {Lohse}, {L{\'o}pez-Coto}, {Lorentz}, {Lypova}, {Marandon}, {Guillem Mart{\'\i}-Devesa}, {Maurin}, {Mitchell}, {Moderski}, {Mohamed}, {Mohrmann}, {Moulin}, {Murach}, {de Naurois}, {Niederwanger}, {Niemiec}, {Oakes}, {O'Brien}, {Ohm}, {Ostrowski}, {Oya}, {Panter}, {Parsons}, {Perennes}, {Piel}, {Pita}, {Poireau}, {Priyana Noel}, {Prokoph}, {P{\"u}hlhofer}, {Quirrenbach}, {Raab}, {Rauth},
  {Renaud}, {Rieger}, {Rinchiuso}, {Romoli}, {Rowell}, {Rudak}, {Sasaki}, {Sanchez}, {Schlickeiser}, {Sch{\"u}ssler}, {Schulz}, {Schwanke}, {Seglar-Arroyo}, {Shafi}, {Simoni}, {Sol}, {Stegmann}, {Steppa}, {Tavernier}, {Taylor}, {Tiziani}, {Trichard}, {Tsirou}, {van Eldik}, {van Rensburg}, {van Soelen}, {Veh}, {Vincent}, {Voisin}, {Wagner}, {Wagner}, {Wierzcholska}, {Zanin}, {Zdziarski}, {Zech}, {Ziegler}, {Zorn}, {{\.Z}ywucka}, {INTEGRAL Team}, {Savchenko}, {Ferrigno}, {Bazzano}, {Diehl}, {Kuulkers}, {Laurent}, {Mereghetti}, {Natalucci}, {Panessa}, {Rodi}, {Ubertini}, {Kanata}, Teams, {Morokuma}, {Ohta}, {Tanaka}, {Mori}, {Yamanaka}, {Kawabata}, {Utsumi}, {Nakaoka}, {Kawabata}, {Nagashima}, {Yoshida}, {Matsuoka}, {Itoh}, {Kapteyn Team}, {Keel}, {Liverpool Telescope Team}, {Copperwheat}, {Steele}, {Swift/NuSTAR Team}, {Cenko}, {Cowen}, {DeLaunay}, {Evans}, {Fox}, {Keivani}, {Kennea}, {Marshall}, {Osborne}, {Santander}, {Tohuvavohu}, {Turley}, {VERITAS Collaboration}, {Abeysekara}, {Archer}, {Benbow}, {Bird},
  {Brill}, {Brose}, {Buchovecky}, {Buckley}, {Bugaev}, {Christiansen}, {Connolly}, {Cui}, {Daniel}, {Errando}, {Falcone}, {Feng}, {Finley}, {Fortson}, {Furniss}, {Gueta}, {H{\"u}tten}, {Hervet}, {Hughes}, {Humensky}, {Johnson}, {Kaaret}, {Kar}, {Kelley-Hoskins}, {Kertzman}, {Kieda}, {Krause}, {Krennrich}, {Kumar}, {Lang}, {Lin}, {Maier}, {McArthur}, {Moriarty}, {Mukherjee}, {Nieto}, {O'Brien}, {Ong}, {Otte}, {Park}, {Petrashyk}, {Pohl}, {Popkow}, {Pueschel}, {Quinn}, {Ragan}, {Reynolds}, {Richards}, {Roache}, {Rulten}, {Sadeh}, {Santander}, {Scott}, {Sembroski}, {Shahinyan}, {Sushch}, {Tr{\'e}panier}, {Tyler}, {Vassiliev}, {Wakely}, {Weinstein}, {Wells}, {Wilcox}, {Wilhelm}, {Williams}, {Zitzer}, {VLA/B Team}, {Tetarenko}, {Kimball}, {Miller-Jones}, \& {Sivakoff}}]{icecube18b}
{IceCube Collaboration}, {Aartsen}, M.~G., {Ackermann}, M., {et~al.} 2018{\natexlab{a}}, Science, 361, eaat1378

\bibitem[{{IceCube Collaboration} {et~al.}(2018{\natexlab{b}}){IceCube Collaboration}, {Aartsen}, {Ackermann}, {Adams}, {Aguilar}, {Ahlers}, {Ahrens}, {Samarai}, {Altmann}, {Andeen}, {Anderson}, {Ansseau}, {Anton}, {Arg{\"u}elles}, {Arsioli}, {Auffenberg}, {Axani}, {Bagherpour}, {Bai}, {Barron}, {Barwick}, {Baum}, {Bay}, {Beatty}, {Becker Tjus}, {Becker}, {BenZvi}, {Berley}, {Bernardini}, {Besson}, {Binder}, {Bindig}, {Blaufuss}, {Blot}, {Bohm}, {B{\"o}rner}, {Bos}, {B{\"o}ser}, {Botner}, {Bourbeau}, {Bourbeau}, {Bradascio}, {Braun}, {Brenzke}, {Bretz}, {Bron}, {Brostean-Kaiser}, {Burgman}, {Busse}, {Carver}, {Cheung}, {Chirkin}, {Christov}, {Clark}, {Classen}, {Coenders}, {Collin}, {Conrad}, {Coppin}, {Correa}, {Cowen}, {Cross}, {Dave}, {Day}, {de Andr{\'e}}, {De Clercq}, {DeLaunay}, {Dembinski}, {DeRidder}, {Desiati}, {de Vries}, {de Wasseige}, {de With}, {DeYoung}, {D{\'\i}az-V{\'e}lez}, {di Lorenzo}, {Dujmovic}, {Dumm}, {Dunkman}, {Dvorak}, {Eberhardt}, {Ehrhardt}, {Eichmann}, {Eller}, {Evenson}, {Fahey},
  {Fazely}, {Felde}, {Filimonov}, {Finley}, {Flis}, {Franckowiak}, {Friedman}, {Fritz}, {Gaisser}, {Gallagher}, {Gerhardt}, {Ghorbani}, {Giommi}, {Glauch}, {Gl{\"u}senkamp}, {Goldschmidt}, {Gonzalez}, {Grant}, {Griffith}, {Haack}, {Hallgren}, {Halzen}, {Hanson}, {Hebecker}, {Heereman}, {Helbing}, {Hellauer}, {Hickford}, {Hignight}, {Hill}, {Hoffman}, {Hoffmann}, {Hoinka}, {Hokanson-Fasig}, {Hoshina}, {Huang}, {Huber}, {Hultqvist}, {H{\"u}nnefeld}, {Hussain}, {In}, {Iovine}, {Ishihara}, {Jacobi}, {Japaridze}, {Jeong}, {Jero}, {Jones}, {Kalaczynski}, {Kang}, {Kappes}, {Kappesser}, {Karg}, {Karle}, {Katz}, {Kauer}, {Keivani}, {Kelley}, {Kheirandish}, {Kim}, {Kim}, {Kintscher}, {Kiryluk}, {Kittler}, {Klein}, {Koirala}, {Kolanoski}, {K{\"o}pke}, {Kopper}, {Kopper}, {Koschinsky}, {Koskinen}, {Kowalski}, {Krammer}, {Krings}, {Kroll}, {Kr{\"u}ckl}, {Kunwar}, {Kurahashi}, {Kuwabara}, {Kyriacou}, {Labare}, {Lanfranchi}, {Larson}, {Lauber}, {Leonard}, {Lesiak-Bzdak}, {Leuermann}, {Liu}, {Lozano Mariscal}, {Lu},
  {L{\"u}nemann}, {Luszczak}, {Madsen}, {Maggi}, {Mahn}, {Mancina}, {Maruyama}, {Mase}, {Maunu}, {Meagher}, {Medici}, {Meier}, {Menne}, {Merino}, {Meures}, {Miarecki}, {Micallef}, {Moment{\'e}}, {Montaruli}, {Moore}, {Morse}, {Moulai}, {Nahnhauer}, {Nakarmi}, {Naumann}, {Neer}, {Niederhausen}, {Nowicki}, {Nygren}, {Obertacke Pollmann}, {Olivas}, {O'Murchadha}, {O'Sullivan}, {Padovani}, {Palczewski}, {Pandya}, {Pankova}, {Peiffer}, {Pepper}, {P{\'e}rez de los Heros}, {Pieloth}, {Pinat}, {Plum}, {Price}, {Przybylski}, {Raab}, {R{\"a}del}, {Rameez}, {Rawlins}, {Rea}, {Reimann}, {Relethford}, {Relich}, {Resconi}, {Rhode}, {Richman}, {Robertson}, {Rongen}, {Rott}, {Ruhe}, {Ryckbosch}, {Rysewyk}, {Safa}, {Sahakyan}, {S{\"a}lzer}, {Sanchez Herrera}, {Sandrock}, {Sandroos}, {Santander}, {Sarkar}, {Sarkar}, {Satalecka}, {Schlunder}, {Schmidt}, {Schneider}, {Schoenen}, {Sch{\"o}neberg}, {Schumacher}, {Sclafani}, {Seckel}, {Seunarine}, {Soedingrekso}, {Soldin}, {Song}, {Spiczak}, {Spiering}, {Stachurska}, {Stamatikos},
  {Stanev}, {Stasik}, {Stettner}, {Steuer}, {Stezelberger}, {Stokstad}, {St{\"o}{\ss}l}, {Strotjohann}, {Stuttard}, {Sullivan}, {Sutherland}, {Taboada}, {Tatar}, {Tenholt}, {Ter-Antonyan}, {Terliuk}, {Tilav}, {Toale}, {Tobin}, {Toennis}, {Toscano}, {Tosi}, {Tselengidou}, {Tung}, {Turcati}, {Turley}, {Ty}, {Unger}, {Usner}, {Vandenbroucke}, {Van Driessche}, {van Eijk}, {van Eijndhoven}, {Vanheule}, {van Santen}, {Vogel}, {Vraeghe}, {Walck}, {Wallace}, {Wallraff}, {Wandler}, {Wandkowsky}, {Waza}, {Weaver}, {Weiss}, {Wendt}, {Werthebach}, {Westerhoff}, {Whelan}, {Whitehorn}, {Wiebe}, {Wiebusch}, {Wille}, {Williams}, {Wills}, {Wolf}, {Wood}, {Wood}, {Woschnagg}, {Xu}, {Xu}, {Xu}, {Yanez}, {Yodh}, {Yoshida}, \& {Yuan}}]{icecube18a}
{IceCube Collaboration}, {Aartsen}, M.~G., {Ackermann}, M., {et~al.} 2018{\natexlab{b}}, Science, 361, 147

\bibitem[{{Jaffe} {et~al.}(2004){Jaffe}, {Meisenheimer}, {R{\"o}ttgering}, {Leinert}, {Richichi}, {Chesneau}, {Fraix-Burnet}, {Glazenborg-Kluttig}, {Granato}, {Graser}, {Heijligers}, {K{\"o}hler}, {Malbet}, {Miley}, {Paresce}, {Pel}, {Perrin}, {Przygodda}, {Schoeller}, {Sol}, {Waters}, {Weigelt}, {Woillez}, \& {de Zeeuw}}]{jaffe04}
{Jaffe}, W., {Meisenheimer}, K., {R{\"o}ttgering}, H.~J.~A., {et~al.} 2004, \nat, 429, 47

\bibitem[{{Jeong} {et~al.}(2023){Jeong}, {Lee}, {Cheong}, {Kim}, {Lee}, {Kang}, {Kim}, {Rani}, {Park}, \& {Gurwell}}]{jeong23}
{Jeong}, H.-W., {Lee}, S.-S., {Cheong}, W.~Y., {et~al.} 2023, \mnras, 523, 5703

\bibitem[{{Jorstad} \& {Marscher}(2016)}]{jorstad16}
{Jorstad}, S. \& {Marscher}, A. 2016, Galaxies, 4, 47

\bibitem[{{Jorstad} {et~al.}(2017){Jorstad}, {Marscher}, {Morozova}, {Troitsky}, {Agudo}, {Casadio}, {Foord}, {G{\'o}mez}, {MacDonald}, {Molina}, {L{\"a}hteenm{\"a}ki}, {Tammi}, \& {Tornikoski}}]{jorstad17}
{Jorstad}, S.~G., {Marscher}, A.~P., {Morozova}, D.~A., {et~al.} 2017, \apj, 846, 98

\bibitem[{{Kadler} {et~al.}(2008){Kadler}, {Ros}, {Perucho}, {Kovalev}, {Homan}, {Agudo}, {Kellermann}, {Aller}, {Aller}, {Lister}, \& {Zensus}}]{kadler08}
{Kadler}, M., {Ros}, E., {Perucho}, M., {et~al.} 2008, \apj, 680, 867

\bibitem[{{Kim} {et~al.}(2020){Kim}, {Krichbaum}, {Broderick}, {Wielgus}, {Blackburn}, {G{\'o}mez}, {Johnson}, {Bouman}, {Chael}, {Akiyama}, {Jorstad}, {Marscher}, {Issaoun}, {Janssen}, {Chan}, {Savolainen}, {Pesce}, {{\"O}zel}, {Alberdi}, {Alef}, {Asada}, {Azulay}, {Baczko}, {Ball}, {Balokovi{\'c}}, {Barrett}, {Bintley}, {Boland}, {Bower}, {Bremer}, {Brinkerink}, {Brissenden}, {Britzen}, {Broguiere}, {Bronzwaer}, {Byun}, {Carlstrom}, {Chatterjee}, {Chatterjee}, {Chen}, {Chen}, {Cho}, {Christian}, {Conway}, {Cordes}, {Crew}, {Cui}, {Davelaar}, {De Laurentis}, {Deane}, {Dempsey}, {Desvignes}, {Dexter}, {Doeleman}, {Eatough}, {Falcke}, {Fish}, {Fomalont}, {Fraga-Encinas}, {Friberg}, {Fromm}, {Galison}, {Gammie}, {Garc{\'\i}a}, {Gentaz}, {Georgiev}, {Goddi}, {Gold}, {G{\'o}mez-Ruiz}, {Gu}, {Gurwell}, {Hada}, {Hecht}, {Hesper}, {Ho}, {Ho}, {Honma}, {Huang}, {Huang}, {Hughes}, {Ikeda}, {Inoue}, {James}, {Jannuzi}, {Jeter}, {Jiang}, {Jimenez-Rosales}, {Jung}, {Karami}, {Karuppusamy}, {Kawashima}, {Keating},
  {Kettenis}, {Kim}, {Kim}, {Kino}, {Koay}, {Koch}, {Koyama}, {Kramer}, {Kramer}, {Kuo}, {Lauer}, {Lee}, {Li}, {Li}, {Lindqvist}, {Lico}, {Liu}, {Liuzzo}, {Lo}, {Lobanov}, {Loinard}, {Lonsdale}, {Lu}, {MacDonald}, {Mao}, {Markoff}, {Marrone}, {Mart{\'\i}-Vidal}, {Matsushita}, {Matthews}, {Medeiros}, {Menten}, {Mizuno}, {Mizuno}, {Moran}, {Moriyama}, {Moscibrodzka}, {Musoke}, {M{\"u}ller}, {Nagai}, {Nagar}, {Nakamura}, {Narayan}, {Narayanan}, {Natarajan}, {Neri}, {Ni}, {Noutsos}, {Okino}, {Olivares}, {Ortiz-Le{\'o}n}, {Oyama}, {Palumbo}, {Park}, {Patel}, {Pen}, {Pi{\'e}tu}, {Plambeck}, {PopStefanija}, {Porth}, {Prather}, {Preciado-L{\'o}pez}, {Psaltis}, {Pu}, {Ramakrishnan}, {Rao}, {Rawlings}, {Raymond}, {Rezzolla}, {Ripperda}, {Roelofs}, {Rogers}, {Ros}, {Rose}, {Roshanineshat}, {Rottmann}, {Roy}, {Ruszczyk}, {Ryan}, {Rygl}, {S{\'a}nchez}, {S{\'a}nchez-Arguelles}, {Sasada}, {Schloerb}, {Schuster}, {Shao}, {Shen}, {Small}, {Sohn}, {SooHoo}, {Tazaki}, {Tiede}, {Tilanus}, {Titus}, {Toma}, {Torne}, {Trent},
  {Traianou}, {Trippe}, {Tsuda}, {van Bemmel}, {van Langevelde}, {van Rossum}, {Wagner}, {Wardle}, {Ward-Thompson}, {Weintroub}, {Wex}, {Wharton}, {Wong}, {Wu}, {Yoon}, {Young}, {Young}, {Younsi}, {Yuan}, {Yuan}, {Zensus}, {Zhao}, {Zhao}, {Zhu}, {Algaba}, {Allardi}, {Amestica}, {Anczarski}, {Bach}, {Baganoff}, {Beaudoin}, {Benson}, {Berthold}, {Blanchard}, {Blundell}, {Bustamente}, {Cappallo}, {Castillo-Dom{\'\i}nguez}, {Chang}, {Chang}, {Chang}, {Chen}, {Chilson}, {Chuter}, {Rosado}, {Coulson}, {Crowley}, {Derome}, {Dexter}, {Dornbusch}, {Dudevoir}, {Dzib}, {Eckart}, {Eckert}, {Erickson}, {Everett}, {Faber}, {Farah}, {Fath}, {Folkers}, {Forbes}, {Freund}, {Gale}, {Gao}, {Geertsema}, {Graham}, {Greer}, {Grosslein}, {Gueth}, {Haggard}, {Halverson}, {Han}, {Han}, {Hao}, {Hasegawa}, {Henning}, {Hern{\'a}ndez-G{\'o}mez}, {Herrero-Illana}, {Heyminck}, {Hirota}, {Hoge}, {Huang}, {Violette Impellizzeri}, {Jiang}, {John}, {Kamble}, {Keisler}, {Kimura}, {Kono}, {Kubo}, {Kuroda}, {Lacasse}, {Laing}, {Leitch}, {Li},
  {Lin}, {Liu}, {Liu}, {Lu}, {Marson}, {Martin-Cocher}, {Massingill}, {Matulonis}, {McColl}, {McWhirter}, {Messias}, {Meyer-Zhao}, {Michalik}, {Monta{\~n}a}, {Montgomerie}, {Mora-Klein}, {Muders}, {Nadolski}, {Navarro}, {Neilsen}, {Nguyen}, {Nishioka}, {Norton}, {Nowak}, {Nystrom}, {Ogawa}, {Oshiro}, {Oyama}, {Parsons}, {Pe{\~n}alver}, {Phillips}, {Poirier}, {Pradel}, {Primiani}, {Raffin}, {Rahlin}, {Reiland}, {Risacher}, {Ruiz}, {S{\'a}ez-Mada{\'\i}n}, {Sassella}, {Schellart}, {Shaw}, {Silva}, {Shiokawa}, {Smith}, {Snow}, {Souccar}, {Sousa}, {Sridharan}, {Srinivasan}, {Stahm}, {Stark}, {Story}, {Timmer}, {Vertatschitsch}, {Walther}, {Wei}, {Whitehorn}, {Whitney}, {Woody}, {Wouterloot}, {Wright}, {Yamaguchi}, {Yu}, {Zeballos}, {Zhang}, {Ziurys}, \& {Event Horizon Telescope Collaboration}}]{kim20}
{Kim}, J.-Y., {Krichbaum}, T.~P., {Broderick}, A.~E., {et~al.} 2020, \aap, 640, A69

\bibitem[{{Kim} {et~al.}(2018{\natexlab{a}}){Kim}, {Krichbaum}, {Lu}, {Ros}, {Bach}, {Bremer}, {de Vicente}, {Lindqvist}, \& {Zensus}}]{kim18}
{Kim}, J.~Y., {Krichbaum}, T.~P., {Lu}, R.~S., {et~al.} 2018{\natexlab{a}}, \aap, 616, A188

\bibitem[{{Kim} {et~al.}(2019){Kim}, {Krichbaum}, {Marscher}, {Jorstad}, {Agudo}, {Thum}, {Hodgson}, {MacDonald}, {Ros}, {Lu}, {Bremer}, {de Vicente}, {Lindqvist}, {Trippe}, \& {Zensus}}]{kim19}
{Kim}, J.~Y., {Krichbaum}, T.~P., {Marscher}, A.~P., {et~al.} 2019, \aap, 622, A196

\bibitem[{{Kim} {et~al.}(2018{\natexlab{b}}){Kim}, {Lee}, {Hodgson}, {Algaba}, {Zhao}, {Kino}, {Byun}, \& {Kang}}]{kim18a}
{Kim}, J.-Y., {Lee}, S.-S., {Hodgson}, J.~A., {et~al.} 2018{\natexlab{b}}, \aap, 610, L5

\bibitem[{{Kim} {et~al.}(2023){Kim}, {Savolainen}, {Voitsik}, {Kravchenko}, {Lisakov}, {Kovalev}, {M{\"u}ller}, {Lobanov}, {Sokolovsky}, {Bruni}, {Edwards}, {Reynolds}, {Bach}, {Gurvits}, {Krichbaum}, {Hada}, {Giroletti}, {Orienti}, {Anderson}, {Lee}, {Sohn}, \& {Zensus}}]{kim23}
{Kim}, J.-Y., {Savolainen}, T., {Voitsik}, P., {et~al.} 2023, \apj, 952, 34

\bibitem[{{Kishimoto} {et~al.}(2007){Kishimoto}, {H{\"o}nig}, {Beckert}, \& {Weigelt}}]{kishimoto07}
{Kishimoto}, M., {H{\"o}nig}, S.~F., {Beckert}, T., \& {Weigelt}, G. 2007, \aap, 476, 713

\bibitem[{{Kochanek} {et~al.}(2017){Kochanek}, {Shappee}, {Stanek}, {Holoien}, {Thompson}, {Prieto}, {Dong}, {Shields}, {Will}, {Britt}, {Perzanowski}, \& {Pojma{\'n}ski}}]{kochanek17}
{Kochanek}, C.~S., {Shappee}, B.~J., {Stanek}, K.~Z., {et~al.} 2017, \pasp, 129, 104502

\bibitem[{{Lisakov} {et~al.}(2017){Lisakov}, {Kovalev}, {Savolainen}, {Hovatta}, \& {Kutkin}}]{lisakov17}
{Lisakov}, M.~M., {Kovalev}, Y.~Y., {Savolainen}, T., {Hovatta}, T., \& {Kutkin}, A.~M. 2017, \mnras, 468, 4478

\bibitem[{{Lister} {et~al.}(2018{\natexlab{a}}){Lister}, {Aller}, {Aller}, {Hodge}, {Homan}, {Kovalev}, {Pushkarev}, \& {Savolainen}}]{mojave}
{Lister}, M.~L., {Aller}, M.~F., {Aller}, H.~D., {et~al.} 2018{\natexlab{a}}, \apjs, 234, 12

\bibitem[{{Lister} {et~al.}(2018{\natexlab{b}}){Lister}, {Aller}, {Aller}, {Hodge}, {Homan}, {Kovalev}, {Pushkarev}, \& {Savolainen}}]{lister18}
{Lister}, M.~L., {Aller}, M.~F., {Aller}, H.~D., {et~al.} 2018{\natexlab{b}}, \apjs, 234, 12

\bibitem[{{Lister} {et~al.}(2009){Lister}, {Cohen}, {Homan}, {Kadler}, {Kellermann}, {Kovalev}, {Ros}, {Savolainen}, \& {Zensus}}]{lister09}
{Lister}, M.~L., {Cohen}, M.~H., {Homan}, D.~C., {et~al.} 2009, \aj, 138, 1874

\bibitem[{{Lister} {et~al.}(2019){Lister}, {Homan}, {Hovatta}, {Kellermann}, {Kiehlmann}, {Kovalev}, {Max-Moerbeck}, {Pushkarev}, {Readhead}, {Ros}, \& {Savolainen}}]{lister19}
{Lister}, M.~L., {Homan}, D.~C., {Hovatta}, T., {et~al.} 2019, \apj, 874, 43

\bibitem[{{Lister} {et~al.}(2021){Lister}, {Homan}, {Kellermann}, {Kovalev}, {Pushkarev}, {Ros}, \& {Savolainen}}]{lister21}
{Lister}, M.~L., {Homan}, D.~C., {Kellermann}, K.~I., {et~al.} 2021, \apj, 923, 30

\bibitem[{{Lobanov}(1998)}]{lobanov88}
{Lobanov}, A.~P. 1998, \aap, 330, 79

\bibitem[{{Marscher}(2014)}]{marscher14}
{Marscher}, A.~P. 2014, \apj, 780, 87

\bibitem[{{Marscher} \& {Gear}(1985)}]{marscher85}
{Marscher}, A.~P. \& {Gear}, W.~K. 1985, \apj, 298, 114

\bibitem[{{Marscher} {et~al.}(1992){Marscher}, {Gear}, \& {Travis}}]{marscher92}
{Marscher}, A.~P., {Gear}, W.~K., \& {Travis}, J.~P. 1992, in Variability of Blazars, ed. E.~{Valtaoja} \& M.~{Valtonen}, 85

\bibitem[{{Mertens} {et~al.}(2016){Mertens}, {Lobanov}, {Walker}, \& {Hardee}}]{mertens16}
{Mertens}, F., {Lobanov}, A.~P., {Walker}, R.~C., \& {Hardee}, P.~E. 2016, \aap, 595, A54

\bibitem[{{M{\'e}sz{\'a}ros} {et~al.}(2019){M{\'e}sz{\'a}ros}, {Fox}, {Hanna}, \& {Murase}}]{meszaros19}
{M{\'e}sz{\'a}ros}, P., {Fox}, D.~B., {Hanna}, C., \& {Murase}, K. 2019, Nature Reviews Physics, 1, 585

\bibitem[{{Niinuma} {et~al.}(2015){Niinuma}, {Kino}, {Doi}, {Hada}, {Nagai}, \& {Koyama}}]{niinuma15}
{Niinuma}, K., {Kino}, M., {Doi}, A., {et~al.} 2015, \apjl, 807, L14

\bibitem[{{Nilsson} {et~al.}(2012){Nilsson}, {Pursimo}, {Villforth}, {Lindfors}, {Takalo}, \& {Sillanp{\"a}{\"a}}}]{nilsson12}
{Nilsson}, K., {Pursimo}, T., {Villforth}, C., {et~al.} 2012, \aap, 547, A1

\bibitem[{{Padovani} {et~al.}(2019){Padovani}, {Oikonomou}, {Petropoulou}, {Giommi}, \& {Resconi}}]{padovani19}
{Padovani}, P., {Oikonomou}, F., {Petropoulou}, M., {Giommi}, P., \& {Resconi}, E. 2019, \mnras, 484, L104

\bibitem[{{Paraschos} {et~al.}(2023){Paraschos}, {Mpisketzis}, {Kim}, {Witzel}, {Krichbaum}, {Zensus}, {Gurwell}, {L{\"a}hteenm{\"a}ki}, {Tornikoski}, {Kiehlmann}, \& {Readhead}}]{paraschos23}
{Paraschos}, G.~F., {Mpisketzis}, V., {Kim}, J.~Y., {et~al.} 2023, \aap, 669, A32

\bibitem[{{Perucho} {et~al.}(2004){Perucho}, {Hanasz}, {Mart{\'\i}}, \& {Sol}}]{perucho04}
{Perucho}, M., {Hanasz}, M., {Mart{\'\i}}, J.~M., \& {Sol}, H. 2004, \aap, 427, 415

\bibitem[{{Petkov} {et~al.}(2021){Petkov}, {Novoseltsev}, {Novoseltseva}, \& {Baksan Underground Scintillation Telescope Group}}]{baksan21}
{Petkov}, V.~B., {Novoseltsev}, Y.~F., {Novoseltseva}, R.~V., \& {Baksan Underground Scintillation Telescope Group}. 2021, The Astronomer's Telegram, 15143, 1

\bibitem[{{Planck Collaboration} {et~al.}(2014){Planck Collaboration}, {Ade}, {Aghanim}, {Armitage-Caplan}, {Arnaud}, {Ashdown}, {Atrio-Barandela}, {Aumont}, {Baccigalupi}, {Banday}, {Barreiro}, {Bartlett}, {Battaner}, {Benabed}, {Beno{\^\i}t}, {Benoit-L{\'e}vy}, {Bernard}, {Bersanelli}, {Bielewicz}, {Bobin}, {Bock}, {Bonaldi}, {Bond}, {Borrill}, {Bouchet}, {Bridges}, {Bucher}, {Burigana}, {Butler}, {Calabrese}, {Cappellini}, {Cardoso}, {Catalano}, {Challinor}, {Chamballu}, {Chary}, {Chen}, {Chiang}, {Chiang}, {Christensen}, {Church}, {Clements}, {Colombi}, {Colombo}, {Couchot}, {Coulais}, {Crill}, {Curto}, {Cuttaia}, {Danese}, {Davies}, {Davis}, {de Bernardis}, {de Rosa}, {de Zotti}, {Delabrouille}, {Delouis}, {D{\'e}sert}, {Dickinson}, {Diego}, {Dolag}, {Dole}, {Donzelli}, {Dor{\'e}}, {Douspis}, {Dunkley}, {Dupac}, {Efstathiou}, {Elsner}, {En{\ss}lin}, {Eriksen}, {Finelli}, {Forni}, {Frailis}, {Fraisse}, {Franceschi}, {Gaier}, {Galeotta}, {Galli}, {Ganga}, {Giard}, {Giardino}, {Giraud-H{\'e}raud},
  {Gjerl{\o}w}, {Gonz{\'a}lez-Nuevo}, {G{\'o}rski}, {Gratton}, {Gregorio}, {Gruppuso}, {Gudmundsson}, {Haissinski}, {Hamann}, {Hansen}, {Hanson}, {Harrison}, {Henrot-Versill{\'e}}, {Hern{\'a}ndez-Monteagudo}, {Herranz}, {Hildebrandt}, {Hivon}, {Hobson}, {Holmes}, {Hornstrup}, {Hou}, {Hovest}, {Huffenberger}, {Jaffe}, {Jaffe}, {Jewell}, {Jones}, {Juvela}, {Keih{\"a}nen}, {Keskitalo}, {Kisner}, {Kneissl}, {Knoche}, {Knox}, {Kunz}, {Kurki-Suonio}, {Lagache}, {L{\"a}hteenm{\"a}ki}, {Lamarre}, {Lasenby}, {Lattanzi}, {Laureijs}, {Lawrence}, {Leach}, {Leahy}, {Leonardi}, {Le{\'o}n-Tavares}, {Lesgourgues}, {Lewis}, {Liguori}, {Lilje}, {Linden-V{\o}rnle}, {L{\'o}pez-Caniego}, {Lubin}, {Mac{\'\i}as-P{\'e}rez}, {Maffei}, {Maino}, {Mandolesi}, {Maris}, {Marshall}, {Martin}, {Mart{\'\i}nez-Gonz{\'a}lez}, {Masi}, {Massardi}, {Matarrese}, {Matthai}, {Mazzotta}, {Meinhold}, {Melchiorri}, {Melin}, {Mendes}, {Menegoni}, {Mennella}, {Migliaccio}, {Millea}, {Mitra}, {Miville-Desch{\^e}nes}, {Moneti}, {Montier}, {Morgante},
  {Mortlock}, {Moss}, {Munshi}, {Murphy}, {Naselsky}, {Nati}, {Natoli}, {Netterfield}, {N{\o}rgaard-Nielsen}, {Noviello}, {Novikov}, {Novikov}, {O'Dwyer}, {Osborne}, {Oxborrow}, {Paci}, {Pagano}, {Pajot}, {Paladini}, {Paoletti}, {Partridge}, {Pasian}, {Patanchon}, {Pearson}, {Pearson}, {Peiris}, {Perdereau}, {Perotto}, {Perrotta}, {Pettorino}, {Piacentini}, {Piat}, {Pierpaoli}, {Pietrobon}, {Plaszczynski}, {Platania}, {Pointecouteau}, {Polenta}, {Ponthieu}, {Popa}, {Poutanen}, {Pratt}, {Pr{\'e}zeau}, {Prunet}, {Puget}, {Rachen}, {Reach}, {Rebolo}, {Reinecke}, {Remazeilles}, {Renault}, {Ricciardi}, {Riller}, {Ristorcelli}, {Rocha}, {Rosset}, {Roudier}, {Rowan-Robinson}, {Rubi{\~n}o-Mart{\'\i}n}, {Rusholme}, {Sandri}, {Santos}, {Savelainen}, {Savini}, {Scott}, {Seiffert}, {Shellard}, {Spencer}, {Starck}, {Stolyarov}, {Stompor}, {Sudiwala}, {Sunyaev}, {Sureau}, {Sutton}, {Suur-Uski}, {Sygnet}, {Tauber}, {Tavagnacco}, {Terenzi}, {Toffolatti}, {Tomasi}, {Tristram}, {Tucci}, {Tuovinen}, {T{\"u}rler}, {Umana},
  {Valenziano}, {Valiviita}, {Van Tent}, {Vielva}, {Villa}, {Vittorio}, {Wade}, {Wandelt}, {Wehus}, {White}, {White}, {Wilkinson}, {Yvon}, {Zacchei}, \& {Zonca}}]{planck14}
{Planck Collaboration}, {Ade}, P.~A.~R., {Aghanim}, N., {et~al.} 2014, \aap, 571, A16

\bibitem[{{Plavin} {et~al.}(2021){Plavin}, {Kovalev}, {Kovalev}, \& {Troitsky}}]{plavin21}
{Plavin}, A.~V., {Kovalev}, Y.~Y., {Kovalev}, Y.~A., \& {Troitsky}, S.~V. 2021, \apj, 908, 157

\bibitem[{{Plavin} {et~al.}(2023){Plavin}, {Kovalev}, {Kovalev}, \& {Troitsky}}]{plavin23}
{Plavin}, A.~V., {Kovalev}, Y.~Y., {Kovalev}, Y.~A., \& {Troitsky}, S.~V. 2023, \mnras, 523, 1799

\bibitem[{{Plavin} {et~al.}(2019){Plavin}, {Kovalev}, {Pushkarev}, \& {Lobanov}}]{plavin19}
{Plavin}, A.~V., {Kovalev}, Y.~Y., {Pushkarev}, A.~B., \& {Lobanov}, A.~P. 2019, \mnras, 485, 1822

\bibitem[{{Prince} {et~al.}(2024){Prince}, {Das}, {Gupta}, {Majumdar}, \& {Czerny}}]{prince24}
{Prince}, R., {Das}, S., {Gupta}, N., {Majumdar}, P., \& {Czerny}, B. 2024, \mnras, 527, 8746

\bibitem[{{Pushkarev} {et~al.}(2009){Pushkarev}, {Kovalev}, {Lister}, \& {Savolainen}}]{pushkarev09}
{Pushkarev}, A.~B., {Kovalev}, Y.~Y., {Lister}, M.~L., \& {Savolainen}, T. 2009, \aap, 507, L33

\bibitem[{{Raiteri} {et~al.}(2017){Raiteri}, {Villata}, {Acosta-Pulido}, {Agudo}, {Arkharov}, {Bachev}, {Baida}, {Ben{\'\i}tez}, {Borman}, {Boschin}, {Bozhilov}, {Butuzova}, {Calcidese}, {Carnerero}, {Carosati}, {Casadio}, {Castro-Segura}, {Chen}, {Damljanovic}, {D'Ammando}, {di Paola}, {Echevarr{\'\i}a}, {Efimova}, {Ehgamberdiev}, {Espinosa}, {Fuentes}, {Giunta}, {G{\'o}mez}, {Grishina}, {Gurwell}, {Hiriart}, {Jermak}, {Jordan}, {Jorstad}, {Joshi}, {Kopatskaya}, {Kuratov}, {Kurtanidze}, {Kurtanidze}, {L{\"a}hteenm{\"a}ki}, {Larionov}, {Larionova}, {Larionova}, {L{\'a}zaro}, {Lin}, {Malmrose}, {Marscher}, {Matsumoto}, {McBreen}, {Michel}, {Mihov}, {Minev}, {Mirzaqulov}, {Mokrushina}, {Molina}, {Moody}, {Morozova}, {Nazarov}, {Nikolashvili}, {Ohlert}, {Okhmat}, {Ovcharov}, {Pinna}, {Polakis}, {Protasio}, {Pursimo}, {Redondo-Lorenzo}, {Rizzi}, {Rodriguez-Coira}, {Sadakane}, {Sadun}, {Samal}, {Savchenko}, {Semkov}, {Skiff}, {Slavcheva-Mihova}, {Smith}, {Steele}, {Strigachev}, {Tammi}, {Thum}, {Tornikoski},
  {Troitskaya}, {Troitsky}, {Vasilyev}, \& {Vince}}]{raiteri17}
{Raiteri}, C.~M., {Villata}, M., {Acosta-Pulido}, J.~A., {et~al.} 2017, \nat, 552, 374

\bibitem[{{Ros} {et~al.}(2020){Ros}, {Kadler}, {Perucho}, {Boccardi}, {Cao}, {Giroletti}, {Krau{\ss}}, \& {Ojha}}]{ros20}
{Ros}, E., {Kadler}, M., {Perucho}, M., {et~al.} 2020, \aap, 633, L1

\bibitem[{{Sahakyan} {et~al.}(2023){Sahakyan}, {Giommi}, {Padovani}, {Petropoulou}, {B{\'e}gu{\'e}}, {Boccardi}, \& {Gasparyan}}]{sahakyan23}
{Sahakyan}, N., {Giommi}, P., {Padovani}, P., {et~al.} 2023, \mnras, 519, 1396

\bibitem[{{Schinzel} {et~al.}(2012){Schinzel}, {Lobanov}, {Taylor}, {Jorstad}, {Marscher}, \& {Zensus}}]{schinzel12}
{Schinzel}, F.~K., {Lobanov}, A.~P., {Taylor}, G.~B., {et~al.} 2012, \aap, 537, A70

\bibitem[{{Shappee} {et~al.}(2014){Shappee}, {Prieto}, {Grupe}, {Kochanek}, {Stanek}, {De Rosa}, {Mathur}, {Zu}, {Peterson}, {Pogge}, {Komossa}, {Im}, {Jencson}, {Holoien}, {Basu}, {Beacom}, {Szczygie{\l}}, {Brimacombe}, {Adams}, {Campillay}, {Choi}, {Contreras}, {Dietrich}, {Dubberley}, {Elphick}, {Foale}, {Giustini}, {Gonzalez}, {Hawkins}, {Howell}, {Hsiao}, {Koss}, {Leighly}, {Morrell}, {Mudd}, {Mullins}, {Nugent}, {Parrent}, {Phillips}, {Pojmanski}, {Rosing}, {Ross}, {Sand}, {Terndrup}, {Valenti}, {Walker}, \& {Yoon}}]{shappee14}
{Shappee}, B.~J., {Prieto}, J.~L., {Grupe}, D., {et~al.} 2014, \apj, 788, 48

\bibitem[{{Sharma}(2024)}]{sharma24}
{Sharma}, A. 2024, Universe, 10, 326

\bibitem[{{Shepherd}(1997)}]{shepherd97}
{Shepherd}, M.~C. 1997, in Astronomical Society of the Pacific Conference Series, Vol. 125, Astronomical Data Analysis Software and Systems VI, ed. G.~{Hunt} \& H.~{Payne}, 77

\bibitem[{{Singal}(2016)}]{singal16}
{Singal}, A.~K. 2016, \apj, 827, 66

\bibitem[{{Stroh} \& {Falcone}(2013)}]{stroh13}
{Stroh}, M.~C. \& {Falcone}, A.~D. 2013, \apjs, 207, 28

\bibitem[{{Tavecchio} \& {Ghisellini}(2015)}]{tavecchio15}
{Tavecchio}, F. \& {Ghisellini}, G. 2015, \mnras, 451, 1502

\bibitem[{{Tavecchio} {et~al.}(2014){Tavecchio}, {Ghisellini}, \& {Guetta}}]{tavecchio14}
{Tavecchio}, F., {Ghisellini}, G., \& {Guetta}, D. 2014, \apjl, 793, L18

\bibitem[{{Traianou} {et~al.}(2024){Traianou}, {Krichbaum}, {G{\'o}mez}, {Lico}, {Paraschos}, {Cho}, {Ros}, {Zhao}, {Liodakis}, {Dahale}, {Toscano}, {Fuentes}, {Foschi}, {Casadio}, {MacDonald}, {Kim}, {Hervet}, {Jorstad}, {Lobanov}, {Hodgson}, {Myserlis}, {Agudo}, {Zensus}, \& {Marscher}}]{traianou24}
{Traianou}, E., {Krichbaum}, T.~P., {G{\'o}mez}, J.~L., {et~al.} 2024, \aap, 682, A154

\bibitem[{{Virtanen} {et~al.}(2020){Virtanen}, {Gommers}, {Oliphant}, {Haberland}, {Reddy}, {Cournapeau}, {Burovski}, {Peterson}, {Weckesser}, {Bright}, {van der Walt}, {Brett}, {Wilson}, {Millman}, {Mayorov}, {Nelson}, {Jones}, {Kern}, {Larson}, {Carey}, {Polat}, {Feng}, {Moore}, {VanderPlas}, {Laxalde}, {Perktold}, {Cimrman}, {Henriksen}, {Quintero}, {Harris}, {Archibald}, {Ribeiro}, {Pedregosa}, {van Mulbregt}, \& {SciPy 1. 0 Contributors}}]{scipy}
{Virtanen}, P., {Gommers}, R., {Oliphant}, T.~E., {et~al.} 2020, Nature Methods, 17, 261

\bibitem[{{Weaver} {et~al.}(2022){Weaver}, {Jorstad}, {Marscher}, {Morozova}, {Troitsky}, {Agudo}, {G{\'o}mez}, {L{\"a}hteenm{\"a}ki}, {Tammi}, \& {Tornikoski}}]{weaver22}
{Weaver}, Z.~R., {Jorstad}, S.~G., {Marscher}, A.~P., {et~al.} 2022, \apjs, 260, 12

\end{thebibliography}

\appendix
\section{Supplementary materials}

\begin{figure*}
	\includegraphics[width=1.0\textwidth]{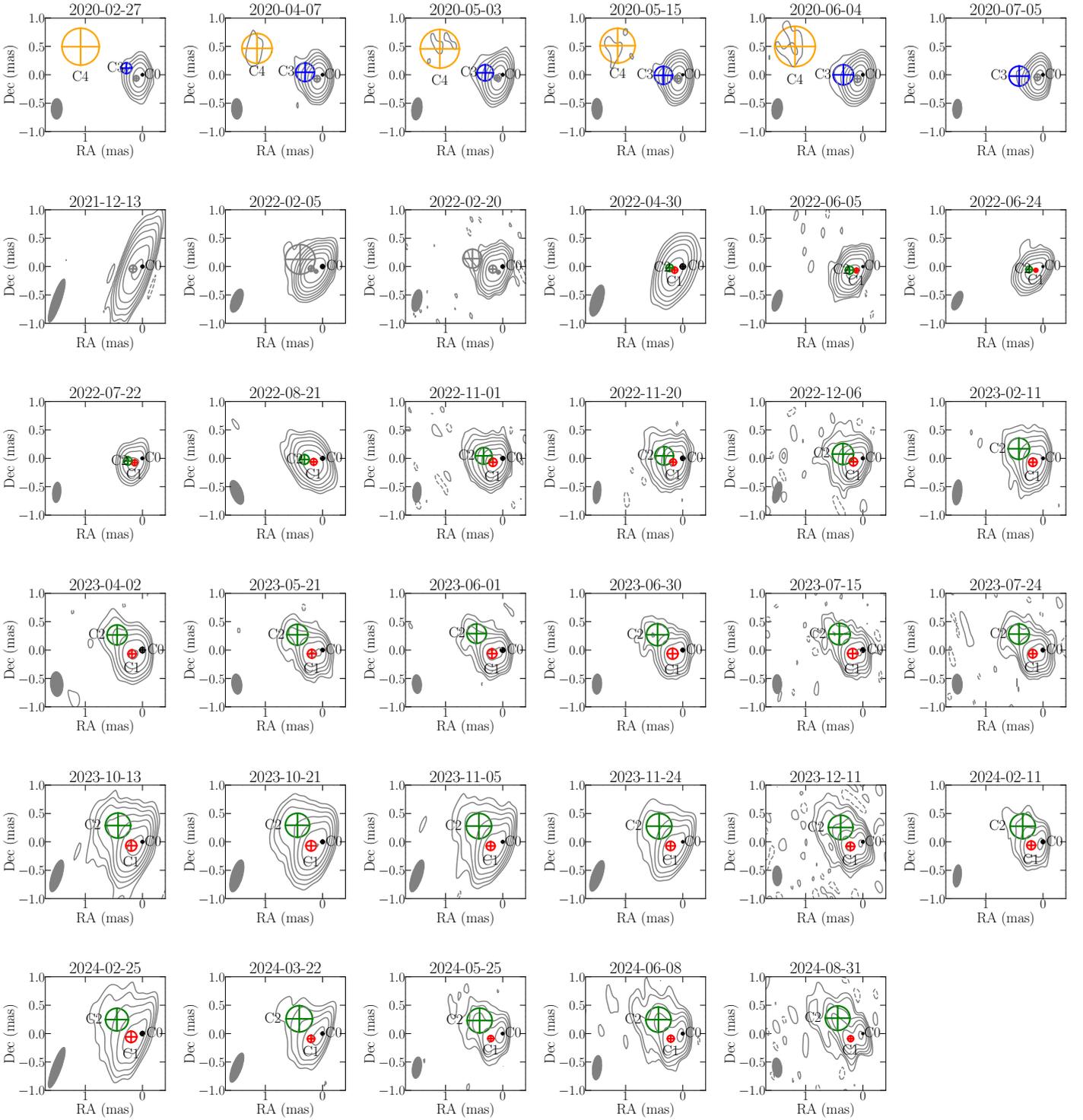}
    \caption{
    Distribution of model-fitted Gaussian components overlaid on the total intensity maps from Difmap using the modelfit with uniform weighting.
    In each panel, the contours represent the 43\,GHz intensity, crossed circles the location and FWHM sizes of the Gaussians, the labels C0 to C4 the component IDs, the grey ellipses the corresponding beam sizes, and the titles the observing epoch in year-month-day format.
    Contours start at three times the image rms noise level and increase by factor 2.
    }
    \label{fig:modelfit}
\end{figure*}

In Fig. \ref{fig:modelfit} we show, for all the analyzed epochs, the positions and FWHM sizes of all the components overlaid on the VLBA 43\,GHz total intensity maps of \pks from the model-fitting. 
In Table \ref{tab:properties_all_maps}, basic properties of all the CLEAN images, especially those used to obtain radio lightcurves and spectral indices in Fig. \ref{fig:lightcurve}, \ref{fig:lightcurve_zoom}, and \ref{fig:spectral} are presented.
We also show in Table \ref{tab:model_comp_all} detailed information of the Gaussian model components for the VLBA 43\,GHz for all the analyzed epochs.

\onecolumn
\begin{longtable}{c c c c c c c c c c }
\caption{Detailed properties of the analyzed VLBI images of \pks at 15 and 43\,GHz from the Difmap CLEAN and provided by the MOJAVE and BU databases. 
From left, each column shows
1) observing epoch in year-month-date format,
2) same as 1) in MJD,
3) flux densities of the nuclear region ($\leq$0.4\,mas within the peak), in Jy, that were used to estimate the spectral indices in Fig. \ref{fig:spectral}, 
4) total flux densities for the entire image regions in Jy,
5) and 6) FWHM sizes of the major and minor axes of the beam in mas, 
7) beam position angle in degrees, 
8) peak flux density in Jy/beam, and
9) the image rms in mJy/beam.
} \\
\hline
Epoch & Epoch & $S_{\rm core}$ & $S_{\rm tot}$ & $B_{\rm maj}$ & $B_{\rm min}$ & PA & Peak & RMS \\
(yyyy/mm/dd) & (MJD) & (Jy) & (Jy) & (mas) & (mas) & (deg) & (Jy/beam) & (mJy/beam) \\
%\hline
\endfirsthead

\multicolumn{9}{c}%
{{\tablename\ \thetable{} -- continued from previous page}} \\
\hline
Epoch & Epoch & $S_{\rm core}$ & $S_{\rm tot}$ & $B_{\rm maj}$ & $B_{\rm min}$ & PA & Peak & RMS \\
(yyyy/mm/dd) & (MJD) & (Jy) & (Jy) & (mas) & (mas) & (deg) & (Jy/beam) & (mJy/beam) \\
\hline
\endhead
\hline \multicolumn{9}{r}{{Continued on next page}} \\ 
%\hline
\endfoot

\endlastfoot

\hline
\multicolumn{9}{c}{15\,GHz} \\
\hline
2007/03/02 & 54161.0 & 0.391 $\pm$ 0.039 & 0.719 $\pm$ 0.072 & 1.087 & 0.583 & $-$5.356 & 0.391 $\pm$ 0.039 & 0.281 \\ 
2007/08/09 & 54321.0 & 0.491 $\pm$ 0.049 & 0.740 $\pm$ 0.074 & 1.050 & 0.530 & $-$7.939 & 0.442 $\pm$ 0.044 & 0.175 \\ 
2008/11/26 & 54796.0 & 0.378 $\pm$ 0.038 & 0.615 $\pm$ 0.062 & 1.474 & 0.575 & $-$17.619 & 0.349 $\pm$ 0.035 & 0.111 \\ 
2009/06/25 & 55007.0 & 0.350 $\pm$ 0.035 & 0.617 $\pm$ 0.062 & 1.023 & 0.555 & $-$3.749 & 0.322 $\pm$ 0.032 & 0.183 \\ 
2010/07/12 & 55389.0 & 0.326 $\pm$ 0.033 & 0.547 $\pm$ 0.055 & 1.186 & 0.484 & $-$8.162 & 0.315 $\pm$ 0.031 & 0.163 \\ 
2011/06/24 & 55736.0 & 0.353 $\pm$ 0.035 & 0.528 $\pm$ 0.053 & 1.128 & 0.536 & $-$7.966 & 0.335 $\pm$ 0.034 & 0.131 \\ 
2011/12/12 & 55907.0 & 0.422 $\pm$ 0.042 & 0.578 $\pm$ 0.058 & 1.153 & 0.628 & 8.509 & 0.415 $\pm$ 0.041 & 0.177 \\ 
2012/09/02 & 56172.0 & 0.730 $\pm$ 0.073 & 0.859 $\pm$ 0.086 & 1.196 & 0.597 & $-$0.335 & 0.704 $\pm$ 0.070 & 0.131 \\ 
2013/03/31 & 56382.0 & 0.881 $\pm$ 0.088 & 1.013 $\pm$ 0.101 & 1.063 & 0.600 & $-$4.343 & 0.830 $\pm$ 0.083 & 0.177 \\ 
2021/12/10 & 59558.0 & 0.816 $\pm$ 0.082 & 0.898 $\pm$ 0.090 & 1.228 & 0.598 & $-$2.431 & 0.757 $\pm$ 0.076 & 0.070 \\ 
2022/03/18 & 59656.0 & 1.068 $\pm$ 0.107 & 1.167 $\pm$ 0.117 & 1.174 & 0.590 & $-$3.229 & 0.988 $\pm$ 0.099 & 0.087 \\ 
2022/06/11 & 59741.0 & 1.110 $\pm$ 0.111 & 1.199 $\pm$ 0.120 & 1.126 & 0.558 & $-$6.074 & 1.019 $\pm$ 0.102 & 0.077 \\ 
2022/12/31 & 59944.0 & 0.996 $\pm$ 0.100 & 1.074 $\pm$ 0.107 & 1.138 & 0.527 & $-$9.980 & 0.854 $\pm$ 0.085 & 0.104 \\ 
2023/05/27 & 60091.0 & 0.859 $\pm$ 0.086 & 1.029 $\pm$ 0.103 & 1.395 & 0.608 & $-$9.768 & 0.802 $\pm$ 0.080 & 0.062 \\ 
2023/08/18 & 60174.0 & 0.866 $\pm$ 0.087 & 1.042 $\pm$ 0.104 & 1.415 & 0.592 & $-$13.726 & 0.800 $\pm$ 0.080 & 0.055 \\ 
2023/12/15 & 60293.0 & 1.082 $\pm$ 0.108 & 1.322 $\pm$ 0.132 & 1.194 & 0.598 & $-$5.985 & 0.989 $\pm$ 0.099 & 0.076 \\ 
2024/05/17 & 60447.0 & 0.801 $\pm$ 0.080 & 1.078 $\pm$ 0.108 & 1.158 & 0.639 & 3.107 & 0.785 $\pm$ 0.079 & 0.116 \\ 
\hline
\multicolumn{9}{c}{43\,GHz} \\
\hline
2007/06/13 & 54264.0 & 0.449 $\pm$ 0.045 & 1.552 $\pm$ 0.155 & 0.350 & 0.210 & $-$10.000 & 0.390 $\pm$ 0.039 & 4.145 \\ 
2007/07/12 & 54293.0 & 0.318 $\pm$ 0.032 & 1.170 $\pm$ 0.117 & 0.350 & 0.210 & $-$10.000 & 0.234 $\pm$ 0.023 & 3.791 \\ 
2007/08/30 & 54342.0 & 0.308 $\pm$ 0.031 & 0.719 $\pm$ 0.072 & 0.350 & 0.210 & $-$10.000 & 0.231 $\pm$ 0.023 & 3.490 \\ 
2007/09/29 & 54372.0 & 0.327 $\pm$ 0.033 & 0.560 $\pm$ 0.056 & 0.350 & 0.210 & $-$10.000 & 0.243 $\pm$ 0.024 & 5.020 \\ 
2007/11/01 & 54405.0 & 0.000 $\pm$ 0.000 & 0.423 $\pm$ 0.042 & 0.350 & 0.210 & 0.000 & 0.270 $\pm$ 0.027 & 3.630 \\ 
2008/02/28 & 54524.0 & 0.239 $\pm$ 0.024 & 0.486 $\pm$ 0.049 & 0.350 & 0.210 & $-$10.000 & 0.175 $\pm$ 0.018 & 1.568 \\ 
2008/06/12 & 54629.0 & 0.375 $\pm$ 0.038 & 0.621 $\pm$ 0.062 & 0.350 & 0.210 & $-$10.000 & 0.225 $\pm$ 0.023 & 2.540 \\ 
2008/07/06 & 54653.0 & 0.410 $\pm$ 0.041 & 0.835 $\pm$ 0.084 & 0.350 & 0.210 & $-$10.000 & 0.344 $\pm$ 0.034 & 5.448 \\ 
2008/08/15 & 54693.0 & 0.306 $\pm$ 0.031 & 0.717 $\pm$ 0.072 & 0.350 & 0.210 & $-$10.000 & 0.213 $\pm$ 0.021 & 1.717 \\ 
2008/09/10 & 54719.0 & 0.297 $\pm$ 0.030 & 0.563 $\pm$ 0.056 & 0.364 & 0.161 & $-$9.255 & 0.180 $\pm$ 0.018 & 0.775 \\ 
2008/11/16 & 54786.0 & 0.286 $\pm$ 0.029 & 0.435 $\pm$ 0.043 & 0.350 & 0.210 & $-$10.000 & 0.195 $\pm$ 0.019 & 1.009 \\ 
2008/12/21 & 54821.0 & 0.151 $\pm$ 0.015 & 0.262 $\pm$ 0.026 & 0.350 & 0.210 & $-$10.000 & 0.130 $\pm$ 0.013 & 0.790 \\ 
2009/01/24 & 54855.0 & 0.237 $\pm$ 0.024 & 0.483 $\pm$ 0.048 & 0.350 & 0.210 & $-$10.000 & 0.179 $\pm$ 0.018 & 0.532 \\ 
2009/02/22 & 54884.0 & 0.257 $\pm$ 0.026 & 0.583 $\pm$ 0.058 & 0.350 & 0.210 & $-$10.000 & 0.194 $\pm$ 0.019 & 0.720 \\ 
2009/04/01 & 54922.0 & 0.215 $\pm$ 0.021 & 0.453 $\pm$ 0.045 & 0.350 & 0.210 & $-$10.000 & 0.172 $\pm$ 0.017 & 0.570 \\ 
2009/05/30 & 54981.0 & 0.151 $\pm$ 0.015 & 0.300 $\pm$ 0.030 & 0.420 & 0.179 & $-$14.056 & 0.124 $\pm$ 0.012 & 1.043 \\ 
2009/06/21 & 55003.0 & 0.253 $\pm$ 0.025 & 0.523 $\pm$ 0.052 & 0.371 & 0.241 & $-$1.991 & 0.188 $\pm$ 0.019 & 1.106 \\ 
2009/07/26 & 55038.0 & 0.224 $\pm$ 0.022 & 0.461 $\pm$ 0.046 & 0.363 & 0.161 & $-$4.534 & 0.181 $\pm$ 0.018 & 1.360 \\ 
2009/08/16 & 55059.0 & 0.208 $\pm$ 0.021 & 0.386 $\pm$ 0.039 & 0.319 & 0.156 & $-$1.464 & 0.133 $\pm$ 0.013 & 0.815 \\ 
2009/09/16 & 55090.0 & 0.264 $\pm$ 0.026 & 0.435 $\pm$ 0.043 & 0.350 & 0.210 & $-$10.000 & 0.228 $\pm$ 0.023 & 0.658 \\ 
2009/10/14 & 55118.0 & 0.190 $\pm$ 0.019 & 0.341 $\pm$ 0.034 & 0.350 & 0.210 & $-$10.000 & 0.165 $\pm$ 0.016 & 1.599 \\ 
2009/10/16 & 55120.0 & 0.265 $\pm$ 0.027 & 0.476 $\pm$ 0.048 & 0.350 & 0.210 & $-$10.000 & 0.218 $\pm$ 0.022 & 0.458 \\ 
2009/10/20 & 55124.0 & 0.252 $\pm$ 0.025 & 0.377 $\pm$ 0.038 & 0.350 & 0.210 & $-$10.000 & 0.223 $\pm$ 0.022 & 0.878 \\ 
2009/10/25 & 55129.0 & 0.157 $\pm$ 0.016 & 0.246 $\pm$ 0.025 & 0.350 & 0.210 & $-$10.000 & 0.134 $\pm$ 0.013 & 0.591 \\ 
2009/11/28 & 55163.0 & 0.243 $\pm$ 0.024 & 0.487 $\pm$ 0.049 & 0.400 & 0.200 & $-$10.000 & 0.213 $\pm$ 0.021 & 0.561 \\ 
2010/01/10 & 55206.0 & 0.188 $\pm$ 0.019 & 0.450 $\pm$ 0.045 & 0.350 & 0.210 & $-$10.000 & 0.169 $\pm$ 0.017 & 0.819 \\ 
2010/02/10 & 55237.0 & 0.137 $\pm$ 0.014 & 0.346 $\pm$ 0.035 & 0.350 & 0.210 & $-$10.000 & 0.128 $\pm$ 0.013 & 0.612 \\ 
2010/03/06 & 55261.0 & 0.237 $\pm$ 0.024 & 0.413 $\pm$ 0.041 & 0.350 & 0.210 & $-$10.000 & 0.206 $\pm$ 0.021 & 0.485 \\ 
2010/05/19 & 55335.0 & 0.249 $\pm$ 0.025 & 0.568 $\pm$ 0.057 & 0.350 & 0.210 & $-$10.000 & 0.202 $\pm$ 0.020 & 0.998 \\ 
2010/06/14 & 55361.0 & 0.230 $\pm$ 0.023 & 0.371 $\pm$ 0.037 & 0.350 & 0.210 & $-$10.000 & 0.195 $\pm$ 0.020 & 0.600 \\ 
2010/08/01 & 55409.0 & 0.236 $\pm$ 0.024 & 0.525 $\pm$ 0.053 & 0.350 & 0.210 & $-$10.000 & 0.195 $\pm$ 0.020 & 1.579 \\ 
2010/08/21 & 55429.0 & 0.209 $\pm$ 0.021 & 0.489 $\pm$ 0.049 & 0.350 & 0.210 & $-$10.000 & 0.177 $\pm$ 0.018 & 1.188 \\ 
2010/09/18 & 55457.0 & 0.271 $\pm$ 0.027 & 0.613 $\pm$ 0.061 & 0.350 & 0.210 & $-$10.000 & 0.220 $\pm$ 0.022 & 0.884 \\ 
2010/10/24 & 55493.0 & 0.276 $\pm$ 0.028 & 0.404 $\pm$ 0.040 & 0.350 & 0.210 & $-$10.000 & 0.233 $\pm$ 0.023 & 0.821 \\ 
2010/11/01 & 55501.0 & 0.357 $\pm$ 0.036 & 0.755 $\pm$ 0.075 & 0.350 & 0.210 & $-$10.000 & 0.281 $\pm$ 0.028 & 1.351 \\ 
2010/11/06 & 55506.0 & 0.437 $\pm$ 0.044 & 0.860 $\pm$ 0.086 & 0.350 & 0.210 & $-$10.000 & 0.321 $\pm$ 0.032 & 1.399 \\ 
2010/11/12 & 55512.0 & 0.452 $\pm$ 0.045 & 0.978 $\pm$ 0.098 & 0.350 & 0.210 & $-$10.000 & 0.348 $\pm$ 0.035 & 1.706 \\ 
2010/12/04 & 55534.0 & 0.256 $\pm$ 0.026 & 0.379 $\pm$ 0.038 & 0.350 & 0.210 & $-$10.000 & 0.209 $\pm$ 0.021 & 0.774 \\ 
2011/01/02 & 55563.0 & 0.273 $\pm$ 0.027 & 0.577 $\pm$ 0.058 & 0.350 & 0.210 & $-$10.000 & 0.233 $\pm$ 0.023 & 1.434 \\ 
2011/02/04 & 55596.0 & 0.229 $\pm$ 0.023 & 0.330 $\pm$ 0.033 & 0.350 & 0.210 & $-$10.000 & 0.197 $\pm$ 0.020 & 0.981 \\ 
2011/03/01 & 55621.0 & 0.318 $\pm$ 0.032 & 0.605 $\pm$ 0.061 & 0.350 & 0.210 & $-$10.000 & 0.216 $\pm$ 0.022 & 0.972 \\ 
2011/04/21 & 55672.0 & 0.292 $\pm$ 0.029 & 0.504 $\pm$ 0.050 & 0.350 & 0.210 & $-$10.000 & 0.236 $\pm$ 0.024 & 0.496 \\ 
2011/05/22 & 55703.0 & 0.337 $\pm$ 0.034 & 0.596 $\pm$ 0.060 & 0.350 & 0.210 & $-$10.000 & 0.228 $\pm$ 0.023 & 1.761 \\ 
2011/06/12 & 55724.0 & 0.252 $\pm$ 0.025 & 0.440 $\pm$ 0.044 & 0.350 & 0.210 & $-$10.000 & 0.206 $\pm$ 0.021 & 0.470 \\ 
2011/07/21 & 55763.0 & 0.279 $\pm$ 0.028 & 0.408 $\pm$ 0.041 & 0.350 & 0.210 & $-$10.000 & 0.215 $\pm$ 0.021 & 0.479 \\ 
2011/08/23 & 55796.0 & 0.303 $\pm$ 0.030 & 0.487 $\pm$ 0.049 & 0.350 & 0.210 & $-$10.000 & 0.235 $\pm$ 0.023 & 0.744 \\ 
2011/09/16 & 55820.0 & 0.211 $\pm$ 0.021 & 0.268 $\pm$ 0.027 & 0.350 & 0.210 & $-$10.000 & 0.176 $\pm$ 0.018 & 0.419 \\ 
2011/09/20 & 55824.0 & 0.332 $\pm$ 0.033 & 0.537 $\pm$ 0.054 & 0.350 & 0.210 & $-$10.000 & 0.255 $\pm$ 0.026 & 0.412 \\ 
2011/10/16 & 55850.0 & 0.362 $\pm$ 0.036 & 0.495 $\pm$ 0.050 & 0.350 & 0.210 & $-$10.000 & 0.302 $\pm$ 0.030 & 0.717 \\ 
2011/12/02 & 55897.0 & 0.257 $\pm$ 0.026 & 0.385 $\pm$ 0.039 & 0.350 & 0.210 & $-$10.000 & 0.214 $\pm$ 0.021 & 0.525 \\ 
2012/01/27 & 55953.0 & 0.339 $\pm$ 0.034 & 0.609 $\pm$ 0.061 & 0.350 & 0.210 & $-$10.000 & 0.267 $\pm$ 0.027 & 0.704 \\ 
2012/03/05 & 55991.0 & 0.273 $\pm$ 0.027 & 0.721 $\pm$ 0.072 & 0.311 & 0.177 & $-$6.515 & 0.190 $\pm$ 0.019 & 1.990 \\ 
2012/04/02 & 56019.0 & 0.410 $\pm$ 0.041 & 0.552 $\pm$ 0.055 & 0.320 & 0.160 & $-$10.000 & 0.290 $\pm$ 0.029 & 1.723 \\ 
2012/05/26 & 56073.0 & 0.537 $\pm$ 0.054 & 0.582 $\pm$ 0.058 & 0.350 & 0.210 & $-$10.000 & 0.413 $\pm$ 0.041 & 3.696 \\ 
2012/07/04 & 56112.0 & 0.503 $\pm$ 0.050 & 0.790 $\pm$ 0.079 & 0.320 & 0.160 & $-$10.000 & 0.354 $\pm$ 0.035 & 0.821 \\ 
2012/08/13 & 56152.0 & 0.493 $\pm$ 0.049 & 0.667 $\pm$ 0.067 & 0.350 & 0.210 & $-$10.000 & 0.365 $\pm$ 0.036 & 0.525 \\ 
2012/10/28 & 56228.0 & 0.787 $\pm$ 0.079 & 0.868 $\pm$ 0.087 & 0.350 & 0.210 & $-$10.000 & 0.542 $\pm$ 0.054 & 0.727 \\ 
2012/12/21 & 56282.0 & 0.687 $\pm$ 0.069 & 0.830 $\pm$ 0.083 & 0.330 & 0.150 & $-$10.000 & 0.433 $\pm$ 0.043 & 1.309 \\ 
2013/01/15 & 56307.0 & 0.663 $\pm$ 0.066 & 0.842 $\pm$ 0.084 & 0.350 & 0.210 & $-$10.000 & 0.456 $\pm$ 0.046 & 0.553 \\ 
2013/02/26 & 56349.0 & 0.579 $\pm$ 0.058 & 0.637 $\pm$ 0.064 & 0.350 & 0.210 & $-$10.000 & 0.438 $\pm$ 0.044 & 1.324 \\ 
2013/04/16 & 56398.0 & 0.559 $\pm$ 0.056 & 0.656 $\pm$ 0.066 & 0.350 & 0.210 & $-$10.000 & 0.378 $\pm$ 0.038 & 1.474 \\ 
2013/05/30 & 56442.0 & 0.828 $\pm$ 0.083 & 0.919 $\pm$ 0.092 & 0.350 & 0.210 & $-$10.000 & 0.541 $\pm$ 0.054 & 1.425 \\ 
2013/06/30 & 56473.0 & 0.775 $\pm$ 0.078 & 1.009 $\pm$ 0.101 & 0.350 & 0.210 & $-$10.000 & 0.517 $\pm$ 0.052 & 0.399 \\ 
2013/07/28 & 56501.0 & 0.695 $\pm$ 0.070 & 1.120 $\pm$ 0.112 & 0.387 & 0.158 & $-$12.869 & 0.391 $\pm$ 0.039 & 0.920 \\ 
2013/08/26 & 56530.0 & 0.688 $\pm$ 0.069 & 0.993 $\pm$ 0.099 & 0.350 & 0.210 & $-$10.000 & 0.435 $\pm$ 0.044 & 1.346 \\ 
2013/11/18 & 56614.0 & 0.995 $\pm$ 0.099 & 1.285 $\pm$ 0.128 & 0.350 & 0.210 & $-$10.000 & 0.581 $\pm$ 0.058 & 0.875 \\ 
2013/12/16 & 56642.0 & 0.895 $\pm$ 0.089 & 1.209 $\pm$ 0.121 & 0.350 & 0.210 & $-$10.000 & 0.616 $\pm$ 0.062 & 0.725 \\ 
2014/01/19 & 56676.0 & 0.716 $\pm$ 0.072 & 0.835 $\pm$ 0.083 & 0.350 & 0.210 & $-$10.000 & 0.455 $\pm$ 0.045 & 0.574 \\ 
2014/02/24 & 56712.0 & 0.701 $\pm$ 0.070 & 0.925 $\pm$ 0.092 & 0.350 & 0.210 & $-$10.000 & 0.448 $\pm$ 0.045 & 0.723 \\ 
2014/05/03 & 56780.0 & 0.660 $\pm$ 0.066 & 0.796 $\pm$ 0.080 & 0.350 & 0.210 & $-$10.000 & 0.409 $\pm$ 0.041 & 0.549 \\ 
2014/06/20 & 56828.0 & 0.885 $\pm$ 0.089 & 1.487 $\pm$ 0.149 & 0.350 & 0.210 & $-$10.000 & 0.665 $\pm$ 0.066 & 1.569 \\ 
2014/07/28 & 56866.0 & 0.511 $\pm$ 0.051 & 0.707 $\pm$ 0.071 & 0.350 & 0.210 & $-$10.000 & 0.389 $\pm$ 0.039 & 0.543 \\ 
2014/09/23 & 56923.0 & 0.898 $\pm$ 0.090 & 1.122 $\pm$ 0.112 & 0.350 & 0.210 & $-$10.000 & 0.649 $\pm$ 0.065 & 0.519 \\ 
2014/11/15 & 56976.0 & 0.659 $\pm$ 0.066 & 0.877 $\pm$ 0.088 & 0.350 & 0.210 & $-$10.000 & 0.465 $\pm$ 0.046 & 0.565 \\ 
2014/12/05 & 56996.0 & 0.588 $\pm$ 0.059 & 0.741 $\pm$ 0.074 & 0.350 & 0.210 & $-$10.000 & 0.452 $\pm$ 0.045 & 0.308 \\ 
2014/12/29 & 57020.0 & 0.805 $\pm$ 0.080 & 1.087 $\pm$ 0.109 & 0.350 & 0.210 & $-$10.000 & 0.582 $\pm$ 0.058 & 0.354 \\ 
2015/02/14 & 57067.0 & 0.684 $\pm$ 0.068 & 1.015 $\pm$ 0.102 & 0.350 & 0.210 & $-$10.000 & 0.525 $\pm$ 0.053 & 0.547 \\ 
2015/04/11 & 57123.0 & 0.783 $\pm$ 0.078 & 0.992 $\pm$ 0.099 & 0.350 & 0.210 & $-$10.000 & 0.539 $\pm$ 0.054 & 0.374 \\ 
2015/05/11 & 57153.0 & 0.564 $\pm$ 0.056 & 0.689 $\pm$ 0.069 & 0.350 & 0.210 & $-$10.000 & 0.390 $\pm$ 0.039 & 0.304 \\ 
2015/06/09 & 57182.0 & 0.732 $\pm$ 0.073 & 1.084 $\pm$ 0.108 & 0.350 & 0.210 & $-$10.000 & 0.526 $\pm$ 0.053 & 0.422 \\ 
2015/07/02 & 57205.0 & 0.494 $\pm$ 0.049 & 0.645 $\pm$ 0.065 & 0.350 & 0.210 & $-$10.000 & 0.337 $\pm$ 0.034 & 0.260 \\ 
2015/08/01 & 57235.0 & 0.434 $\pm$ 0.043 & 0.580 $\pm$ 0.058 & 0.350 & 0.210 & $-$10.000 & 0.296 $\pm$ 0.030 & 0.311 \\ 
2015/09/22 & 57287.0 & 0.518 $\pm$ 0.052 & 0.805 $\pm$ 0.080 & 0.350 & 0.210 & $-$10.000 & 0.369 $\pm$ 0.037 & 0.685 \\ 
2015/12/05 & 57361.0 & 0.422 $\pm$ 0.042 & 0.576 $\pm$ 0.058 & 0.350 & 0.210 & $-$10.000 & 0.292 $\pm$ 0.029 & 0.261 \\ 
2016/01/01 & 57388.0 & 0.372 $\pm$ 0.037 & 0.546 $\pm$ 0.055 & 0.350 & 0.210 & $-$10.000 & 0.262 $\pm$ 0.026 & 0.560 \\ 
2016/01/31 & 57418.0 & 0.466 $\pm$ 0.047 & 0.683 $\pm$ 0.068 & 0.350 & 0.210 & $-$10.000 & 0.338 $\pm$ 0.034 & 0.656 \\ 
2016/03/18 & 57465.0 & 0.436 $\pm$ 0.044 & 0.571 $\pm$ 0.057 & 0.350 & 0.210 & $-$10.000 & 0.307 $\pm$ 0.031 & 0.259 \\ 
2016/04/22 & 57500.0 & 0.430 $\pm$ 0.043 & 0.565 $\pm$ 0.056 & 0.350 & 0.210 & $-$10.000 & 0.316 $\pm$ 0.032 & 0.294 \\ 
2016/06/10 & 57549.0 & 0.404 $\pm$ 0.040 & 0.549 $\pm$ 0.055 & 0.350 & 0.210 & $-$10.000 & 0.287 $\pm$ 0.029 & 0.380 \\ 
2016/07/04 & 57573.0 & 0.386 $\pm$ 0.039 & 0.527 $\pm$ 0.053 & 0.350 & 0.210 & $-$10.000 & 0.265 $\pm$ 0.027 & 0.438 \\ 
2016/07/31 & 57600.0 & 0.579 $\pm$ 0.058 & 0.807 $\pm$ 0.081 & 0.350 & 0.210 & $-$10.000 & 0.393 $\pm$ 0.039 & 0.707 \\ 
2016/09/05 & 57636.0 & 0.512 $\pm$ 0.051 & 0.687 $\pm$ 0.069 & 0.350 & 0.210 & $-$10.000 & 0.364 $\pm$ 0.036 & 0.456 \\ 
2016/10/06 & 57667.0 & 0.333 $\pm$ 0.033 & 0.465 $\pm$ 0.046 & 0.350 & 0.210 & $-$10.000 & 0.296 $\pm$ 0.030 & 0.443 \\ 
2016/10/23 & 57684.0 & 0.356 $\pm$ 0.036 & 0.477 $\pm$ 0.048 & 0.350 & 0.210 & $-$10.000 & 0.251 $\pm$ 0.025 & 0.375 \\ 
2016/11/28 & 57720.0 & 0.316 $\pm$ 0.032 & 0.428 $\pm$ 0.043 & 0.350 & 0.210 & $-$10.000 & 0.232 $\pm$ 0.023 & 0.428 \\ 
2016/12/23 & 57745.0 & 0.584 $\pm$ 0.058 & 0.931 $\pm$ 0.093 & 0.350 & 0.210 & $-$10.000 & 0.437 $\pm$ 0.044 & 0.769 \\ 
2017/01/14 & 57767.0 & 0.518 $\pm$ 0.052 & 0.691 $\pm$ 0.069 & 0.350 & 0.210 & $-$10.000 & 0.359 $\pm$ 0.036 & 0.440 \\ 
2017/02/04 & 57788.0 & 0.495 $\pm$ 0.050 & 0.652 $\pm$ 0.065 & 0.350 & 0.210 & $-$10.000 & 0.350 $\pm$ 0.035 & 0.347 \\ 
2017/03/19 & 57831.0 & 0.349 $\pm$ 0.035 & 0.486 $\pm$ 0.049 & 0.350 & 0.210 & $-$10.000 & 0.254 $\pm$ 0.025 & 0.469 \\ 
2017/04/16 & 57859.0 & 0.289 $\pm$ 0.029 & 0.391 $\pm$ 0.039 & 0.350 & 0.210 & $-$10.000 & 0.205 $\pm$ 0.020 & 0.248 \\ 
2017/05/13 & 57886.0 & 0.501 $\pm$ 0.050 & 0.694 $\pm$ 0.069 & 0.350 & 0.210 & $-$10.000 & 0.368 $\pm$ 0.037 & 0.734 \\ 
2017/06/08 & 57912.0 & 0.464 $\pm$ 0.046 & 0.626 $\pm$ 0.063 & 0.350 & 0.210 & $-$10.000 & 0.331 $\pm$ 0.033 & 0.396 \\ 
2017/07/03 & 57937.0 & 0.454 $\pm$ 0.045 & 0.652 $\pm$ 0.065 & 0.350 & 0.210 & $-$10.000 & 0.306 $\pm$ 0.031 & 0.384 \\ 
2017/08/06 & 57971.0 & 0.314 $\pm$ 0.031 & 0.410 $\pm$ 0.041 & 0.350 & 0.210 & $-$10.000 & 0.216 $\pm$ 0.022 & 0.228 \\ 
2017/09/04 & 58000.0 & 0.297 $\pm$ 0.030 & 0.409 $\pm$ 0.041 & 0.350 & 0.210 & $-$10.000 & 0.202 $\pm$ 0.020 & 0.297 \\ 
2017/11/06 & 58063.0 & 0.280 $\pm$ 0.028 & 0.375 $\pm$ 0.037 & 0.350 & 0.210 & $-$10.000 & 0.204 $\pm$ 0.020 & 0.182 \\ 
2017/12/23 & 58110.0 & 0.262 $\pm$ 0.026 & 0.372 $\pm$ 0.037 & 0.350 & 0.210 & $-$10.000 & 0.233 $\pm$ 0.023 & 0.160 \\ 
2018/02/17 & 58166.0 & 0.224 $\pm$ 0.022 & 0.315 $\pm$ 0.031 & 0.350 & 0.210 & $-$10.000 & 0.171 $\pm$ 0.017 & 0.244 \\ 
2018/03/10 & 58187.0 & 0.247 $\pm$ 0.025 & 0.364 $\pm$ 0.036 & 0.350 & 0.210 & $-$10.000 & 0.185 $\pm$ 0.018 & 0.223 \\ 
2018/04/19 & 58227.0 & 0.225 $\pm$ 0.023 & 0.314 $\pm$ 0.031 & 0.350 & 0.210 & $-$10.000 & 0.178 $\pm$ 0.018 & 0.138 \\ 
2018/05/11 & 58249.0 & 0.262 $\pm$ 0.026 & 0.382 $\pm$ 0.038 & 0.350 & 0.210 & $-$10.000 & 0.206 $\pm$ 0.021 & 0.296 \\ 
2018/06/16 & 58285.0 & 0.246 $\pm$ 0.025 & 0.330 $\pm$ 0.033 & 0.350 & 0.210 & $-$10.000 & 0.196 $\pm$ 0.020 & 0.279 \\ 
2018/07/16 & 58315.0 & 0.263 $\pm$ 0.026 & 0.414 $\pm$ 0.041 & 0.350 & 0.210 & $-$10.000 & 0.198 $\pm$ 0.020 & 0.321 \\ 
2018/08/26 & 58356.0 & 0.242 $\pm$ 0.024 & 0.327 $\pm$ 0.033 & 0.350 & 0.210 & $-$10.000 & 0.191 $\pm$ 0.019 & 0.262 \\ 
2018/10/15 & 58406.0 & 0.201 $\pm$ 0.020 & 0.319 $\pm$ 0.032 & 0.350 & 0.210 & $-$10.000 & 0.172 $\pm$ 0.017 & 0.282 \\ 
2018/12/08 & 58460.0 & 0.226 $\pm$ 0.023 & 0.309 $\pm$ 0.031 & 0.350 & 0.210 & $-$10.000 & 0.186 $\pm$ 0.019 & 0.186 \\ 
2019/01/10 & 58493.0 & 0.206 $\pm$ 0.021 & 0.310 $\pm$ 0.031 & 0.350 & 0.210 & $-$10.000 & 0.170 $\pm$ 0.017 & 0.257 \\ 
2019/02/03 & 58517.0 & 0.219 $\pm$ 0.022 & 0.346 $\pm$ 0.035 & 0.350 & 0.210 & $-$10.000 & 0.165 $\pm$ 0.016 & 0.172 \\ 
2019/02/08 & 58522.0 & 0.212 $\pm$ 0.021 & 0.282 $\pm$ 0.028 & 0.350 & 0.210 & $-$10.000 & 0.181 $\pm$ 0.018 & 0.281 \\ 
2019/03/31 & 58573.0 & 0.185 $\pm$ 0.018 & 0.265 $\pm$ 0.026 & 0.350 & 0.210 & $-$10.000 & 0.153 $\pm$ 0.015 & 0.177 \\ 
2019/05/12 & 58615.0 & 0.150 $\pm$ 0.015 & 0.214 $\pm$ 0.021 & 0.350 & 0.210 & $-$10.000 & 0.126 $\pm$ 0.013 & 0.426 \\ 
2019/06/18 & 58652.0 & 0.351 $\pm$ 0.035 & 0.557 $\pm$ 0.056 & 0.350 & 0.210 & $-$10.000 & 0.305 $\pm$ 0.030 & 1.393 \\ 
2019/07/01 & 58665.0 & 0.362 $\pm$ 0.036 & 0.781 $\pm$ 0.078 & 0.350 & 0.210 & $-$10.000 & 0.331 $\pm$ 0.033 & 1.706 \\ 
2019/08/03 & 58698.0 & 0.443 $\pm$ 0.044 & 0.787 $\pm$ 0.079 & 0.350 & 0.210 & $-$10.000 & 0.366 $\pm$ 0.037 & 1.324 \\ 
2019/10/06 & 58762.0 & 0.252 $\pm$ 0.025 & 0.320 $\pm$ 0.032 & 0.350 & 0.210 & $-$10.000 & 0.190 $\pm$ 0.019 & 0.404 \\ 
2019/10/19 & 58775.0 & 0.244 $\pm$ 0.024 & 0.316 $\pm$ 0.032 & 0.350 & 0.210 & $-$10.000 & 0.191 $\pm$ 0.019 & 0.155 \\ 
2019/11/03 & 58790.0 & 0.341 $\pm$ 0.034 & 0.473 $\pm$ 0.047 & 0.350 & 0.210 & $-$10.000 & 0.265 $\pm$ 0.026 & 0.487 \\ 
2019/12/03 & 58820.0 & 0.202 $\pm$ 0.020 & 0.294 $\pm$ 0.029 & 0.350 & 0.210 & $-$10.000 & 0.170 $\pm$ 0.017 & 0.227 \\ 
2020/01/03 & 58851.0 & 0.222 $\pm$ 0.022 & 0.302 $\pm$ 0.030 & 0.350 & 0.210 & $-$10.000 & 0.170 $\pm$ 0.017 & 0.363 \\ 
2020/02/27 & 58906.0 & 0.245 $\pm$ 0.024 & 0.333 $\pm$ 0.033 & 0.350 & 0.210 & $-$10.000 & 0.198 $\pm$ 0.020 & 0.449 \\ 
2020/04/07 & 58946.0 & 0.323 $\pm$ 0.032 & 0.411 $\pm$ 0.041 & 0.350 & 0.210 & $-$10.000 & 0.238 $\pm$ 0.024 & 0.417 \\ 
2020/05/03 & 58972.0 & 0.296 $\pm$ 0.030 & 0.386 $\pm$ 0.039 & 0.350 & 0.210 & $-$10.000 & 0.231 $\pm$ 0.023 & 0.186 \\ 
2020/05/15 & 58984.0 & 0.308 $\pm$ 0.031 & 0.393 $\pm$ 0.039 & 0.350 & 0.210 & $-$10.000 & 0.215 $\pm$ 0.022 & 0.263 \\ 
2020/06/04 & 59004.0 & 0.306 $\pm$ 0.031 & 0.388 $\pm$ 0.039 & 0.350 & 0.210 & $-$10.000 & 0.224 $\pm$ 0.022 & 0.233 \\ 
2020/07/05 & 59035.0 & 0.256 $\pm$ 0.026 & 0.333 $\pm$ 0.033 & 0.350 & 0.210 & $-$10.000 & 0.199 $\pm$ 0.020 & 0.240 \\ 
2021/12/13 & 59561.0 & 0.760 $\pm$ 0.076 & 0.858 $\pm$ 0.086 & 0.350 & 0.150 & $-$10.000 & 0.386 $\pm$ 0.039 & 0.490 \\ 
2022/02/05 & 59615.0 & 0.946 $\pm$ 0.095 & 1.022 $\pm$ 0.102 & 0.350 & 0.150 & $-$10.000 & 0.433 $\pm$ 0.043 & 0.296 \\ 
2022/02/20 & 59630.0 & 0.964 $\pm$ 0.096 & 1.049 $\pm$ 0.105 & 0.350 & 0.150 & $-$10.000 & 0.438 $\pm$ 0.044 & 0.447 \\ 
2022/04/30 & 59699.0 & 1.751 $\pm$ 0.175 & 1.937 $\pm$ 0.194 & 0.350 & 0.150 & $-$10.000 & 0.831 $\pm$ 0.083 & 0.549 \\ 
2022/06/05 & 59735.0 & 1.372 $\pm$ 0.137 & 1.499 $\pm$ 0.150 & 0.350 & 0.150 & $-$10.000 & 0.643 $\pm$ 0.064 & 0.529 \\ 
2022/06/24 & 59754.0 & 1.227 $\pm$ 0.123 & 1.402 $\pm$ 0.140 & 0.350 & 0.150 & $-$10.000 & 0.575 $\pm$ 0.058 & 0.526 \\ 
2022/07/15 & 59775.0 & 1.193 $\pm$ 0.119 & 1.392 $\pm$ 0.139 & 0.350 & 0.150 & $-$10.000 & 0.566 $\pm$ 0.057 & 0.770 \\ 
2022/07/22 & 59782.0 & 1.158 $\pm$ 0.116 & 1.317 $\pm$ 0.132 & 0.350 & 0.150 & $-$10.000 & 0.552 $\pm$ 0.055 & 0.562 \\ 
2022/08/21 & 59812.0 & 1.683 $\pm$ 0.168 & 1.947 $\pm$ 0.195 & 0.350 & 0.150 & $-$10.000 & 0.772 $\pm$ 0.077 & 0.784 \\ 
2022/11/01 & 59884.0 & 1.061 $\pm$ 0.106 & 1.181 $\pm$ 0.118 & 0.350 & 0.150 & $-$10.000 & 0.451 $\pm$ 0.045 & 0.316 \\ 
2022/11/20 & 59903.0 & 1.027 $\pm$ 0.103 & 1.172 $\pm$ 0.117 & 0.350 & 0.150 & $-$10.000 & 0.459 $\pm$ 0.046 & 0.536 \\ 
2022/12/06 & 59919.0 & 0.949 $\pm$ 0.095 & 1.090 $\pm$ 0.109 & 0.350 & 0.150 & $-$10.000 & 0.432 $\pm$ 0.043 & 0.359 \\ 
2023/02/11 & 59986.0 & 1.159 $\pm$ 0.116 & 1.380 $\pm$ 0.138 & 0.350 & 0.150 & $-$10.000 & 0.516 $\pm$ 0.052 & 0.671 \\ 
2023/04/02 & 60036.0 & 0.869 $\pm$ 0.087 & 1.232 $\pm$ 0.123 & 0.350 & 0.150 & $-$10.000 & 0.370 $\pm$ 0.037 & 1.099 \\ 
2023/05/21 & 60085.0 & 1.127 $\pm$ 0.113 & 1.412 $\pm$ 0.141 & 0.350 & 0.150 & $-$10.000 & 0.475 $\pm$ 0.047 & 0.755 \\ 
2023/06/01 & 60096.0 & 0.859 $\pm$ 0.086 & 1.063 $\pm$ 0.106 & 0.350 & 0.150 & $-$10.000 & 0.347 $\pm$ 0.035 & 0.538 \\ 
2023/06/30 & 60125.0 & 0.914 $\pm$ 0.091 & 1.128 $\pm$ 0.113 & 0.350 & 0.150 & $-$10.000 & 0.385 $\pm$ 0.039 & 0.442 \\ 
2023/07/15 & 60140.0 & 1.146 $\pm$ 0.115 & 1.481 $\pm$ 0.148 & 0.350 & 0.150 & $-$10.000 & 0.487 $\pm$ 0.049 & 0.702 \\ 
2023/07/24 & 60149.0 & 1.075 $\pm$ 0.108 & 1.433 $\pm$ 0.143 & 0.350 & 0.150 & $-$10.000 & 0.499 $\pm$ 0.050 & 0.889 \\ 
2023/10/13 & 60230.0 & 1.231 $\pm$ 0.123 & 1.507 $\pm$ 0.151 & 0.350 & 0.150 & $-$10.000 & 0.528 $\pm$ 0.053 & 0.375 \\ 
2023/10/21 & 60238.0 & 1.439 $\pm$ 0.144 & 1.712 $\pm$ 0.171 & 0.350 & 0.150 & $-$10.000 & 0.599 $\pm$ 0.060 & 0.296 \\ 
2023/11/05 & 60253.0 & 1.312 $\pm$ 0.131 & 1.623 $\pm$ 0.162 & 0.350 & 0.150 & $-$10.000 & 0.559 $\pm$ 0.056 & 0.770 \\ 
2023/11/24 & 60272.0 & 0.989 $\pm$ 0.099 & 1.246 $\pm$ 0.125 & 0.350 & 0.150 & $-$10.000 & 0.390 $\pm$ 0.039 & 0.453 \\ 
2023/12/11 & 60289.0 & 0.857 $\pm$ 0.086 & 1.100 $\pm$ 0.110 & 0.350 & 0.150 & $-$10.000 & 0.344 $\pm$ 0.034 & 0.406 \\ 
2024/02/11 & 60351.0 & 1.189 $\pm$ 0.119 & 1.704 $\pm$ 0.170 & 0.350 & 0.150 & $-$10.000 & 0.524 $\pm$ 0.052 & 1.383 \\ 
2024/02/25 & 60365.0 & 0.913 $\pm$ 0.091 & 1.193 $\pm$ 0.119 & 0.350 & 0.150 & $-$10.000 & 0.390 $\pm$ 0.039 & 0.361 \\ 
2024/03/22 & 60391.0 & 0.568 $\pm$ 0.057 & 0.806 $\pm$ 0.081 & 0.350 & 0.150 & $-$10.000 & 0.260 $\pm$ 0.026 & 0.487 \\ 
2024/05/25 & 60455.0 & 0.710 $\pm$ 0.071 & 1.087 $\pm$ 0.109 & 0.350 & 0.150 & $-$10.000 & 0.328 $\pm$ 0.033 & 0.754 \\ 
2024/06/08 & 60469.0 & 0.958 $\pm$ 0.096 & 1.374 $\pm$ 0.137 & 0.350 & 0.150 & $-$10.000 & 0.431 $\pm$ 0.043 & 0.620 \\ 
2024/07/03 & 60494.0 & 1.837 $\pm$ 0.184 & 2.643 $\pm$ 0.264 & 0.350 & 0.150 & $-$10.000 & 0.840 $\pm$ 0.084 & 2.932 \\ 
2024/08/31 & 60553.0 & 0.880 $\pm$ 0.088 & 1.383 $\pm$ 0.138 & 0.350 & 0.150 & $-$10.000 & 0.421 $\pm$ 0.042 & 0.776 \\ 
\hline
\label{tab:properties_all_maps}
\end{longtable}

\onecolumn
\begin{longtable}{c c c c c c}
\caption{
Detailed properties of the Gaussian components from model-fitting of the VLBA 43\,GHz data of \pks.
From left, each column shows 
1) component ID,
2) flux density of the component in Jy,
3) radial distance of the component from the core (C0) in mas,
4) position angle of the component in degrees,
5) FWHM size in mas, and
6) logarithm of the observed brightness temperature in K.
}\\
\hline
ID & Flux density & Radial Distance & P.A. & FWHM & $\log T_{\mathrm{B}}$ \\
 & (Jy) & (mas) & (deg) & (mas) & ($K$) \\
\hline
\endfirsthead

\multicolumn{6}{c}%
{{\tablename\ \thetable{} -- continued from previous page}} \\
\hline
\textbf{ID} & \textbf{Flux (Jy)} & \textbf{Radial Distance (mas)} & \textbf{P.A. ($^\circ$)} & \textbf{FWHM (mas)} & \textbf{$\log T_{\mathrm{B}}$ (K)} \\
\endhead

\hline \multicolumn{6}{r}{{Continued on next page}} \\ 
%\hline
\endfoot

\hline
\endlastfoot
    \multicolumn{6}{c}{\textbf{MJD : 58906.0}} \\ \hline
C0 & 0.162 $\pm$ 0.018 & 0.000 $\pm$ 0.000 & 0.000 $\pm$ 0.000 & 0.032 $\pm$ 0.011 & 11.171 $\pm$ 0.291 \\ \hline
C3 & 0.007 $\pm$ 0.013 & 0.305 $\pm$ 0.172 & 67.467 $\pm$ 50.662 & 0.189 $\pm$ 0.057 & 8.264 $\pm$ 0.874 \\ \hline
C4 & 0.022 $\pm$ 0.015 & 1.190 $\pm$ 0.391 & 65.417 $\pm$ 28.525 & 0.651 $\pm$ 0.084 & 7.687 $\pm$ 0.326 \\ \hline
N0 & 0.077 $\pm$ 0.012 & 0.126 $\pm$ 0.019 & 120.027 $\pm$ 12.043 & 0.094 $\pm$ 0.022 & 9.912 $\pm$ 0.215 \\ \hline
\multicolumn{6}{c}{\textbf{MJD : 58946.0}} \\ \hline
C0 & 0.136 $\pm$ 0.015 & 0.000 $\pm$ 0.000 & 0.000 $\pm$ 0.000 & 0.033 $\pm$ 0.011 & 11.068 $\pm$ 0.300 \\ \hline
C3 & 0.024 $\pm$ 0.013 & 0.309 $\pm$ 0.137 & 81.634 $\pm$ 48.658 & 0.328 $\pm$ 0.056 & 8.320 $\pm$ 0.286 \\ \hline
C4 & 0.020 $\pm$ 0.015 & 1.238 $\pm$ 0.311 & 67.932 $\pm$ 22.841 & 0.527 $\pm$ 0.076 & 7.829 $\pm$ 0.349 \\ \hline
N0 & 0.158 $\pm$ 0.018 & 0.119 $\pm$ 0.017 & 126.098 $\pm$ 9.763 & 0.112 $\pm$ 0.021 & 10.072 $\pm$ 0.167 \\ \hline
\multicolumn{6}{c}{\textbf{MJD : 58972.0}} \\ \hline
C0 & 0.149 $\pm$ 0.016 & 0.000 $\pm$ 0.000 & 0.000 $\pm$ 0.000 & 0.035 $\pm$ 0.011 & 11.057 $\pm$ 0.285 \\ \hline
C3 & 0.021 $\pm$ 0.013 & 0.307 $\pm$ 0.124 & 83.454 $\pm$ 45.007 & 0.285 $\pm$ 0.054 & 8.384 $\pm$ 0.319 \\ \hline
C4 & 0.024 $\pm$ 0.016 & 1.192 $\pm$ 0.383 & 67.395 $\pm$ 28.899 & 0.682 $\pm$ 0.085 & 7.684 $\pm$ 0.301 \\ \hline
N0 & 0.129 $\pm$ 0.016 & 0.115 $\pm$ 0.014 & 118.131 $\pm$ 10.091 & 0.096 $\pm$ 0.020 & 10.118 $\pm$ 0.187 \\ \hline
\multicolumn{6}{c}{\textbf{MJD : 58984.0}} \\ \hline
C0 & 0.122 $\pm$ 0.014 & 0.000 $\pm$ 0.000 & 0.000 $\pm$ 0.000 & 0.024 $\pm$ 0.010 & 11.298 $\pm$ 0.359 \\ \hline
C3 & 0.012 $\pm$ 0.014 & 0.339 $\pm$ 0.194 & 91.690 $\pm$ 65.551 & 0.319 $\pm$ 0.065 & 8.043 $\pm$ 0.543 \\ \hline
C4 & 0.017 $\pm$ 0.016 & 1.250 $\pm$ 0.422 & 65.772 $\pm$ 29.487 & 0.612 $\pm$ 0.086 & 7.629 $\pm$ 0.418 \\ \hline
N0 & 0.143 $\pm$ 0.017 & 0.116 $\pm$ 0.018 & 124.509 $\pm$ 11.287 & 0.117 $\pm$ 0.021 & 9.991 $\pm$ 0.168 \\ \hline
\multicolumn{6}{c}{\textbf{MJD : 59004.0}} \\ \hline
C0 & 0.139 $\pm$ 0.016 & 0.000 $\pm$ 0.000 & 0.000 $\pm$ 0.000 & 0.035 $\pm$ 0.012 & 11.027 $\pm$ 0.290 \\ \hline
C3 & 0.018 $\pm$ 0.014 & 0.342 $\pm$ 0.173 & 90.168 $\pm$ 57.963 & 0.355 $\pm$ 0.063 & 8.127 $\pm$ 0.371 \\ \hline
C4 & 0.023 $\pm$ 0.016 & 1.289 $\pm$ 0.414 & 67.216 $\pm$ 28.792 & 0.711 $\pm$ 0.088 & 7.630 $\pm$ 0.315 \\ \hline
N0 & 0.145 $\pm$ 0.017 & 0.120 $\pm$ 0.020 & 126.203 $\pm$ 11.692 & 0.126 $\pm$ 0.022 & 9.932 $\pm$ 0.162 \\ \hline
N0 & 0.021 $\pm$ 0.019 & 3.243 $\pm$ 1.114 & 71.610 $\pm$ 33.243 & 1.624 $\pm$ 0.149 & 6.873 $\pm$ 0.393 \\ \hline
\multicolumn{6}{c}{\textbf{MJD : 59035.0}} \\ \hline
C0 & 0.128 $\pm$ 0.015 & 0.000 $\pm$ 0.000 & 0.000 $\pm$ 0.000 & 0.036 $\pm$ 0.012 & 10.966 $\pm$ 0.292 \\ \hline
C3 & 0.018 $\pm$ 0.014 & 0.418 $\pm$ 0.171 & 93.157 $\pm$ 46.772 & 0.351 $\pm$ 0.063 & 8.136 $\pm$ 0.370 \\ \hline
N0 & 0.117 $\pm$ 0.015 & 0.109 $\pm$ 0.018 & 113.749 $\pm$ 14.153 & 0.114 $\pm$ 0.022 & 9.926 $\pm$ 0.178 \\ \hline
\multicolumn{6}{c}{\textbf{MJD : 59561.0}} \\ \hline
C0 & 0.341 $\pm$ 0.035 & 0.000 $\pm$ 0.000 & 0.000 $\pm$ 0.000 & 0.045 $\pm$ 0.011 & 11.198 $\pm$ 0.209 \\ \hline
N0 & 0.421 $\pm$ 0.043 & 0.171 $\pm$ 0.008 & 103.548 $\pm$ 5.090 & 0.129 $\pm$ 0.018 & 10.375 $\pm$ 0.128 \\ \hline
\multicolumn{6}{c}{\textbf{MJD : 59615.0}} \\ \hline
C0 & 0.347 $\pm$ 0.035 & 0.000 $\pm$ 0.000 & 0.000 $\pm$ 0.000 & 0.066 $\pm$ 0.013 & 10.873 $\pm$ 0.175 \\ \hline
N0 & 0.174 $\pm$ 0.019 & 0.142 $\pm$ 0.007 & 126.384 $\pm$ 3.429 & 0.057 $\pm$ 0.014 & 10.701 $\pm$ 0.219 \\ \hline
N0 & 0.403 $\pm$ 0.041 & 0.202 $\pm$ 0.006 & 100.254 $\pm$ 3.280 & 0.100 $\pm$ 0.016 & 10.577 $\pm$ 0.142 \\ \hline
N0 & 0.035 $\pm$ 0.014 & 0.412 $\pm$ 0.208 & 72.181 $\pm$ 49.361 & 0.521 $\pm$ 0.067 & 8.082 $\pm$ 0.211 \\ \hline
\multicolumn{6}{c}{\textbf{MJD : 59630.0}} \\ \hline
C0 & 0.357 $\pm$ 0.036 & 0.000 $\pm$ 0.000 & 0.000 $\pm$ 0.000 & 0.052 $\pm$ 0.011 & 11.092 $\pm$ 0.193 \\ \hline
N0 & 0.017 $\pm$ 0.010 & 0.124 $\pm$ 0.035 & 142.853 $\pm$ 13.591 & 0.058 $\pm$ 0.025 & 9.675 $\pm$ 0.450 \\ \hline
N0 & 0.577 $\pm$ 0.058 & 0.180 $\pm$ 0.007 & 104.808 $\pm$ 4.204 & 0.135 $\pm$ 0.017 & 10.472 $\pm$ 0.118 \\ \hline
N0 & 0.016 $\pm$ 0.014 & 0.548 $\pm$ 0.187 & 75.203 $\pm$ 34.926 & 0.332 $\pm$ 0.062 & 8.134 $\pm$ 0.412 \\ \hline
\multicolumn{6}{c}{\textbf{MJD : 59699.0}} \\ \hline
C0 & 0.649 $\pm$ 0.065 & 0.000 $\pm$ 0.000 & 0.000 $\pm$ 0.000 & 0.089 $\pm$ 0.013 & 10.885 $\pm$ 0.135 \\ \hline
C1 & 0.678 $\pm$ 0.068 & 0.155 $\pm$ 0.005 & 113.925 $\pm$ 3.104 & 0.104 $\pm$ 0.014 & 10.769 $\pm$ 0.126 \\ \hline
C2 & 0.411 $\pm$ 0.042 & 0.232 $\pm$ 0.008 & 95.686 $\pm$ 3.757 & 0.126 $\pm$ 0.018 & 10.385 $\pm$ 0.129 \\ \hline
\multicolumn{6}{c}{\textbf{MJD : 59735.0}} \\ \hline
C0 & 0.260 $\pm$ 0.027 & 0.000 $\pm$ 0.000 & 0.000 $\pm$ 0.000 & 0.023 $\pm$ 0.008 & 11.663 $\pm$ 0.304 \\ \hline
C1 & 0.439 $\pm$ 0.045 & 0.132 $\pm$ 0.006 & 122.005 $\pm$ 3.430 & 0.083 $\pm$ 0.014 & 10.776 $\pm$ 0.151 \\ \hline
C2 & 0.668 $\pm$ 0.067 & 0.243 $\pm$ 0.007 & 105.007 $\pm$ 3.050 & 0.143 $\pm$ 0.017 & 10.486 $\pm$ 0.113 \\ \hline
\multicolumn{6}{c}{\textbf{MJD : 59754.0}} \\ \hline
C0 & 0.334 $\pm$ 0.034 & 0.000 $\pm$ 0.000 & 0.000 $\pm$ 0.000 & 0.022 $\pm$ 0.007 & 11.811 $\pm$ 0.293 \\ \hline
C1 & 0.446 $\pm$ 0.045 & 0.140 $\pm$ 0.004 & 117.663 $\pm$ 2.418 & 0.064 $\pm$ 0.012 & 11.009 $\pm$ 0.168 \\ \hline
C2 & 0.451 $\pm$ 0.046 & 0.248 $\pm$ 0.007 & 101.853 $\pm$ 2.879 & 0.113 $\pm$ 0.016 & 10.520 $\pm$ 0.132 \\ \hline
\multicolumn{6}{c}{\textbf{MJD : 59782.0}} \\ \hline
C0 & 0.282 $\pm$ 0.029 & 0.000 $\pm$ 0.000 & 0.000 $\pm$ 0.000 & 0.039 $\pm$ 0.010 & 11.240 $\pm$ 0.233 \\ \hline
C1 & 0.511 $\pm$ 0.052 & 0.158 $\pm$ 0.007 & 119.655 $\pm$ 3.417 & 0.102 $\pm$ 0.015 & 10.663 $\pm$ 0.135 \\ \hline
C2 & 0.371 $\pm$ 0.038 & 0.259 $\pm$ 0.010 & 100.226 $\pm$ 4.201 & 0.145 $\pm$ 0.020 & 10.218 $\pm$ 0.125 \\ \hline
\multicolumn{6}{c}{\textbf{MJD : 59812.0}} \\ \hline
C0 & 0.405 $\pm$ 0.041 & 0.000 $\pm$ 0.000 & 0.000 $\pm$ 0.000 & 0.060 $\pm$ 0.012 & 11.023 $\pm$ 0.176 \\ \hline
C1 & 0.933 $\pm$ 0.094 & 0.167 $\pm$ 0.005 & 112.119 $\pm$ 2.710 & 0.115 $\pm$ 0.014 & 10.820 $\pm$ 0.115 \\ \hline
C2 & 0.359 $\pm$ 0.037 & 0.320 $\pm$ 0.012 & 93.588 $\pm$ 4.279 & 0.171 $\pm$ 0.022 & 10.061 $\pm$ 0.119 \\ \hline
\multicolumn{6}{c}{\textbf{MJD : 59884.0}} \\ \hline
C0 & 0.294 $\pm$ 0.030 & 0.000 $\pm$ 0.000 & 0.000 $\pm$ 0.000 & 0.061 $\pm$ 0.013 & 10.869 $\pm$ 0.188 \\ \hline
C1 & 0.585 $\pm$ 0.059 & 0.185 $\pm$ 0.009 & 112.262 $\pm$ 4.251 & 0.144 $\pm$ 0.018 & 10.422 $\pm$ 0.115 \\ \hline
C2 & 0.214 $\pm$ 0.024 & 0.337 $\pm$ 0.032 & 82.664 $\pm$ 10.565 & 0.294 $\pm$ 0.033 & 9.365 $\pm$ 0.109 \\ \hline
\multicolumn{6}{c}{\textbf{MJD : 59903.0}} \\ \hline
C0 & 0.297 $\pm$ 0.031 & 0.000 $\pm$ 0.000 & 0.000 $\pm$ 0.000 & 0.071 $\pm$ 0.014 & 10.742 $\pm$ 0.176 \\ \hline
C1 & 0.526 $\pm$ 0.053 & 0.183 $\pm$ 0.007 & 112.091 $\pm$ 3.559 & 0.117 $\pm$ 0.016 & 10.556 $\pm$ 0.127 \\ \hline
C2 & 0.255 $\pm$ 0.028 & 0.329 $\pm$ 0.033 & 82.486 $\pm$ 10.981 & 0.325 $\pm$ 0.034 & 9.354 $\pm$ 0.102 \\ \hline
\multicolumn{6}{c}{\textbf{MJD : 59919.0}} \\ \hline
C0 & 0.246 $\pm$ 0.026 & 0.000 $\pm$ 0.000 & 0.000 $\pm$ 0.000 & 0.043 $\pm$ 0.011 & 11.096 $\pm$ 0.230 \\ \hline
C1 & 0.557 $\pm$ 0.056 & 0.178 $\pm$ 0.009 & 110.066 $\pm$ 4.598 & 0.144 $\pm$ 0.018 & 10.401 $\pm$ 0.116 \\ \hline
C2 & 0.213 $\pm$ 0.024 & 0.357 $\pm$ 0.044 & 77.872 $\pm$ 13.072 & 0.371 $\pm$ 0.038 & 9.161 $\pm$ 0.102 \\ \hline
\multicolumn{6}{c}{\textbf{MJD : 59986.0}} \\ \hline
C0 & 0.321 $\pm$ 0.033 & 0.000 $\pm$ 0.000 & 0.000 $\pm$ 0.000 & 0.036 $\pm$ 0.010 & 11.366 $\pm$ 0.234 \\ \hline
C1 & 0.714 $\pm$ 0.072 & 0.191 $\pm$ 0.008 & 111.468 $\pm$ 3.719 & 0.146 $\pm$ 0.017 & 10.497 $\pm$ 0.111 \\ \hline
C2 & 0.220 $\pm$ 0.025 & 0.455 $\pm$ 0.049 & 68.175 $\pm$ 9.788 & 0.374 $\pm$ 0.038 & 9.168 $\pm$ 0.101 \\ \hline
\multicolumn{6}{c}{\textbf{MJD : 60036.0}} \\ \hline
C0 & 0.317 $\pm$ 0.033 & 0.000 $\pm$ 0.000 & 0.000 $\pm$ 0.000 & 0.092 $\pm$ 0.016 & 10.545 $\pm$ 0.155 \\ \hline
C1 & 0.494 $\pm$ 0.050 & 0.195 $\pm$ 0.009 & 110.384 $\pm$ 4.494 & 0.144 $\pm$ 0.018 & 10.349 $\pm$ 0.119 \\ \hline
C2 & 0.153 $\pm$ 0.019 & 0.518 $\pm$ 0.063 & 59.074 $\pm$ 9.401 & 0.352 $\pm$ 0.039 & 9.063 $\pm$ 0.111 \\ \hline
\multicolumn{6}{c}{\textbf{MJD : 60085.0}} \\ \hline
C0 & 0.412 $\pm$ 0.042 & 0.000 $\pm$ 0.000 & 0.000 $\pm$ 0.000 & 0.052 $\pm$ 0.011 & 11.155 $\pm$ 0.187 \\ \hline
C1 & 0.625 $\pm$ 0.063 & 0.199 $\pm$ 0.009 & 109.073 $\pm$ 4.143 & 0.153 $\pm$ 0.018 & 10.398 $\pm$ 0.111 \\ \hline
C2 & 0.218 $\pm$ 0.024 & 0.517 $\pm$ 0.055 & 58.523 $\pm$ 7.994 & 0.368 $\pm$ 0.038 & 9.178 $\pm$ 0.102 \\ \hline
\multicolumn{6}{c}{\textbf{MJD : 60096.0}} \\ \hline
C0 & 0.313 $\pm$ 0.032 & 0.000 $\pm$ 0.000 & 0.000 $\pm$ 0.000 & 0.075 $\pm$ 0.014 & 10.717 $\pm$ 0.169 \\ \hline
C1 & 0.502 $\pm$ 0.051 & 0.197 $\pm$ 0.011 & 108.343 $\pm$ 5.393 & 0.169 $\pm$ 0.020 & 10.217 $\pm$ 0.112 \\ \hline
C2 & 0.140 $\pm$ 0.018 & 0.540 $\pm$ 0.066 & 57.635 $\pm$ 9.055 & 0.341 $\pm$ 0.039 & 9.052 $\pm$ 0.115 \\ \hline
\multicolumn{6}{c}{\textbf{MJD : 60125.0}} \\ \hline
C0 & 0.314 $\pm$ 0.032 & 0.000 $\pm$ 0.000 & 0.000 $\pm$ 0.000 & 0.058 $\pm$ 0.012 & 10.942 $\pm$ 0.190 \\ \hline
C1 & 0.541 $\pm$ 0.055 & 0.190 $\pm$ 0.012 & 109.674 $\pm$ 6.115 & 0.190 $\pm$ 0.021 & 10.147 $\pm$ 0.106 \\ \hline
C2 & 0.173 $\pm$ 0.021 & 0.518 $\pm$ 0.066 & 58.581 $\pm$ 9.744 & 0.387 $\pm$ 0.041 & 9.034 $\pm$ 0.105 \\ \hline
\multicolumn{6}{c}{\textbf{MJD : 60140.0}} \\ \hline
C0 & 0.434 $\pm$ 0.044 & 0.000 $\pm$ 0.000 & 0.000 $\pm$ 0.000 & 0.072 $\pm$ 0.013 & 10.895 $\pm$ 0.161 \\ \hline
C1 & 0.637 $\pm$ 0.064 & 0.189 $\pm$ 0.010 & 108.818 $\pm$ 5.138 & 0.177 $\pm$ 0.020 & 10.280 $\pm$ 0.106 \\ \hline
C2 & 0.235 $\pm$ 0.026 & 0.499 $\pm$ 0.056 & 55.610 $\pm$ 8.044 & 0.380 $\pm$ 0.038 & 9.183 $\pm$ 0.099 \\ \hline
\multicolumn{6}{c}{\textbf{MJD : 60149.0}} \\ \hline
C0 & 0.462 $\pm$ 0.047 & 0.000 $\pm$ 0.000 & 0.000 $\pm$ 0.000 & 0.057 $\pm$ 0.011 & 11.125 $\pm$ 0.175 \\ \hline
C1 & 0.547 $\pm$ 0.055 & 0.191 $\pm$ 0.009 & 109.573 $\pm$ 4.517 & 0.149 $\pm$ 0.018 & 10.363 $\pm$ 0.115 \\ \hline
C2 & 0.204 $\pm$ 0.023 & 0.506 $\pm$ 0.057 & 56.121 $\pm$ 8.132 & 0.360 $\pm$ 0.038 & 9.169 $\pm$ 0.104 \\ \hline
\multicolumn{6}{c}{\textbf{MJD : 60230.0}} \\ \hline
C0 & 0.438 $\pm$ 0.044 & 0.000 $\pm$ 0.000 & 0.000 $\pm$ 0.000 & 0.041 $\pm$ 0.009 & 11.388 $\pm$ 0.205 \\ \hline
C1 & 0.719 $\pm$ 0.072 & 0.207 $\pm$ 0.010 & 108.872 $\pm$ 4.808 & 0.192 $\pm$ 0.020 & 10.262 $\pm$ 0.101 \\ \hline
C2 & 0.265 $\pm$ 0.029 & 0.514 $\pm$ 0.062 & 56.094 $\pm$ 8.741 & 0.442 $\pm$ 0.041 & 9.104 $\pm$ 0.093 \\ \hline
\multicolumn{6}{c}{\textbf{MJD : 60238.0}} \\ \hline
C0 & 0.542 $\pm$ 0.055 & 0.000 $\pm$ 0.000 & 0.000 $\pm$ 0.000 & 0.057 $\pm$ 0.011 & 11.194 $\pm$ 0.169 \\ \hline
C1 & 0.792 $\pm$ 0.080 & 0.213 $\pm$ 0.009 & 109.201 $\pm$ 4.108 & 0.181 $\pm$ 0.019 & 10.355 $\pm$ 0.101 \\ \hline
C2 & 0.287 $\pm$ 0.031 & 0.527 $\pm$ 0.056 & 56.370 $\pm$ 7.730 & 0.423 $\pm$ 0.039 & 9.177 $\pm$ 0.093 \\ \hline
\multicolumn{6}{c}{\textbf{MJD : 60253.0}} \\ \hline
C0 & 0.487 $\pm$ 0.049 & 0.000 $\pm$ 0.000 & 0.000 $\pm$ 0.000 & 0.038 $\pm$ 0.009 & 11.500 $\pm$ 0.208 \\ \hline
C1 & 0.722 $\pm$ 0.073 & 0.224 $\pm$ 0.008 & 109.326 $\pm$ 3.403 & 0.154 $\pm$ 0.018 & 10.455 $\pm$ 0.108 \\ \hline
C2 & 0.292 $\pm$ 0.031 & 0.501 $\pm$ 0.059 & 56.151 $\pm$ 8.520 & 0.444 $\pm$ 0.040 & 9.142 $\pm$ 0.092 \\ \hline
\multicolumn{6}{c}{\textbf{MJD : 60272.0}} \\ \hline
C0 & 0.350 $\pm$ 0.036 & 0.000 $\pm$ 0.000 & 0.000 $\pm$ 0.000 & 0.028 $\pm$ 0.008 & 11.621 $\pm$ 0.258 \\ \hline
C1 & 0.563 $\pm$ 0.057 & 0.225 $\pm$ 0.010 & 108.355 $\pm$ 4.181 & 0.162 $\pm$ 0.019 & 10.303 $\pm$ 0.111 \\ \hline
C2 & 0.240 $\pm$ 0.026 & 0.498 $\pm$ 0.067 & 56.470 $\pm$ 9.719 & 0.446 $\pm$ 0.042 & 9.053 $\pm$ 0.095 \\ \hline
\multicolumn{6}{c}{\textbf{MJD : 60289.0}} \\ \hline
C0 & 0.334 $\pm$ 0.034 & 0.000 $\pm$ 0.000 & 0.000 $\pm$ 0.000 & 0.038 $\pm$ 0.010 & 11.336 $\pm$ 0.226 \\ \hline
C1 & 0.448 $\pm$ 0.046 & 0.233 $\pm$ 0.010 & 110.067 $\pm$ 4.163 & 0.149 $\pm$ 0.019 & 10.277 $\pm$ 0.119 \\ \hline
C2 & 0.219 $\pm$ 0.025 & 0.464 $\pm$ 0.067 & 56.961 $\pm$ 10.613 & 0.431 $\pm$ 0.042 & 9.043 $\pm$ 0.097 \\ \hline
\multicolumn{6}{c}{\textbf{MJD : 60351.0}} \\ \hline
C0 & 0.509 $\pm$ 0.051 & 0.000 $\pm$ 0.000 & 0.000 $\pm$ 0.000 & 0.057 $\pm$ 0.011 & 11.167 $\pm$ 0.172 \\ \hline
C1 & 0.544 $\pm$ 0.055 & 0.216 $\pm$ 0.009 & 106.420 $\pm$ 4.163 & 0.152 $\pm$ 0.018 & 10.344 $\pm$ 0.114 \\ \hline
C2 & 0.368 $\pm$ 0.038 & 0.448 $\pm$ 0.053 & 52.338 $\pm$ 7.937 & 0.440 $\pm$ 0.039 & 9.251 $\pm$ 0.088 \\ \hline
\multicolumn{6}{c}{\textbf{MJD : 60366.0}} \\ \hline
C0 & 0.359 $\pm$ 0.037 & 0.000 $\pm$ 0.000 & 0.000 $\pm$ 0.000 & 0.071 $\pm$ 0.013 & 10.824 $\pm$ 0.168 \\ \hline
C1 & 0.470 $\pm$ 0.048 & 0.207 $\pm$ 0.012 & 105.376 $\pm$ 5.931 & 0.183 $\pm$ 0.021 & 10.119 $\pm$ 0.110 \\ \hline
C2 & 0.273 $\pm$ 0.029 & 0.488 $\pm$ 0.054 & 60.809 $\pm$ 8.714 & 0.416 $\pm$ 0.039 & 9.170 $\pm$ 0.094 \\ \hline
\multicolumn{6}{c}{\textbf{MJD : 60391.0}} \\ \hline
C0 & 0.257 $\pm$ 0.027 & 0.000 $\pm$ 0.000 & 0.000 $\pm$ 0.000 & 0.049 $\pm$ 0.012 & 11.001 $\pm$ 0.214 \\ \hline
C1 & 0.231 $\pm$ 0.025 & 0.222 $\pm$ 0.013 & 114.721 $\pm$ 5.185 & 0.126 $\pm$ 0.020 & 10.135 $\pm$ 0.146 \\ \hline
C2 & 0.195 $\pm$ 0.023 & 0.484 $\pm$ 0.078 & 57.492 $\pm$ 11.911 & 0.462 $\pm$ 0.044 & 8.932 $\pm$ 0.098 \\ \hline
\multicolumn{6}{c}{\textbf{MJD : 60455.0}} \\ \hline
C0 & 0.322 $\pm$ 0.033 & 0.000 $\pm$ 0.000 & 0.000 $\pm$ 0.000 & 0.038 $\pm$ 0.010 & 11.320 $\pm$ 0.228 \\ \hline
C1 & 0.280 $\pm$ 0.029 & 0.225 $\pm$ 0.009 & 112.698 $\pm$ 3.702 & 0.105 $\pm$ 0.017 & 10.376 $\pm$ 0.151 \\ \hline
C2 & 0.305 $\pm$ 0.032 & 0.466 $\pm$ 0.054 & 60.301 $\pm$ 9.046 & 0.438 $\pm$ 0.040 & 9.173 $\pm$ 0.091 \\ \hline
\multicolumn{6}{c}{\textbf{MJD : 60469.0}} \\ \hline
C0 & 0.437 $\pm$ 0.044 & 0.000 $\pm$ 0.000 & 0.000 $\pm$ 0.000 & 0.041 $\pm$ 0.009 & 11.387 $\pm$ 0.206 \\ \hline
C1 & 0.396 $\pm$ 0.040 & 0.228 $\pm$ 0.009 & 112.968 $\pm$ 3.449 & 0.119 $\pm$ 0.017 & 10.418 $\pm$ 0.133 \\ \hline
C2 & 0.404 $\pm$ 0.042 & 0.496 $\pm$ 0.047 & 60.126 $\pm$ 7.393 & 0.449 $\pm$ 0.038 & 9.274 $\pm$ 0.087 \\ \hline
\multicolumn{6}{c}{\textbf{MJD : 60494.0}} \\ \hline
C0 & 0.868 $\pm$ 0.087 & 0.000 $\pm$ 0.000 & 0.000 $\pm$ 0.000 & 0.043 $\pm$ 0.008 & 11.643 $\pm$ 0.173 \\ \hline
C1 & 0.589 $\pm$ 0.059 & 0.241 $\pm$ 0.003 & 111.448 $\pm$ 0.983 & 0.053 $\pm$ 0.010 & 11.293 $\pm$ 0.171 \\ \hline
C2 & 0.653 $\pm$ 0.066 & 0.484 $\pm$ 0.035 & 57.919 $\pm$ 5.471 & 0.441 $\pm$ 0.035 & 9.498 $\pm$ 0.082 \\ \hline
\multicolumn{6}{c}{\textbf{MJD : 60553.0}} \\ \hline
C0 & 0.396 $\pm$ 0.040 & 0.000 $\pm$ 0.000 & 0.000 $\pm$ 0.000 & 0.035 $\pm$ 0.009 & 11.481 $\pm$ 0.226 \\ \hline
C1 & 0.374 $\pm$ 0.038 & 0.233 $\pm$ 0.008 & 112.714 $\pm$ 2.941 & 0.103 $\pm$ 0.016 & 10.519 $\pm$ 0.143 \\ \hline
C2 & 0.423 $\pm$ 0.044 & 0.518 $\pm$ 0.044 & 58.335 $\pm$ 6.470 & 0.431 $\pm$ 0.037 & 9.329 $\pm$ 0.087 \\ \hline
\label{tab:model_comp_all}
\end{longtable}

\end{document}